\documentclass[aps,prd,nofootinbib,superscriptaddress,reprint,onecolumn,notitlepage,11pt,twoside]{revtex4-1}

\usepackage[includeheadfoot,paperheight = 269mm,paperwidth = 190mm,top=18mm, left = 28mm, right=18mm, bottom=22mm]{geometry}
\usepackage{graphicx}
\usepackage{amsfonts,amssymb,amsmath}
\usepackage{hyperref}
\usepackage{color}
\usepackage{eimodels}
\usepackage{eivars}
\usepackage{xspace}
\usepackage{fancyhdr}
\usepackage{bm}

\hypersetup{
    colorlinks=true,       
    linkcolor=red,          
    citecolor=cyan,        
    filecolor=magenta,      
    urlcolor=blue,           
}

\newcommand{\BibitemShut}[1]{}

\newcommand{\ie}{i.e.\xspace}

\newlength{\wsingfig}
\newlength{\wdblefig}
\newlength{\wfull}
\newlength{\hfull}
\newcommand{\sss}[1]{{\scriptscriptstyle{#1}}}

\newcommand{\usssREF}{\sss{\mathrm{REF}}}

\newcommand{\calP}{\mathcal{P}}

\newcommand{\calO}{\mathcal{O}}

\newcommand{\calL}{\mathcal{L}}

\newcommand{\dd}{\mathrm{d}} 

\newcommand{\eV}{\mathrm{eV}}

\newcommand{\MeV}{\mathrm{MeV}}
\newcommand{\GeV}{\mathrm{GeV}}

\newcommand{\TeV}{\mathrm{TeV}}

\newcommand{\GN}{G_{_{\mathrm N}}}

\newcommand{\mpl}{m_\usssPl}
\newcommand{\Mp}{M_\usssPl}

\def\beq{\begin{equation}}
\def\eeq{\end{equation}} 
\def\bea{\begin{eqnarray}}
\def\eea{\end{eqnarray}}
\def\benu{\begin{enumerate}}
\def\eenu{\end{enumerate}}

\newcommand{\fnl}{f_{_\uNL}}

\newcommand{\BayesFactor}[2]{B^{#1}_{#2}}
\newcommand{\Bref}[1]{\BayesFactor{#1}{\usssREF}}

\newcounter{contribution}
\numberwithin{equation}{contribution}

\begin{document}
\setcounter{page}{3}


\title{Cosmic Inflation: Trick or Treat?}

\author{J\'er\^ome Martin\footnote{E-mail: jmartin@iap.fr}}
\affiliation{Institut d'Astrophysique de Paris, UMR 7095-CNRS, 98bis
  boulevard Arago, 75014 Paris, France}

\begin{abstract}
  Discovered almost forty years ago, inflation has become the leading
  paradigm for the early universe. Originally invented to avoid the
  fine-tuning puzzles of the standard model of cosmology, the
  so-called hot Big Bang phase, inflation has always been the subject
  of intense debates. In this article, after a brief review of the
  theoretical and observational status of inflation, we discuss the
  criticisms that have been expressed against it and attempt to assess
  whether inflation can really be viewed as a successful solution to
  the above mentioned issues.
\end{abstract}

\maketitle


\tableofcontents

\section{Introduction}
\label{sec:intro}

The theory of cosmic inflation was invented to solve fine-tuning
problems~\cite{Starobinsky:1979ty,Starobinsky:1980te,
  Guth:1980zm,Linde:1981mu,Mukhanov:1981xt,Mukhanov:1982nu,
  Starobinsky:1982ee}. Indeed, the pre-inflationary standard model of
cosmology, the hot Big Bang
model~\cite{Weinberg:100595,Peter:1208401}, suffers from a number of
issues all related to a fragile adjustment of the initial conditions
needed to make it work. For instance, it is well-known that, in a
cosmological model without inflation, when one looks at the last
scattering surface (lss) where the Cosmic Microwave Background (CMB)
radiation was emitted, one looks at different causally disconnected
patches of the universe. But, despite being causally disconnected,
they all share, approximately, the same temperature. Unless one
fine tunes artificially the initial conditions, this fact is not
understandable.

Soon after its advent, it was also realized that inflation provides a
mechanism for structure
formation~\cite{Mukhanov:1981xt,Mukhanov:1982nu,
  Starobinsky:1982ee}. In brief, the unavoidable vacuum quantum
fluctuations of the gravitational and inflaton fields are stretched
over cosmological distances by the inflationary cosmic expansion and
are amplified by gravitational instability to eventually give rise to
the large scale structures observed in our universe and to the CMB
temperature anisotropy. This simple idea implies a series of
remarkable predictions among which is the fact that the cosmological
perturbations spend time outside the Hubble radius, implying the
disappearance of the decaying mode and the presence of coherent
oscillations in the CMB power spectrum, or the fact that the two-point
correlation function of the inflationary fluctuations should be close
to scale invariance.

In $1992$ the CMB anisotropies were discovered by the COsmic
Background Explorer (COBE)
satellite~\cite{Smoot:1992td,Bennett:1996ce} and this marked the
beginning of a very important experimental effort by the international
community to measure, with a high accuracy, these anisotropies in
order to constrain the physics of the early universe. This culminated
recently with the publication of the Planck data which is a cosmic
variance limited
experiment~\cite{Ade:2013zuv,Ade:2013sjv,Planck:2013jfk,Ade:2013ydc,Ade:2015tva,Ade:2015xua,Ade:2015lrj,Aghanim:2018eyx,Akrami:2018odb,Akrami:2018vks}. The
results of these $30$ years of experimental work is consistent with
the predictions of single field slow-roll inflation with a minimal
kinetic term. It is worth emphasizing that, in some cases, what has
been confirmed are predictions and not postdictions. In particular,
the prediction that the scalar spectral index should be close but not
equal to one has been shown to be true at more than five sigmas by the
Planck experiment since $\nS=0.9645\pm 0.0049$~\cite{Ade:2015lrj}.

Despite these important successes and despite the fact that it has
become the leading paradigm for the early universe, inflation has
always been the subject of doubts and
criticisms~\cite{Penrose:1988mg,Peebles:1999xc,Ijjas:2013vea,Ijjas:2014nta}. Soon
after its invention, two questions were mainly discussed, the choice
of the inflationary parameters (for instance the coupling constant in
the potential) needed to match the level of CMB anisotropies, a
question related to model building and to the physical nature of the
inflaton field, and the question of initial conditions at the
beginning of inflation. Another issue, the graceful exit or how to
stop inflation, was also a hot topic but, apparently, the theory of
reheating (and then preheating) gave a satisfactory
answer~\cite{Turner:1983he,Traschen:1990sw,Kofman:1997yn,Amin:2014eta}. But
the two first questions remain debated. In addition, in conjunction
with the experimental efforts mentioned above, various theoretical
developments also took place. In particular, it was realized that
single field slow-roll models are not the only way to realize
inflation and, gradually, a large zoo of models started to appear on
stage~\cite{Starobinsky:1992ts,ArmendarizPicon:2000ah,Wands:2007bd,Avila:2013ela,Martin:2013tda}. Importantly,
some of these scenarios make different predictions that single field
slow-roll inflation. For instance, the level of Non-Gaussianity (NG),
which is negligible for single field slow-roll models, can be
significant for a model with a non-minimal kinetic term.

Another major theoretical development is the claim that inflation can
be
eternal~\cite{Linde:1982ur,Steinhardt:1982kg,Linde:1986fd,Linde:1993xx,Guth:2007ng,Guth:2000ka,Linde:2015edk}. This
is based on the fact that, due to quantum fluctuations, the various
causally disconnected patches that are produced during inflation can
be such that the value of the inflaton field is different from one
patch to another. In particular, there can be patches where, due to
quantum fluctuations, the field climbs its potential instead of
rolling it down as it does classically. And, as a consequence, this
means that there are patches where inflation never stops. This idea,
coupled to the concept of a string landscape, leads to the multiverse,
an idea which is nowadays the subject of hot discussions.

The aim of this article is to review the present status of cosmic
inflation and to assess whether it can be considered as successful
given the assumptions on which it rests and given what it has
achieved. In particular, we discuss whether, driven out by the door,
fine-tuning problems do not simply slip in again by the window under a
different name. A warning is also in order at this stage. In this
manuscript, we will use the word ``fine-tuning'' in a loose sense and
will not attempt to define this concept very rigorously. In fact, this
question is related to a more general one, namely what are the
measures relevant for inflation and how they can be justified. This is
important, for instance, for the flatness problem or for the problem
of initial conditions. However, here, we will say very little about it
and we refer the reader to Ref.~\cite{Chowdhury:2019otk} where these
issues are discussed in great detail.

The article is organized as follows. In the next section,
Sec.~\ref{sec:standard}, we briefly present the, pre-inflationary,
standard model of cosmology, namely the hot Big Bang model. We first
discuss its theoretical foundations in Sec.~\ref{subsec:rela} and,
then, in Sec.~\ref{subsec:realuniverse}, how astrophysical
observations can constrain it. In Sec.~\ref{sec:puzzles}, we review
the difficulties of this model, in particular the horizon problem, see
Sec.~\ref{subsec:horizon} and the flatness problem, see
Sec.~\ref{subsec:flatness}. In Sec.~\ref{sec:inf}, we introduce
inflation and discuss how it can solve the above mentioned puzzles in
Sec.~\ref{subsec:solving}. In Sec.~\ref{subsec:realizing}, we study
how it can be realized in practice and show that the presence of a
scalar field dominating the energy budget of the universe is a likely
possibility. In Sec.~\ref{subsec:pert}, we present the theory of
inflationary cosmological perturbations of quantum-mechanical origin
which is at the heart of the calculation of CMB anisotropy. In
Sec.~\ref{subsec:constraints}, we briefly review the consequences for
inflation of the recently released Planck data. In Sec.~\ref{sec:ft},
we discuss whether inflation is a fine-tuned scenario, in particular
we address the question of whether the choices of the parameters
needed in order to have a satisfactory model of inflation is
``natural''. Then, in Sec.~\ref{sec:ini}, we discuss the initial
conditions at the beginning of inflation, first in an homogeneous and
isotropic situation in Sec.~\ref{subsec:homoini}, then in an
homogeneous but anisotropic situation in Sec.~\ref{subsec:aniini} and,
finally, in a general inhomogeneous situation in
Sec.~\ref{subsec:inhomoini}. We also consider the question of initial
conditions for the quantum perturbations, the so-called
trans-Planckian problem of inflation in Sec.~\ref{subsec:inipert}. In
Sec.~\ref{sec:multi}, we discuss various aspects of the multiverse
question. In Sec.~\ref{subsec:stochainf}, we explain stochastic
inflation and in Sec.~\ref{subsec:eternal}, we show how the
backreaction is usually taken into account leading to the concept of
an eternal inflating universe. In Sec.~\ref{subsec:avoid}, we point
out that there are models where inflation is not eternal and in
Sec.~\ref{subsec:threat?}, we discuss the consequences of the possible
existence of a multiverse for inflation itself. Finally, in
Sec.~\ref{sec:conclusion}, we present our conclusions.

\section{The Standard Model of Cosmology}
\label{sec:standard}

\subsection{Relativistic Cosmology}
\label{subsec:rela}

Inflation is supposed to be a solution to some issues of the standard
model of cosmology. In order to understand why this is the case,
clearly, it is necessary to start with a presentation of the standard
model itself. Only after having understood its main features, will it
be possible to appreciate its unsatisfactory aspects.

The shape of the Universe is controlled by gravity which, in General
Relativity, is described by a metric tensor
$g_{\mu \nu}\left(x^{\kappa}\right)$. The action of the system is
given by
\begin{equation}
\label{eq:EHaction}
S=-\frac{c^4}{16\pi \GN}\int {\rm d}^4 x\sqrt{-g}
\left(R+2\Lambda_{_{\rm B}} \right)+S_{\rm matter}.
\end{equation}
This so-called Einstein-Hilbert action involves two fundamental
constants, the speed of light
$c=3\times 10^8 \, \mbox{m}\cdot \mbox{s}^{-1}$ and the Newton
constant
$\GN=6.67\times 10^{-11} \mbox{m}^3\cdot \mbox{kg}^{-1}\cdot
\mbox{s}^{-2}$,
as appropriate for a relativistic theory of the gravitational
field. Quantum effects, which are controlled by the Planck constant,
$\hbar=1.05\times 10^{-34}\mbox{m}^2\cdot \mbox{kg}\cdot
\mbox{s}^{-1}$,
are not needed to describe the dynamics of background spacetime. But,
as we will see, they play a fundamental role at the perturbative
level. In the following, we will work in terms of natural units for
which $\hbar=c=1$. In this system of units, everything can be
expressed in terms of energy, in particular $\mpl=1/\sqrt{\GN}$ where
$\mpl$ is known as the Planck mass,
$\mpl\equiv \sqrt{\hbar c/\GN}=2.17 \times 10^{-8}\mbox{kg}$. We will
also use the reduced Planck mass defined by
$\Mp\equiv \mpl/\sqrt{8\pi}=2.43\times 10^{18}\mbox{GeV}$.

Let us now describe the quantities appearing in the
action~(\ref{eq:EHaction}). $g$ denotes the determinant of the metric
tensor $g_{\mu \nu}\left(x^{\kappa}\right)$.
$R\equiv g^{\mu \nu}R_{\mu \nu}$ is the scalar curvature where
$R_{\mu \nu}=R^{\alpha }{}_{\mu \alpha \nu }$ denotes the Ricci tensor
which is a contraction of the Riemann tensor. Finally, the quantity
$\Lambda _{_{\rm B}}$ is the bare cosmological constant. Clearly $R$
and, therefore the cosmological constant $\Lambda_{_{\rm B}}$ are of
dimension two, $[R]=[\Lambda_{_{\rm B}}]=2$ (writing the natural
dimension of a quantity within square bracket).

One can then obtain the equation of motion by varying the
action~(\ref{eq:EHaction}) with respect to the metric tensor. The
result reads
\begin{equation}
  G_{\mu \nu}+\Lambda_{_{\rm B}} g_{\mu \nu}=R_{\mu \nu}-\frac{1}{2}Rg_{\mu \nu}
+\Lambda_{_{\rm B}} g_{\mu \nu}
  =\frac{1}{\Mp^2}T_{\mu \nu}\, ,
\end{equation}
where we have defined the stress-energy tensor which describes the
matter distribution responsible for the curvature of spacetime by the
following expression
\begin{equation}
T_{\mu \nu}\equiv-\frac{2}{\sqrt{-g}}
\frac{\delta S_{\rm matter}}{\delta g^{\mu \nu}}.
\end{equation}
Conservation of energy amounts to $\nabla_{\alpha}T^{\alpha \mu}=0$,
where $\nabla_{\alpha}$ denotes the covariant derivative.
Let us notice that energy conservation is compatible with the Bianchi
identities, $\nabla _{\alpha}G^{\alpha \mu}=0$ and the fact that the
metric tensor has also a vanishing covariant derivative. We see that
the Einstein equations are a priori very complicated since they are
partial, second order and non linear differential equations for the
metric tensor.

However, the cosmological principle states that the Universe is, on
large scales, homogeneous and isotropic. Of course, this assumption is
not obvious a priori and must be carefully observationally checked. We
refer the reader to Ref.~\cite{Maartens:2011yx} where this point is
discussed in details. Moreover, it must also be explained, rather than
postulated, since it would be rather contrived to assume that the
initial state was so peculiar. We will of course come back to this
question at length in the following sections since inflation is a
scenario where this question can, in principle, be addressed. As a
consequence of the cosmological principle, the metric tensor takes the
Friedmann-Lema\^{\i}tre-Robertson-Walker (FLRW) form, namely
\begin{equation}
  \label{eq:metric}
  {\rm d}s^2=g_{\mu \nu}{\rm d}x^{\mu }{\rm d}x^{\nu }
=-{\rm d}t^2+a^2(t)\gamma _{ij}^{(3)}{\rm d}x^i{\rm d}x^j,
\end{equation}
where $t$ is the cosmic time and $x^i$ are space-like coordinates. The
quantity $\gamma _{ij}^{(3)}$ is the metric of the three-dimensional
spacelike sections which have a constant scalar curvature. From the
above equation, we have the relation
$g_{ij}=a^2(t)\gamma_{ij}^{(3)}$. In polar coordinates, the
three-dimensional metric can be written as
\begin{equation}
\label{eq:threemetricpolar}
\gamma _{ij}^{(3)}{\rm d}x^i{\rm d}x^j=\biggl[\frac{{\rm d}r^2}
{1-{\cal K}r^2} 
+r^2\left({\rm d}\theta ^2+\sin ^2\theta {\rm d}
\varphi ^2\right)\biggr]\, , 
\end{equation}
while in Cartesian coordinates, it reads
\begin{equation}
\label{eq:threemetriccartesian}
\gamma _{ij}^{(3)}=\delta _{ij}
\biggl[1+\frac{{\cal K}}{4}\left(x^2+y^2+z^2\right)\biggr]^{-2}\, .
\end{equation}
The constant ${\cal K}$ describes the curvature of the spacelike
sections [since ${}^{(3)}R=6{\cal K}$, see below] and, without loss of
generality, can be chosen to be ${\cal K}=0,\pm 1$. As is apparent
from the previous equations, there is only one unknown function left,
the scale factor $a(t)$ and, moreover, this function is a function of
time only.

\par

On the other hand, matter is assumed to be a collection of $N$
perfect fluids and, as a consequence, its stress-energy tensor is
given by the following expression
\begin{equation}
T_{\mu \nu }=\sum _{i=1}^{i=N}T_{\mu \nu}^{(i)}=
\sum _{i=1}^{i=N}\left\{\left[\rho _i(t) +p_i(t)\right]u_{\mu
}u_{\nu}+p_i(t)g_{\mu \nu}\right\},
\end{equation}
where $\rho _i(t)$ and $p_i(t)$ are respectively the energy density
and pressure of the fluid ``$i$''. The vector $u_{\mu }$ is the four
velocity and satisfies the relation $u_{\mu }u^{\mu }=-1$. In terms of
cosmic time this means that $u^{\mu }=(1,0)$ and $u_{\mu }=(-1,0)$. In
accordance with the cosmological principle, the quantities
$\rho _i(t)$ and $p_i(t)$ only depend on time. In order to close the
system of equations, the relation between energy density and pressure,
namely the equation of state $p_i=w_i \left(\rho _i\right)$, must also
be provided.

We are now in a position to explicit Einstein equations. In the case
of a FLRW metric, one arrives at
\begin{align}
\label{eq:fried}
\frac{\dot{a}^2}{a^2}+\frac{{\cal K}}{a^2} &=
\frac{1}{3\Mp^2}\sum _{i=1}^{N}\rho _i+\frac{\Lambda _{_{\rm B}}}{3}\, , 
\\
-\biggl(2\frac{\ddot{a}}{a}+\frac{\dot{a}^2}{a^2}
+\frac{{\cal K}}{a^2}\biggr) 
& =\frac{1}{\Mp^2} \sum _{i=1}^Np_i-\Lambda_{_{\rm B}} \, .
\end{align}
We see that one has obtained ordinary, non linear, second order
differential equation for the scale factor $a(t)$. The fact that we
now deal with ordinary differential equation is of course due to the
cosmological principle and to the fact that the only unknown function
in the metric, the scale factor, is a function of time only.
Combining the two equations of motion obtained above, one gets an
equation which gives the acceleration of the scale factor, namely
\begin{equation}
\label{eq:accela}
\frac{\ddot{a}}{a}=-\frac{1}{6\Mp^2}\sum _{i=1}^N(\rho _i
+3p_i)+\frac{1}{3}\Lambda_{_{\rm B}}  .
\end{equation}
This equation is especially interesting because it provides the
condition leading to an accelerated expansion, namely
\begin{equation}
\label{eq:condddapositive}
\rho _{_{\rm T}}+3p_{_{\rm T}}<0 \, ,
\end{equation}
where $\rho_{_{\rm T}}=\sum_{i=1}^N\rho_i $ and
$p_{_{\rm T}}=\sum_{i=1}^Np_i $ denote the total energy density and
pressure (assuming a vanishing cosmological constant or including its
contribution in an extra fluid, see below). Since the energy density
of matter must be positive, we see that the above condition requires a
negative pressure, i.e. some exotic form of matter.

Even if the Einstein equations have been considerably simplified by
the use of the cosmological principle, they remain difficult to solve
analytically. However, it turns out that, if the curvature term
vanishes and if there is only one fluid with a constant equation of
state, an exact solution to the Einstein equations is available. Of
course, one can always solve these equations numerically, but exact
solutions will be interesting when we discuss the puzzles of the hot
Big Bang phase in the next sections. For this reason, we briefly
present them. Since the equation of state is supposed to be constant,
the conservation equation, which can be written as
\begin{equation}
\dot{\rho}+3H(1+w)\rho=0,
\end{equation}
can be integrated exactly and the solution reads
\begin{equation}
\rho(t) =\rho_{_{\rm f}}\left(\frac{a_{_{\rm f}}}{a}\right)^{3(1+w)},
\end{equation}
where $\rho_{_{\rm f}}$ and $a_{_{\rm f}}$ are the energy density and
the scale factor expressed at a fiducial time $t_{_{\rm f}}$ that
can be chosen arbitrarily. Then, one inserts the above result in the
Friedmann equation, namely
\begin{equation}
\left(\frac{1}{a}\frac{{\rm d}a}{{\rm d}t}\right)^2=
\frac{\rho_{_{\rm f}}}{3\Mp^2}\left(\frac{a_{_{\rm f}}}{a}\right)^{3(1+w)},
\end{equation}
whose solution can also be found and reads
\begin{equation}
\left(\frac{a}{a_{_{\rm f}}}\right)^{\frac{3(1+w)}{2}}
=\frac{3(1+w)}{2}\frac{\rho_{_{\rm f}}^{1/2}}{\sqrt{3}\Mp}t+C.
\end{equation}
In this expression $C$ is an integration constant. Requiring 
that $a=a_{_{\rm f}}$ when $t=t_{_{\rm f}}$, one finds that
$C=-3(1+w)\rho_{_{\rm f}}^{1/2}t_{_{\rm f}}/(2\sqrt{3}\Mp)+1$. Finally
noticing that $H_{_{\rm f}}=\rho_{_{\rm f}}^{1/2}/(\sqrt{3}\Mp)$, one
arrives at
\begin{equation}
\label{eq:scalefactorsol}
a(t)=a_{_{\rm f}}\left[\frac32(1+w)H_{_{\rm f}}\left(t-t_{_{\rm f}}\right)
+1\right]^{\frac{2}{3(1+w)}}.
\end{equation}
The corresponding Hubble parameter can be expressed as
$H(t)=H_{_{\rm f}}/[3(1+w)H_{_{\rm f}}\left(t-t_{_{\rm
      f}}\right)/2+1]$.
We notice that the scale factor vanishes when $t=t_{_{\rm BB}}$ with
$t_{_{\rm BB}}=t_{_{\rm f}}-2/[3(1+w)H_{_{\rm f}}]$. In some sense,
``time begins'' at $t_{_{\rm BB}}$ and it would be meaningless to
consider times such that $t<t_{_{\rm BB}}$. This is of course the
famous Big Bang point where the classical analysis breaks down. This
singularity is of course a serious problem for the hot Big Bang
model. However, it is not considered as a problem for inflation simply
because inflation does not aim at addressing it. It could be solved
if, prior to inflation, there is a
bounce~\cite{Battefeld:2014uga,Brandenberger:2016vhg} or if quantum
gravitational effects take over and somehow regularize the singularity
as done, for instance, in quantum cosmology~\cite{Hartle:1983ai}. We
see that the singularity problem can be treated separately and does
not involve the inflationary scenario.

For future convenience, it is also interesting to rewrite the scale
factor in terms of $t_{_{\rm BB}}$ and one obtains
$a(t)=a_{_{\rm f}}\left[\frac32(1+w)H_{_{\rm f}}\left(t-t_{_{\rm
        BB}}\right) \right]^{\frac{2}{3(1+w)}}$,
and $ H(t)=2/[3(1+w)\left(t-t_{_{\rm BB}}\right)]$. If, in addition,
one chooses $t_{_{\rm BB}}=0$ (which can always be done), then the
scale factor takes the form (using that, with this parameterization,
$H_{_{\rm f}}=2/\left[3(1+w)t_{_{\rm f}}\right]$)
\begin{equation}
\label{eq:scalefactorpower}
a(t)=a_{_{\rm f}}\left(\frac{t}{t_{_{\rm f}}}\right)^{\frac{2}{3(1+w)}},
\end{equation}
that is to say a power-law function. For radiation, $w=1/3$, the scale
factor behaves as $a(t)\propto t^{1/2}$ and for pressure-less matter,
$w=0$, one has $a(t)\propto t^{2/3}$. We also notice that the previous
expressions are ill-defined if $w=-1$.  This is just because in that
case we have an exponential solution, namely
$a(t)=a_{_{\rm f}}\exp\left[H_{_{\rm f}}\left(t-t_{_{\rm
        f}}\right)\right]$, known as the de Sitter solution.

Putting aside the particular case $w=-1$, let us finally come back to
the fact that, for $t=t_{_{\rm BB}}$, the scale factor vanishes. This
is clearly not an artifact of the coordinate system used, as is
confirmed by a calculation of the scalar curvature
\begin{equation}
R=\frac{4(1-3w)}{3(1+w)^2}\frac{1}{\left(t-t_{_{\rm BB}}\right)^2},
\end{equation}
which blows up when $t\rightarrow t_{_{\rm BB}}$. This confirms the
fact that $t=t_{_{\rm BB}}$ corresponds to a real
singularity\footnote{Notice also that for radiation $R$ is identically
  zero. Of course, this does not mean that there is no singularity in
  a radiation-dominated epoch. This can be shown by computing another
  invariant, for instance $R_{\mu \nu}R^{\mu \nu}$ which reads
\begin{align}
R_{\mu \nu}R^{\mu \nu}& =R_{00}R^{00}+R_{ij}R^{ij}
=9\left(\frac{\ddot{a}}{a}\right)^2
+\left(\frac{\ddot{a}}{a}+2\frac{\dot{a}^2}{a^2}\right)^2
g_{ij}g^{ij}
\nonumber \\ 
&=12\left(\frac{\ddot{a}}{a}\right)^2
+12\frac{\ddot{a}}{a}\frac{\dot{a}^2}{a^2}
+12\left(\frac{\dot{a}}{a}\right)^4
\nonumber \\
& =\frac{48(3w^2+1)}{27(1+w)^4}
\frac{1}{\left(t-t_{_{\rm BB}}\right)^4}.
\end{align}
Clearly, $R_{\mu \nu}R^{\mu \nu}$ blows up as
$t\rightarrow t_{_{\rm BB}}$ even if $w=1/3$}. 

Having introduced the theoretical tools needed in order to understand
the hot Big Bang model, we now discuss the parameters that describe
the model and how their values can be inferred from cosmological data.

\subsection{The Real Universe}
\label{subsec:realuniverse}

In order to describe our Universe, we need to know its energy budget,
namely the contribution of the different forms of energy density
present in the Universe. Our Universe is made of photons, with energy
density $\rho_{\gamma}$, neutrinos with energy density $\rho_{\nu}$,
baryons with energy density $\rho_{\rm b}$, cold dark matter with
energy density $\rho_{\rm c}$ and dark energy with energy density
$\rho_{_{\Lambda}}$ (here assumed to be a cosmological
constant). Photons and neutrinos have an equation of state $1/3$,
baryons and cold dark matter have a vanishing equation of state and,
finally, dark energy has an equation of state $-1$. We have therefore
three types of fluids, radiation
$\rho_{\rm r}=\rho_{\gamma}+\rho_{\nu}$, matter
$\rho_{\rm m}=\rho_{\rm b}+\rho_{\rm cdm}$ and dark energy
$\rho_{_{\Lambda}}$. Their relative importance must be inferred from
observations. In order to describe the results of those observations,
it is convenient to introduce new quantities. Let us first define the
critical energy density: in order to do so, we rewrite the Friedmann
equation, Eq.~(\ref{eq:fried}), as
\begin{equation}
H^2+\frac{{\cal K}}{a^2}=\frac{1}{3\Mp^2}\left(
\rho_{_{\Lambda}}+\sum_{i=1}^{i=N}\rho_i\right)
\end{equation}
with $\rho_{_{\Lambda}}=\Lambda_{_{\rm B}} \Mp^2$ the vacuum energy
density. We then define the critical energy density by
$\rho_{_{\rm cri}}\equiv 3H^2\Mp^2$, which is clearly a time-dependent
quantity. Then, the Friedmann equation can be rewritten as
\begin{align}
1+\frac{{\cal K}}{a^2H^2}=\frac{\rho_{_{\rm T}}}{\rho_{_{\rm cri}}},
\end{align}
where $\rho_{_{\rm T}}=\rho_{_{\Lambda}}+\sum_{i=1}^{i=N}\rho_i$ is
the total energy density [compared to the definition below
Eq.~(\ref{eq:condddapositive}), we have now explicitly included the
contribution of the cosmological constant in the total energy
density]. This means that, if the spatial curvature vanishes then
$\rho_{_{\rm T}}=\rho_{_{\rm cri}}$ and if ${\cal K}>0$ (respectively
${\cal K}<0$) then $\rho_{_{\rm T}}>\rho_{_{\rm cri}}$ (respectively
$\rho_{_{\rm T}}<\rho_{_{\rm cri}}$). One can also express the weight
of a given form of matter by the quantity $\Omega _i$ defined by
\begin{equation}
\Omega_i \equiv \frac{\rho_i}{\rho_{_{\rm cri}}},
\end{equation}
and, as a consequence, the Friedmann equation can be re-written as
\begin{equation}
\label{eq:omegatot}
1+\frac{{\cal K}}{a^2H^2}=\Omega_{_{\Lambda}}+\sum _{i=1}^{i=N}\Omega_i.
\end{equation}
In particular, if the spacelike sections are flat then the sum of all
the $\Omega_i$'s should be one. It follows from the previous
considerations that the contributions of the different forms of energy
density in our Universe are expressed trough
$\Omega _i^0=\rho_i^0/\rho_{_{\rm cri}}^0$, namely the quantity
$\Omega_i$ evaluated at present time.  The critical energy density
today is $\rho_{_{\rm cri}}^0=3H_0^2\Mp^2$ with
$H_0 =100h\, \mbox{km}\cdot \mbox{s}^{-1} \cdot \mbox{Mpc}^{-1}$,
where $h$ takes into account the uncertainty about $H_0$ (recent
measurements indicate that $h\simeq 0.67$~\cite{Ade:2015xua}).  $H_0$
has clearly the dimension of the inverse of a time (is of dimension
one) and the above strange units are used because of the measurement
of $H_0$ was historically performed using the Hubble
diagram~\cite{Perlmutter:1996ds,Perlmutter:1997zf,Kowalski:2008ez,Astier:2005qq}. In
standard units, one has $H_0=3.24h \times 10^{-18}\mbox{s}^{-1}$ while
in natural units $H_0=2.12h\times 10^{-42}\mbox{GeV}$.  Therefore, we
see that, by high energy standards, the current expansion of the
Universe is a low energy phenomenon. Given the value of the reduced
Planck mass, this implies that
$\rho_{_{\rm cri}}^0\simeq 8.0990 h^2 \times 10^{-47} \mbox{GeV}^4$.

Let us now describe the composition of our Universe. Data analysis is
complicated as it depends on which data sets is included in the
analysis. For the moment, let us say that the Planck $2013$ data plus
the WMAP data on large scale polarization imply
that~\cite{Ade:2013zuv,Ade:2013sjv,Planck:2013jfk,Ade:2013ydc}
\begin{equation}
\Omega_{\cal K}=-0.058^{+0.046}_{-0.026}.
\end{equation}
If, in addition, Baryonic Acoustic Oscillations (BAO) data are
included~\cite{Ade:2013zuv,Ade:2013sjv,Planck:2013jfk,Ade:2013ydc},
one obtains $\Omega_{\cal K}=-0.004\pm 0.0036$. The conclusion is that
everything is consistent with a vanishing spatial curvature. The
photon energy density is given by $\pi^2T_0^4/15$ where $T_0$ is the
CMB temperature which has been measured to be
$T_0=2.7255\pm 0.00006~\mbox{K}$~\cite{2009ApJ...707..916F}. This
implies that
\begin{align}
\Omega_{\gamma}^0h^2 &=2.47159\times 10^{-5}.
\end{align}
In the same way, the neutrino energy density is fixed since
$\rho_{\nu}=N_{\rm eff}(7/8)(4/11)^{4/3}\rho_{\gamma}\simeq
0.68132\rho_{\gamma}$ with $N_{\rm eff}=3$. This leads to
\begin{align}
\Omega_{\nu }^0h^2=1.68394 \times 10^{-5}.
\end{align}
For the baryon and cold dark matter energy densities, Planck $2015$
with PlanckTT, TE, EE+lowP has obtained
~\cite{Ade:2015tva,Ade:2015xua,Ade:2015lrj}
\begin{align}
\Omega_{\rm b}^0h^2 &= 0.02225\pm 0.00016, \quad 
\Omega_{\rm cdm}^0h^2=0.1198\pm 0.0015.
\end{align}
Finally, since the curvature is zero, one must have $\Omega_{\rm
  b}^0+\Omega_{\rm
  cdm}^0+\Omega_{\gamma}^0+\Omega_{\nu}^0+\Omega_{_{\Lambda}}^0=1$. from which 
one deduces that
\begin{align}
\Omega _{_{\Lambda}}h^2&=0.306.
\end{align}
The previous considerations describe the current state of our
universe. The model is a six parameter model: $\rho_{\rm b}$,
$\rho_{\rm cdm}$, $\rho_{_{\Lambda}}$, the optical depth $\tau $ that
controls re-ionization~\cite{1965ApJ...142.1633G} and two parameters
that describe the fluctuations, their amplitude $A_{_{\rm S}}$ and
spectral index $\nS$ (we discuss these two parameters in more details
in the section on inflationary perturbations). A priori,
$\rho_{\gamma}$ and $\rho_{\nu}$ are also parameters but they are
usually considered as fully determined given the precision of the
measurement of the CMB temperature and given the fact that we have
only three families of particles. It is impressive that with only six
parameters, one can account for all the astrophysical and cosmological
data.

From those numbers, using the theoretical description presented in the
previous section, one can also infer the past history of the
universe. The scaling of the three different types of energy densities
are given by $\rho_{\gamma}\propto 1/a^4$, $\rho_{\rm m}\propto 1/a^3$
and $\rho_{_{\Lambda}}$ is a constant. As a consequence, equality
between radiation and matter occurs when
\begin{equation}
\left(\rho_{\rm b}^0+\rho_{\rm cdm}^0\right)
\left(\frac{a_0}{a_{\rm eq}}\right)^3=
\left(\rho_{\gamma}^0+\rho_{\nu}^0\right)
\left(\frac{a_0}{a_{\rm eq}}\right)^4,
\end{equation}
that is to say
\begin{equation}
\label{eq:redshiftequality}
1+z_{\rm eq}=\frac{h^2\Omega _{\rm b}^0+h^2\Omega_{\rm cdm}^0}
{h^2\Omega_{\gamma}\left(1+0.68132\right)}\simeq 3417,
\end{equation}
where $z\equiv a_0/a(t)-1$ is the redshift. In the same way, equality
between pressure-less matter and vacuum energy occurs at
\begin{equation}
1+z_{\rm vac}=\left(\frac{h^2\Omega_{_{\Lambda}}^0}
{h^2\Omega_{\rm b}^0+h^2\Omega_{\rm cdm}^0}\right)^{1/3}
\simeq 1.29.
\end{equation}
We thus have three different eras. In the early Universe, radiation
dominates, then matter with vanishing pressure takes over and finally,
recently, the expansion of the universe became dominated by vacuum
energy. During each of these epochs, it is a good approximation to
assume that the equation of state is a constant and, therefore, the
solution of the Einstein equations discussed previously, see
Eqs.~(\ref{eq:scalefactorsol}) and~(\ref{eq:scalefactorpower}), will
be very useful.

The model that we have just described, the hot Big Bang model or, in
its modern incarnation the $\Lambda$CDM model, was the standard model
of cosmology before the $80$'s (of course, the discovery that
$\Lambda_{_{\rm B}}\neq 0$ was in fact made later but, here, we refer
to the description of the universe at very high redshifts). It is a
very successful model since, with a small number of parameters, it can
explain a large number of different observations. Historically, three
observational pillars have been the expansion of the universe, the Big
Bang Nucleosynthesis (BBN)~\cite{2016RvMP...88a5004C} and the presence
of the CMB but, nowadays, the model is supported by a much larger sets
of observations. Nevertheless, as we are now going to explain, it
possesses some undesirable features. It is not that some predictions
of this model are in contradiction with the data; it is rather the
fact that the initial conditions that need to be postulated in order
for the hot Big Bang model to work appears to be very weird. In the
next section, we turn to this question.

\section{Fine-Tuning Puzzles of the Standard Model}
\label{sec:puzzles}

\subsection{The Horizon problem}
\label{subsec:horizon}

The first puzzle that the hot Big Bang model faces is the horizon
problem. As the name indicates, it is has something to do with the
causality of initial conditions. A first question is ``when'' should
we fix the initial conditions. A priori, this should be done at the
earliest time available in the model, namely just after the Big Bang,
say at Planck time where the concept of a background spacetime becomes
well-defined. But, in practice, can we ``see'' what happens just after
the Big Bang? The answer is no because, prior to recombination, the
Universe was opaque and became transparent only after. Recombination
is the process by which free electrons and protons combine to form
Hydrogen atoms~\cite{1968ApJ...153....1P}. Before recombination, light
could not propagate freely because the cross-section between photons
and free electrons was very large (Compton scattering). However, the
cross-section of photons with Hydrogen atoms is much smaller and this
is the reason why the universe became transparent after
recombination. Recombination is described by the reaction
$p+e^{-}\rightarrow H+\gamma $ which is itself controlled by the Saha
equation~\cite{Kolb:1990vq}
\begin{align}
\frac{1-X_{_{\rm e}}}{X_{_{\rm e}}^2}
=\frac{2\zeta(3)}{\pi ^2}\eta \left(\frac{2\pi T}{m_{_{\rm e}}}\right)^{3/2}
e^{B_{_{\rm H}}/T},
\end{align}
where $X_{_{\rm e}}\equiv n_{_{\rm e}}/n_{_{\rm B}}$ with
$n_{_{\rm e}}$ the free electron number density and $n_{_{\rm B}}$ the
baryons one. $m_{_{\rm e}}=0.511\, \MeV$ is the mass of the electron
and
$B_{_{\rm H}}=m_{_{\rm p}}+m_{_{\rm e}}-m_{_{\rm H}}\simeq 13.6 \,
\eV$,
$m_{_{\rm p}}$ being the proton mass and $m_{_{\rm H}}$ the Hydrogen
atom mass, is the binding energy. Finally
$\eta\equiv n_{_{\rm B}}/n_{\gamma}$ where $n_{\gamma}$ is the photons
number density. If we require $X_{_{\rm e}}\simeq 0.1$, namely $90\%$
of the free electrons have formed Hydrogen atoms, then we find
$T_{\rm rec}=0.3\, \eV$ which corresponds to $z_{\rm rec}\simeq 1300$.
This is the furthest redshift we can reach or observe by traditional
means. We see that this event takes place after equality between
radiation and matter, see Eq.~(\ref{eq:redshiftequality}), and during
the matter dominated era.

Let us now recall the definition of an horizon in cosmology. For this
purpose, let us first rewrite the metric in polar coordinates, see
Eq.~(\ref{eq:threemetricpolar}). One has, assuming no spatial
curvature, namely ${\cal K}=0$
\begin{align}
{\rm d}s^2=-{\rm d}t^2+a^2(t)
\left[{\rm d}r^2+r^2\left({\rm d}\theta^2
+\sin^2\theta{\rm d}\varphi^2\right)\right].
\end{align}
The horizon problem comes from the fact that information propagates
with a finite speed given by the speed of light. A photon follows a
null geodesic and satisfies ${\rm d}s^2=0$ which implies that its
radial comoving coordinate can be written as
\begin{equation}
r(t)=r_{_{\rm E}}-\int_{t_{_{\rm E}}}^t\frac{{\rm d}\tau}{a(\tau)},
\end{equation}
where $r_{_{\rm E}}$ is the comoving radial coordinate of the source
and $t_{_{\rm E}}$ the emission time (there is a minus sign in the
above equation because the ``distance'' between the observer of the
photon is decreasing with time as it is heading towards the
telescope). Then, at time $t$, the proper distance is defined to be
$d_{_{\rm P}}(t)=a(t)r(t)$. If, without loss of generality, we put the
origin of the coordinates on Earth, then, at reception at time
$t=t_{_{\rm R}}$, one has by definition
$d_{_{\rm P}}\left(t_{_{\rm R}}\right)=0$, which allows us to estimate
the comoving radial coordinate at emission, namely
$r_{_{\rm E}}=\int_{t_{_{\rm E}}}^{t_{_{\rm R}}}{\rm d}\tau/a(\tau)$.
Clearly, this means that the radial coordinate of the furthest event
one can, in principle, observe from Earth is obtained by taking the
emission time to be the Big Bang time, namely
$t_{_{\rm E}}\rightarrow 0$. This defines the size of the horizon a
time $t_{_{\rm R}}$
\begin{align}
d_{_{\rm H}}\left(t_{_{\rm R}}\right)=a\left(t_{_{\rm R}}\right)
\int_{0}^{t_{_{\rm R}}}
\frac{{\rm d}\tau}{a(\tau)}.
\end{align}
Clearly, the horizon increases as $t_{_{\rm R}}$ increases since there
is more time for light to travel and, hence, we have access to more
and more remote regions of our Universe.

Then, since we have seen that recombination is the earliest event one
can observe in practice, let us calculate the angular size of the
horizon at that time. From the metric we know that the apparent size
$D$ of a source is given by
$D^2=a^2\left(t_{_{\rm E}}\right) r_{_{\rm E}}^2{\rm d}\theta^2$,
which implies that its angular size is given by
$\delta \theta =D/\left[a\left(t_{_{\rm E}}\right)r_{_{\rm
      E}}\right]$.
As a consequence, the angular size of the horizon at recombination (or
on the lss) is given by
\begin{align}
\label{eq:angularsize}
\delta \theta=
\left[\int_{t_{_{\rm lss}}}^{t_0}\frac{{\rm d}\tau}{a(\tau)}\right]^{-1}
\int_{0}^{t_{_{\rm lss}}}\frac{{\rm d}\tau}{a(\tau)}.
\end{align}
We see that one needs to know the behavior of the scale factor $a(t)$
in order to carry out this calculation. Unfortunately, as was already
discussed, an exact, analytic, solution valid at any time is not
available for the hot Big Bang model. This is here that a piece-wise
approximation, where one has several successive epochs with constant
equation of state and a scale factor in each era given by
Eq.~(\ref{eq:scalefactorsol}), will be useful. In accordance with the
description of the hot Big Bang model made before, the first phase
(phase I) is a phase dominated by radiation for which the scale factor
reads $a(t)=a_{\mathrm i}\left(2H_{\mathrm i}t\right)^{1/2}$, see
Eq.~(\ref{eq:scalefactorpower}). The quantities $a_{\mathrm i}$ and
$H_{\mathrm i}$ are free parameters. At $t=0$, the scale factor
vanishes and the scalar curvature blows up; this corresponds to the
Big Bang as already discussed. The scale factor behaves according to
the above equation for times such that $0<t<t_{\mathrm i}$. At
$t=t_{\mathrm i}$, we assume that the behavior of $a(t)$ changes and,
for $t_{\mathrm i}<t<t_{\rm end}$, we assume it is given by (phase II)
\begin{equation}
a(t)=a_{\mathrm i}\left[\frac{3}{2}\left(1+w\right)
H_{\mathrm i}\left(t-t_{\mathrm i}\right)+1\right]^{\frac{2}{3(1+w)}},
\end{equation}
in accordance with Eq.~(\ref{eq:scalefactorsol}). Notice that, here,
we are using Eq.~(\ref{eq:scalefactorsol}) and not
Eq.~(\ref{eq:scalefactorpower}). Usually, this difference is not
important but it is relevant when one considers a piece-wise solution
for the scale factor. The ``normalization'' of time has been chosen by
using $a(t)\propto t^{1/2}$ during the initial radiation dominated era
and, then, it can no longer be modified hence the use of
Eq.~(\ref{eq:scalefactorsol}). The scale factor and its derivative
(and therefore the Hubble parameter $H=\dot{a}/a$) are continuous at
the transition. The quantity $w$ is a free parameter describing the
equation of state of matter during phase II. Phase II is not part of
the hot Big Bang model and we introduce it just for future
convenience. If we do not want to include it in our description of the
model, we just have to switch it off by taking
$t_{\mathrm i}=t_{\rm end}$. Then, at $t=t_{\rm end}$, phase II is
over and the radiation dominated era starts again (or continues). This
phase III has a scale factor given by
$a(t)=a_{\rm end}\left[2H_{\rm end}\left(t-t_{\rm end}\right)+1
\right]^{1/2}$,
for times such that $t_{\rm end}<t<t_{\rm eq}$. The quantity
$a_{\rm end}$ is the scale factor at $t=t_{\rm end}$ where $a(t)$ and
$H(t)$ are continuous. Again, if one switches off phase II, then there
is of course no need to distinguish phase I and phase III. At equality
between radiation and matter, at time $t=t_{\rm eq}$, the matter
dominated era starts (phase IV) and the scale factor can now be
expressed as
$a(t)=a_{\rm eq}\left[\frac{3}{2} H_{\rm eq}\left(t-t_{\rm eq}\right)
  +1\right]^{2/3}$.
This form is valid for times such that $t_{\rm eq}<t<t_{\rm de}$.
Finally at $t=t_{\rm de}$ starts the phase dominated by the
cosmological constant (phase V) for which $a(t)$ is given by
$a(t)=a_{\rm de}e^{H_0\left(t-t_{\rm de}\right)}$. This form is valid
until present time so for $t_{\rm de}<t<t_0$. During this phase the
Hubble parameter is constant and given by its present value $H_0$. We
stress again that, if phase II is switched off, then the above simple
piece-wise model exactly mimics the behavior of $a(t)$ for the
standard hot Big Bang phase.

One has then to calculate the two integrals appearing at the numerator
and denominator of Eq.~(\ref{eq:angularsize}). This can easily be done
given that the behavior of the piece-wise scale factor described
previously is, during each phase, just a power law. The integral at
the denominator reads
\begin{align}
\int _{t_{\rm lss}}^{t_0}\frac{{\rm d}\tau}{a(\tau)}
& =\int _{t_{\rm lss}}^{t_{\rm de}}\frac{{\rm d}\tau}{a(\tau)}
+\int _{t_{\rm de}}^{t_{0}}\frac{{\rm d}\tau}{a(\tau)} \\
& =\frac{2}{a_{\rm eq}H_{\rm eq}}
\left(\frac{a_0}{a_{\rm eq}}\right)^{1/2}
\left[\left(\frac{a_{\rm de}}{a_0}\right)^{1/2}
-\left(\frac{a_{\rm lss}}{a_0}\right)^{1/2}
\right]+\frac{1}{a_0H_0}\left(\frac{a_0}{a_{\rm de}}-1\right).
\end{align}
But the chain rule gives that
\begin{align}
\frac{2}{a_{\rm eq}H_{\rm eq}}
&=\frac{2}{a_0H_0}\frac{a_0H_0}{a_{\rm de}H_{\rm de}}
\frac{a_{\rm de}H_{\rm de}}{a_{\rm eq}H_{\rm eq}}
=\frac{2}{a_0H_0}\frac{a_0}{a_{\rm de}}
\left(\frac{a_{\rm de}}{a_{\rm eq}}\right)^{-1/2}
\nonumber \\ & 
=\frac{2}{a_0H_0}\frac{a_0}{a_{\rm de}}
\left(\frac{a_{\rm de}}{a_0}\right)^{-1/2}
\left(\frac{a_0}{a_{\rm eq}}\right)^{-1/2},
\end{align}
where we have used that, for power law scale factors, the Hubble
parameter can be expressed as a power law of the scale factor. As a
consequence, it follows that the integral can be expressed, as
expected, only in terms of scale factor ratios at different times,
namely
\begin{align}
\int _{t_{\rm lss}}^{t_0}\frac{{\rm d}\tau}{a(\tau)}
&=
\frac{2}{a_0H_0}
\left(\frac{a_0}{a_{\rm de}}\right)^{3/2}
\left[\left(\frac{a_{\rm de}}{a_0}\right)^{1/2}
-\left(\frac{a_{\rm lss}}{a_0}\right)^{1/2}
\right]+\frac{1}{a_0H_0}\left(\frac{a_0}{a_{\rm de}}-1\right).
\end{align}

\begin{figure}
\begin{center}
\includegraphics[width=12cm]{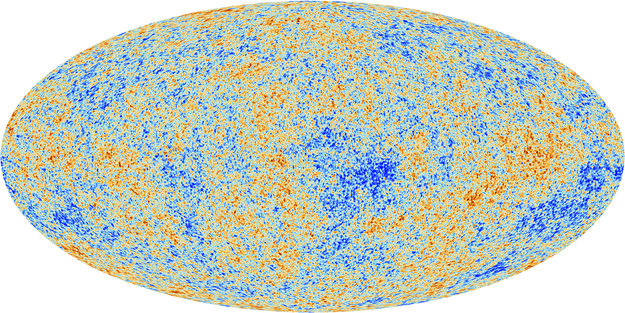}
\end{center}
\caption{Map of the temperature anisotropy measured by the European
  Space Agency (ESA) Planck satellite. The amplitude of the anisotropy
  is very small, of the order of $\sim 10^{-5}$, which means that the
  universe was in fact extremely homogeneous and isotropic on the last
  scattering surface. Figure taken from Ref.~\cite{Ade:2013zuv}.}
\label{fig:TTplanck13}
\end{figure}

The second step consists in calculating the integral appearing at the
numerator of Eq.~(\ref{eq:angularsize}). Following the same procedure
as before, one arrives at
\begin{align}
\int _{0}^{t_{\rm lss}}\frac{{\rm d}\tau}{a(\tau)}
& =\int _{0}^{t_{\rm i}}\frac{{\rm d}\tau}{a(\tau)}
+\int _{t_{\rm i}}^{t_{\rm end}}\frac{{\rm d}\tau}{a(\tau)}
+\int _{t_{\rm end}}^{t_{\rm eq}}\frac{{\rm d}\tau}{a(\tau)}
+\int _{t_{\rm eq}}^{t_{\rm lss}}\frac{{\rm d}\tau}{a(\tau)},
\end{align}
and, using the piece-wise solution described before, one obtains the
following expression
\begin{align}
\int _{0}^{t_{\rm lss}}\frac{{\rm d}\tau}{a(\tau)}
& =\frac{1}{a_{\rm i}H_{\rm i}}
+\frac{1}{a_{\rm i}H_{\rm i}}\frac{2}{1+3w}
\left[\left(\frac{a_{\rm end}}{a_{\rm i}}\right)^{\frac{1+3w}{2}}
-1\right]
+\frac{1}{a_{\rm end}H_{\rm end}}
\left(\frac{a_{\rm eq}}{a_{\rm end}}-1\right)\nonumber \\
& +\frac{2}{a_{\rm eq}H_{\rm eq}}
\left[\left(\frac{a_{\rm lss}}{a_{\rm eq}}\right)^{1/2}-1\right]
\end{align}
Then, using the power law behavior of the scale factor in each phase,
it is easy to show that
$1/(a_{\rm i}H_{\rm i})=1/(a_{\rm end}H_{\rm end}) (a_{\rm i}/a_{\rm
  end})^{(1+3w)/2}$
and
$1/(a_{\rm end}H_{\rm end})=1/(a_{\rm eq}H_{\rm eq}) (a_{\rm
  eq}/a_{\rm end})^{-1}$.
As a consequence, the integral at the numerator takes the form
\begin{align}
\int _{0}^{t_{\rm lss}}\frac{{\rm d}\tau}{a(\tau)}
& =\frac{1}{a_{\rm eq}H_{\rm eq}}
\left[1+\frac{1-3w}{1+3w}\frac{a_{\rm end}}{a_{\rm eq}}
-\frac{1-3w}{1+3w}\frac{a_{\rm end}}{a_{\rm eq}}
\left(\frac{a_{\rm i}}{a_{\rm end}}\right)^{\frac{1+3w}{2}}\right]
\nonumber \\ &
+\frac{2}{a_{\rm eq}H_{\rm eq}}
\left[\left(\frac{a_{\rm lss}}{a_{\rm eq}}\right)^{1/2}-1\right]
\end{align}
Finally, since $1/(a_{\rm eq}H_{\rm eq})=1/(a_0H_0)(a_0/a_{\rm de}) 
(a_{\rm eq}/a_{\rm de})^{1/2}$, one can establish the expression 
of the angular size of the horizon, namely
\begin{align}
\label{eq:dtheta}
\delta \theta &=\left(\frac{a_{\rm eq}}{a_0}\right)^{1/2}
\left(\frac{a_0}{a_{\rm de}}\right)^{3/2}
\left[2\left(\frac{a_{\rm lss}}{a_{\rm eq}}\right)^{1/2}
-1+\frac{1-3w}{1+3w}\frac{a_{\rm end}}{a_{\rm eq}}
-\frac{1-3w}{1+3w}\frac{a_{\rm end}}{a_{\rm eq}}
\left(\frac{a_{\rm i}}{a_{\rm end}}\right)^{\frac{1+3w}{2}}\right]
\nonumber \\ &
\times \left\{2\left(\frac{a_0}{a_{\rm de}}\right)^{3/2}
\left[\left(\frac{a_{\rm de}}{a_0}\right)^{1/2}
-\left(\frac{a_{\rm lss}}{a_0}\right)^{1/2}\right]
+\frac{a_0}{a_{\rm de}}-1\right\}^{-1}
\end{align}
As already emphasized, we have introduced the phase dominated by the
fluid with equation of state $w$ (\ie the phase II) for future
convenience but in the standard model this phase is absent. So we have
to switch it off by assuming $a_{\rm i}=a_{\rm end}$. It is also a
good approximation to take $a_0\simeq a_{\rm de}$ and
$a_{\rm lss}\simeq a_{\rm eq}$. In that case one obtains
\begin{align}
\label{eq:dthetastandard}
  \delta \theta \simeq \frac12 \left(1+z_{\rm lss}\right)^{-1/2}\simeq 0.0138.
\end{align}
(without the simplifying assumptions $a_0\simeq a_{\rm de}$ and
$a_{\rm lss}\simeq a_{\rm eq}$, one easily checks that
$\delta \theta \simeq 0.0153$). This means that we should have about
$40000$ patches on the celestial sphere with completely different
temperatures, meaning, a priori, with temperature fluctuations of
order one. This is clearly not the case as revealed by the impressive
isotropy of the CMB, see Fig.~\ref{fig:TTplanck13}. On the Planck map,
one indeed sees that the temperature anisotropy is everywhere of the
order $10^{-5}$.

Facing this situation, we have two options: either we say that the
initial conditions were the same (meaning were fine-tuned at the
$10^{-5}$ level) on super-causal scales or we say that the expansion
was, in the early Universe, different from that predicted by the
standard model. The first solution corresponds to a fine-tuning
(moreover on super-causal scales) while the other one corresponds to
inflation. Therefore, in some sense, the concept of fine-tuning is at
the heart of inflation: inflation was invented to prevent its
appearance.

\subsection{The Flatness Problem}
\label{subsec:flatness}

We have just discussed the horizon problem. But this problem is not
the only one faced by the hot Big Bang model and we now turn to
another one, namely the flatness problem (also discussed in more details in
Ref.~\cite{Chowdhury:2019otk}). Let us now consider Eq.~(\ref{eq:omegatot})
again. This equations reads
\begin{equation}
\label{eq:omegatot_2version}
1+\frac{{\cal K}}{a^2H^2}=\Omega_{_{\rm T}},
\end{equation}
and we know that observations indicate that
$\vert \Omega_{_{\rm T}}^0-1\vert \lesssim 0.01$. Clearly, this means
that we live in a spatially flat Universe to a very good
approximation. In the context of the standard model of cosmology, this
is problematic. Indeed, using the Friedmann equation, one has in
general
\begin{align}
\Omega _{_{\rm T}}(t)=\frac{\sum_i \Omega_i^0\left(\frac{a_0}{a}\right)^{3(1+w_i)}}
{\sum_i \Omega_i^0\left(\frac{a_0}{a}\right)^{3(1+w_i)}
-\left(\Omega _{_{\rm T}}^0-1\right)\left(\frac{a_0}{a}\right)^2}.
\end{align}
In the case of the hot Big Bang model, we have seen that the universe
is made of radiation and pressure-less matter. As a consequence, the
above expression takes the form
\begin{align}
\Omega _{_{\rm T}}(t)=\frac{\Omega_{\rm m}^0\left(\frac{a_0}{a}\right)^{3}
+\Omega_{\gamma}^0\left(\frac{a_0}{a}\right)^{4}}
{\Omega_{\rm m}^0\left(\frac{a_0}{a}\right)^{3}
+\Omega_{\gamma}^0\left(\frac{a_0}{a}\right)^{4}
-\left(\Omega _{_{\rm T}}^0-1\right)\left(\frac{a_0}{a}\right)^2}.
\end{align}
Then, deep in the radiation era, this equation can be approximately
expressed as
\begin{align}
\label{eq:omegatotapprox}
\Omega_{_{\rm T}}(t)\simeq 1+\frac{\Omega_{_{\rm T}}^0-1}{\Omega_{\gamma}^0}
\left(\frac{a}{a_0}\right)^2+\cdots ,
\end{align}
which implies that
\begin{align}
\label{eq:omegatotapproxbis}
\Omega_{_{\rm T}}^0-1\simeq \Omega _{\gamma}^0
\left[\Omega _{_{\rm T}}(z)-1\right]
(1+z)^2\simeq 2.47h^{-2}\times 10^{-5}\left[\Omega _{_{\rm T}}(z)-1\right]
(1+z)^2.
\end{align}
This equation clearly shows the problem. We know as an observational
fact that $\vert \Omega_{_{\rm T}}^0-1\vert \lesssim 0.01$. As we go
backwards in time, the redshift $z$ increases and, in order to satisfy
$\vert \Omega_{_{\rm T}}^0-1\vert \lesssim 0.01$,
$\Omega _{_{\rm T}}(z)-1$ must be less and less. If, for instance, we
evaluate $\Omega _{_{\rm T}}(z)-1$ at BBN ($z\simeq 10^8$), we obtain
$\vert \Omega_{_{\rm T}}^{_{\rm BBN}}-1\vert \lesssim 10^{-13}{\cal
  O}\left(<0.01\right)$.
Obviously, if we increase $z$ (namely consider even earlier times),
this fine tuning problem becomes even more severe. Going back all the
way down to the Planck scale, one has indeed
$\vert \Omega_{_{\rm T}}^{_{\rm Pl}}-1\vert \lesssim 10^{-57}{\cal
  O}\left(<0.01\right)$.
The question is then why was the Universe so flat in the early stages
of its evolution? 

Another way to see the same question is to notice that
Eq.~(\ref{eq:omegatot_2version}) implies that the solution
$\Omega_{_{\rm T}}=1$ is an unstable point. In presence of a single
fluid with equation of state $w$ (for simplicity), it can indeed be
re-written as
\begin{align}
\Omega _{_{\rm T}}(N)=1+\frac{{\cal K}}{a_{\rm ini}^2H_{\rm ini}^2}
e^{(1+3w)N},
\end{align}
where $N\equiv \ln \left(a/a_{\rm ini}\right)$ is the number of
e-folds.  We see that, if $1+3w>0$, which is always true in a
decelerated Universe, the deviation from $\Omega_{_{\rm T}}=1$
exponentially grows. In order to understand why this is physically
problematic, let us use an analogy with another unstable system,
namely a pencil balancing on its tip. Let us represent the pencil by a
rod, whose moment of inertia is given by $I=m\ell^2/3$ where $m$ is
the mass of the pencil and $\ell $ is length. The pencil is subject to
the force of gravity which acts at its mass center. The equation of
motion is given by $I\ddot{\bm \Omega}={\bm r}\wedge {\bm F}$
with\footnote{For simplicity, and without loss of generality, we
  restrict the motion of the pencil to the two-dimensional plan
  $(y,z)$.}  ${\bm r}=(\ell/2 \sin \theta, \ell /2 \cos \theta) $,
${\bm F}=(0,-mg)$, $\theta $ being the angle between the pencil and
the vertical axis and $g=9.81 \, \mbox{m}\cdot \mbox{s}^{-2}$ the
gravitational acceleration. As a consequence, the equation of motion
reads
\begin{align}
\label{eq:pencileom}
\ddot{\theta}-\frac{3g}{2\ell}\sin \theta =0.
\end{align}
If we assume that, initially, the pencil is vertical
($\theta_\mathrm{ini}=0$), then the solution for small angles reads
\begin{align}
\theta(t)\simeq \frac{\dot{\theta}_\mathrm{ini}}{m}\sinh(\omega t),
\end{align}
with a fundamental frequency given by $\omega^2\equiv 3g/(2\ell)$. We
see that, for any non vanishing initial velocity, the system is
strongly unstable and $\theta(t)$ grows exponentially. Therefore,
finding the pencil still balancing on its tip after some time would be
surprising and would require an explanation. This argument is,
however, sometimes dismissed on the basis that one should first define
a measure in order to assess, in a quantitative way, how unlikely is
this situation (see Ref.~\cite{Chowdhury:2019otk} for a full treatment
of this issue). For instance, if one believes that the measure in
phase space is peaked at $\dot{\theta}_\mathrm{ini}=0$, then one might
be tempted to say that the pencil will never fall. This leads to argue
that, in absence of a well justified measure in the space of initial
conditions, one cannot say whether it is surprising or not to find the
pencil balancing on its tip.

However, this argument ignores a crucial aspect, which is the presence
of unavoidable classical and/or quantum fluctuations. Classical
fluctuations, for instance, could be modeled, by a random force the
components of which in the $(y,z)$ plan are written as
${\bm \eta}=(\eta_1,\eta_2)$ with
\begin{align}
\left \langle \eta_i(t)\eta_j(t')\right \rangle 
=\Gamma \delta_{ij}\delta(t-t'),
\end{align}
where $\Gamma $ is a parameter describing the amplitude of the
correlation function. In presence of this force, the equation of
motion~(\ref{eq:pencileom}) becomes
\begin{align}
\ddot{\theta}-\left(\frac{3g}{2\ell}+\frac{3\eta_2}{2m}\right)\sin \theta
=-\frac{3\eta_1}{2m}\cos \theta.
\end{align}
The point is that, now, even if $\dot{\theta}_\mathrm{ini}=0$, and
contrary to what happened before, $\theta(t)$ will always grow. In
other words, small initial fluctuations will always cause the pencil
to fall. Technically, this can be viewed straightforwardly: if one
takes $\eta_2=0$ (which simplifies the problem with no loss of
generality), then using the Green function method, the motion of the
pencil with $\theta_\uini=\dot{\theta}_\uini=0$ now reads
\begin{align}
\theta (t)=-\frac{3}{2m\omega}
\int _{t_\uini}^{t}\sinh \left[\omega\left (t-t'\right)\right]
\eta_1(t'){\rm d}t'.
\end{align}
In order to obtain this solution, we have assumed small angles (namely
$\sin \theta \simeq \theta$) and, as a consequence, there will a value
of $t>t_\uini$ for which the above solution ceases to apply. But this
is just a technical limitation that can easily be fixed if needed. The
most important property, which would be shared by the exact solution
obtained without the small angle approximation, is that small
fluctuations will always push very quickly the system out of the
unstable equilibrium. Even if those fluctuations are quantum
fluctuations, this is sufficient to insure that a macroscopic pen
falls in a couple of seconds~\cite{2014arXiv1406.1125L}. We conclude
that, even if one manages to obtain a measure which is peaked over the
unstable equilibrium position, finding the pencil balancing on its tip
would remain a physical problem that needs an explanation.

In the case of Cosmology, small fluctuations in the early Universe are
present and, therefore, based on the previous considerations,
observing $\Omega _{_{\rm T}}^0\simeq 1$ requires an explanation. The
flatness problem consists in finding a solution to this problem.

The hot Big Bang model has other puzzles, such as, for instance, the
presence of dangerous relics originating from phase transitions taking
place in the early universe. Rather than describing all these issues
in an exhaustive way, we now turn to a possible solution, namely the
theory of cosmic inflation.

\section{Inflation}
\label{sec:inf}

\subsection{Solving the Standard Model Puzzles}
\label{subsec:solving}

The main idea of inflation is that the puzzles we have described in
the previous sections are an indication that the dynamics of the
universe at very high redshifts was different from that implied by the
hot Big Bang model. According to this model, at very high energies,
the universe was radiation dominated, with a scale factor
$a(t)\propto t^{1/2}$. According to inflation, this was not the
case. Let us now see how it works in practice and let us discuss how
inflation can solve the horizon problem. For this purpose, we switch
on the phase dominated by the fluid with equation of state $w$ (phase
II) and rewrite Eq.~(\ref{eq:dtheta}) as
\begin{align}
\delta \theta \simeq \frac12 \left(1+z_{\rm lss}\right)^{-1/2}
\left\{1+\frac{1-3w}{1+3w}\frac{a_{\rm end}}{a_{\rm lss}}
\left[1-e^{-\frac{1}{2}N_{_{\mathrm T}}\left(1+3w\right)}
\right]\right\},
\end{align}
where we have introduced the total number of e-folds
$N_{_{\rm T}}=\ln \left(a_{\rm end}/a_{\rm i}\right)$ during phase
II. The presence of phase II introduces a correction to the standard
result~(\ref{eq:dthetastandard}), namely the second factor in the
above equation. If we want this correction to play a significant role,
then the exponential term must be non-negligible. And this is the case
if
\begin{align}
\label{eq:sec}
1+3w<0,
\end{align}
or, in other words, using Eq.~(\ref{eq:accela}), if the Universe was
accelerating $\ddot{a}>0$. By definition, a phase of accelerating
expansion is called a phase of inflation. But having a phase of
acceleration is not sufficient, we also need a phase of acceleration
that lasts long enough. Indeed requiring $\delta \theta >2\pi$ gives
$N_{_{\rm T}}\gtrsim \ln \left(1+z_{\rm end}\right)$ (here, we assume
that $w$ is not fine tuned to $\lesssim -1/3$). If we write the energy
scale at the end of inflation as
$\rho_{\rm end}\simeq \left(10^x\right)^4 \GeV^4$, then the previous
condition reduces to $N_{_{\rm T}}\gtrsim 2.3 x+29$. For the Grand
Unified Theory (GUT) scale, namely $x=15$, this gives
$N_{_{\rm T}}\gtrsim 63$.  Therefore, one concludes that the horizon
problem is solved if we have a phase of inflation. If this phase of
inflation takes place at the GUT scale, then it must last more than
$\sim 60$ e-folds. If the energy scale is lower, then we need less
e-folds.

Let us now see what would be the consequence for the flatness
problem. In agreement with what we have discussed before, this means
that we postulate the presence of a new fluid, with an a priori
unknown equation of state $w$. This unknown fluid dominates the
energy density budget of the Universe if $t_{\rm i}<t<t_{\rm end}$,
namely during phase II, and is smoothly connected to the standard Big
Bang phase which takes place for $t>t_{\rm end}$. As a consequence,
this implies that Eq.~(\ref{eq:omegatotapproxbis}) can only be applied
if $z<z_{\rm end}$ since $t_{\rm end}$ is the earliest time where the
standard evolution is valid. In that case, one has
\begin{align}
\Omega_{_{\rm T}}^0-1\simeq \Omega _{\gamma}^0
\left[\Omega _{_{\rm T}}(z_{\rm end})-1\right]
(1+z_{\rm end})^2\simeq 2.47h^{-2}\times 10^{-5}
\left[\Omega _{_{\rm T}}(z_{\rm end})-1\right]
(1+z_{\rm end})^2.
\end{align}
Now our goal is to calculate $\Omega _{_{\rm T}}(z_{\rm end})-1$ in
terms of $\Omega _{_{\rm T}}(z_{\rm ini})-1$, namely in terms of the
initial conditions at the beginning of inflation. During inflation,
one has
\begin{align}
\label{eq:omegatx2}
\Omega _{_{\rm T}}(t)\simeq \frac{\Omega _{_{\rm X}}^{\rm ini}
\left(\frac{a_{\rm ini}}{a}\right)^{3(1+w)}}
{\Omega _{_{\rm X}}^{\rm ini}
\left(\frac{a_{\rm ini}}{a}\right)^{3(1+w)}
-\left(\Omega _{_{\rm T}}^{\rm ini}-1\right)\left(\frac{a_{\rm ini}}{a}\right)^2}.
\end{align}
which implies that
\begin{align}
\label{eq:omegatx2}
\Omega _{_{\rm T}}(z_{\rm end})\simeq \frac{\Omega _{_{\rm X}}^{\rm ini}}
{\Omega _{_{\rm X}}^{\rm ini}
-\left(\Omega _{_{\rm T}}^{\rm ini}-1\right)
\left(\frac{a_{\rm ini}}{a_{\rm end}}\right)^{-1-3w}}.
\end{align}
Clearly the only way to solve the flatness problem is if inflation is
such that $\Omega _{_{\rm T}}(z_{\rm end})\simeq 1$ and the only way
to achieve it is to have $1+3w<0$, that to say the same condition than
the one derived to solve the horizon problem, see
Eq.~(\ref{eq:sec}). In that situation, the above equation takes the
form
\begin{align}
\Omega _{_{\rm T}}(z_{\rm end})\simeq
1-\frac{\Omega _{_{\rm T}}(z_{\rm ini})-1}{\Omega _{_{\rm X}}^{\rm ini}}
e^{-N_{_{\rm T}}\vert 1+3w\vert},
\end{align}
and, as a consequence 
\begin{align}
\Omega_{_{\rm T}}^0-1\simeq 2.47h^{-2}\times 10^{-5}
\frac{\Omega _{_{\rm T}}(z_{\rm ini})-1}{\Omega _{_{\rm X}}^{\rm ini}}
e^{-N_{_{\rm T}}\vert 1+3w\vert}
(1+z_{\rm end})^2.
\end{align}
Requiring $\vert \Omega_{_{\rm T}}^0-1\vert \lesssim 0.01$ without
postulating that $\Omega _{_{\rm T}}(z_{\rm ini})-1$ is very small,
namely without postulating any fine-tuning of the initial conditions
at the beginning of inflation leads to
$N_{_{\rm T}}\gtrsim \ln\left(1+z_{\rm end}\right)$, that is to say,
again, the same condition as for the horizon problem. The fact that
the conditions for solving the horizon and the flatness problems are
the same is very suggestive and is also an argument in favor of
inflation.

We conclude that inflation can solve the fine-tuning puzzles of the
Big Bang model. In addition, we mentioned before the existence of
additional puzzles. One can show that inflation can also fix them. The
next question is then which type of matter can produce such a phase.

\subsection{Realizing a Phase of Inflation}
\label{subsec:realizing}

As explained in detail in the previous sections, a phase of
accelerated expansion in the early universe solves the puzzles of the
standard model of cosmology. Clearly, at very high energies, the
correct framework to describe matter is field theory and its simplest
version, compatible with isotropy and homogeneity, is when a scalar
field dominates the energy budget of the Universe. This scalar field
is called the ``inflaton''. In that case, the energy density and
pressure are given by
\begin{equation}
\rho =\frac{\dot{\phi}^2}{2}+V(\phi), \quad 
 p=\frac{\dot{\phi}^2}{2}-V(\phi). 
\end{equation}
As a consequence, if the potential energy dominates over the kinetic
energy, one obtains a negative pressure and, hence, inflation. This
can be achieved when the field moves slowly or, equivalently, when the
potential is almost flat.

From a field theory perspective, the micro-physics of inflation should
be described by an effective field theory characterized by a cutoff
$\Lambda$. One usually assumes that the gravitational sector is
described by General Relativity, which itself is viewed as an
effective theory with a cutoff at the Planck scale, then
$\Lambda <\Mp$. On the other hand, we will see that the CMB anisotropy
data suggests that inflation could have taken place at energies as
high as the GUT scale and this suggests $\Lambda >10^{15}\GeV$.
Particle physics has been tested in accelerators only up to scales of
$\sim \TeV$ and this implies that our freedom in building models of
inflation will remain very important. A priori, without any further
theoretical guidance, the effective action can therefore be written as
\begin{eqnarray}
\label{eq:generalS}
S &=& \int {\rm d}^4x\sqrt{-g}
\biggl[\Mp^2\Lambda _{_{\rm B}}+\frac{\Mp^2}{2}R+
aR^2+b R_{\mu \nu}R^{\mu \nu}+\frac{c}{\Mp^2}R^3+\cdots
\nonumber \\ & & 
-\frac{1}{2}\sum _i g^{\mu
  \nu}\partial_\mu\phi_i\partial_{\nu }\phi_i
-V(\phi_1, \cdots ,\phi_n)
+\sum_i d_i 
\frac{{\cal O}_i}{\Lambda^{n_i-4}}\biggr]
\nonumber \\ & &
+S_{\rm int}(\phi_1, \cdots, \phi_n,A_{\mu},\Psi)
+\cdots \, .
\end{eqnarray}
In the above equation, the first line represents the effective
Lagrangian for gravity (recall that $\Lambda_{_{\rm B}}$ is the
cosmological constant). In practice, we will mainly work with the
Einstein-Hilbert term only. The second line represents the scalar
field sector and we have postulated that, a priori, several scalar
fields are present. The first two terms represent the canonical
Lagrangian while ${\cal O}_i$ represents a higher order operator of
dimension $n_i>4$, the amplitude of which is determined by the
coefficient $d_i$. Those corrections can modify the potential but also
the (standard) kinetic term~\cite{Chialva:2014rla}. The last term
encodes the interaction between the inflaton fields and the other
fields present in Nature, \ie gauge fields $A_{\mu}$ and fermions
$\Psi$. Those terms are especially important to describe how inflation
ends and is connected to the standard model of cosmology. Finally, the
dots stand for the rest of the terms such as kinetic terms of gauge
bosons $A_{\mu}$, of fermions $\Psi$ etc \dots

\begin{figure}
\begin{center}
\includegraphics[width=10cm]{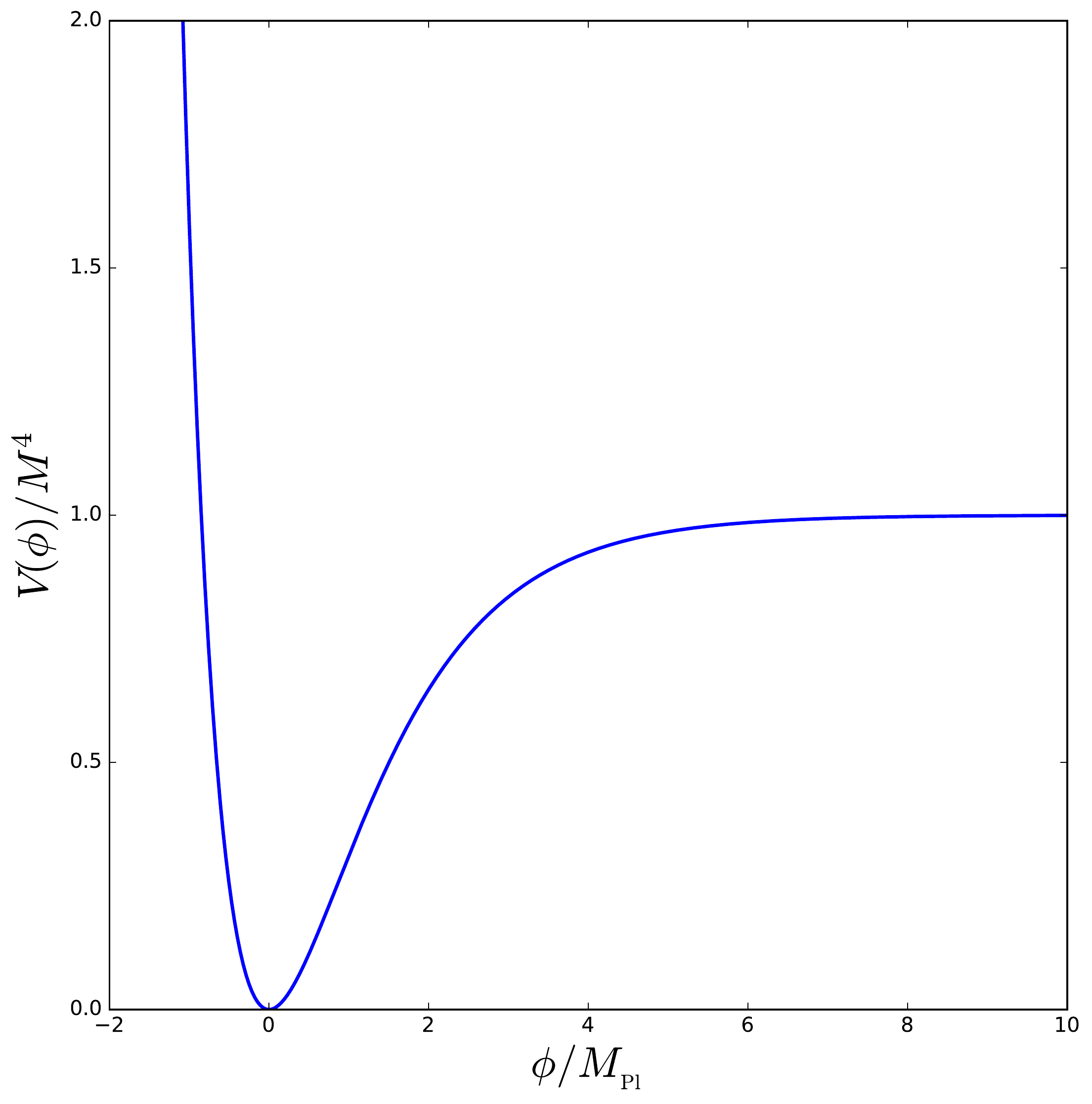}
\end{center}
\caption{Example of a potential [the Starobinsky
  potential~(\ref{eq:potstaro})] that can support inflation. Slow roll
  inflation occurs along the plateau where the potential is almost
  flat and the reheating phase takes place when the field oscillates
  around its minimum, here located at the origin.}
\label{fig:pot}
\end{figure}

Given the complexity of the above Lagrangian, it is clear that it is
impossible to single out a model of inflation from theoretical
considerations only. However, as we will see, the CMB data have given
us precious information. In particular, from the absence of
non-adiabatic perturbations and from the fact that the CMB
fluctuations are Gaussian, models with a single field, a minimal
kinetic term and a smooth potential are favored. This does not mean
that more complicated scenarios are ruled out (as a matter of fact
they are not) but that, for the moment, they are not needed to
describe the data. It is important to emphasize that we are driven to
this class of models, which is clearly easier to investigate than the
more complicated models mentioned above, not because we want to
simplify the analysis but because this is what the CMB data
suggest. Then, the Lagrangian~(\ref{eq:generalS}) can be simplified to
\begin{equation}
\label{eq:normallagrange}
  \calL=-\frac{1}{2}g^{\mu \nu}\partial_{\mu}\phi\partial_{\nu }\phi
-V(\phi)+\calL_{\rm int}(\phi,A_{\mu},\Psi).
\end{equation}
During the accelerated phase, the interaction term is supposed to be
sub-dominant and will be neglected. Then, only one arbitrary function
remains in the Lagrangian, the potential $V(\phi)$. An example of a
potential that supports inflation is given in Fig.~\ref{fig:pot}. From
CMB data, one can constrain this function and this will be discussed
in the following. As already mentioned, the interaction term plays a
crucial role in the process which ends inflation. Indeed, it controls
how the inflaton field decays into particles describing ordinary
matter. These decay products are then supposed to thermalize and the
radiation dominated epoch starts at a temperature which is known as
the reheating temperature $T_{\rm rh}$. This quantity is an important
parameter of any inflationary model and we will see that the CMB data
can also say something about its value.

Following the above considerations, during inflation itself, the
interaction term is neglected and the evolution of the system is
controlled by the Friedmann and Klein-Gordon equations, namely
\begin{align}
\label{eq:friedman}
H^2 &=\frac{1}{3\Mp^2}\left[\frac{\dot{\phi}^2}{2}+V(\phi)\right], \\
\label{eq:kg}
\ddot{\phi} & +3H\dot{\phi}+V_{\phi} = 0,
\end{align}
where a subscript $\phi$ means a derivative with respect to the
inflaton field. For an arbitrary potential, this system of equations
cannot be solved analytically. This means that we have to use either
numerical calculations or a perturbative method. In general, a
perturbative method is based on the presence of a small parameter in
the problem and on an expansion of the relevant quantities of the
theory in terms of this small parameter. In the case of inflation,
there exists such a small parameter which physically expresses the
fact that the potential is flat. So it can be chosen as the curvature
of the potential or, equivalently, as the kinetic to potential energy
ratio or, given that inflation corresponds to an approximately
constant Hubble parameter, as the derivative of $H$. Therefore, we
introduce the Hubble flow functions $\epsilon_n$ defined
by~\cite{Schwarz:2001vv,Leach:2002ar}
\begin{equation}
\label{eq:defhf}
\epsilon_{n+1} \equiv \frac{\dd \ln \left \vert 
\epsilon_n \right \vert}{\dd N}, 
\quad n\ge 0,
\end{equation}
where $\epsilon_0\equiv H_\uini/H$ starts the hierarchy and we remind
that $N\equiv \ln(a/a_\uini)$ is the number of e-folds already
introduced before. From the above expression, the first Hubble flow
parameter can be written as
\begin{equation}
\label{eq:eps1}
\epsilon_1=-\frac{\dot{H}}{H^2}=1-\frac{\ddot{a}}{aH^2}
=\frac{3\dot{\phi}^2}{2}\frac{1}{\dot{\phi}^2/2+V(\phi)},
\end{equation}
and, therefore, inflation ($\ddot{a}>0$) occurs if $\epsilon_1<1$. In
terms of the Hubble flow parameters, the Friedmann and Klein-Gordon
equations take the form
\begin{eqnarray}
\label{eq:exactfried}
H^2 &=& \frac{V}{\Mp^2(3-\epsilon_1)}, \\
\left(1 + \dfrac{\epsilon_2}{6 - 2\epsilon_1}\right)
\dfrac{\ud \phi}{\ud N} &=& - \Mp^2 \dfrac{\ud \ln V}{\ud \phi}\,.
\end{eqnarray}
It is worth stressing the point that these expressions are exact. The
condition $\epsilon_1<1$ during $\sim 60$ e-folds is sufficient to
solve the fine-tuning problems of the standard model, as discussed
above. But, if one wants to describe properly the CMB anisotropy (see
the discussion below), one needs $\epsilon_n\ll 1$, which is called
the slow-roll regime. In this situation, the first three Hubble flow
parameters can be approximated as~\cite{Liddle:1994dx}
\begin{align} 
\label{eq:epsfirst}
\epsilon_1 & \simeq
\frac{\Mp^2}{2}\left(\frac{V_\phi}{V}\right)^2, \\ 
\label{eq:eps2}
\epsilon_2 & \simeq
2\Mp^2\left[\left(\frac{V_\phi}{V}\right)^2
-\frac{V_{\phi \phi}}{V}\right], \\
\label{eq:eps3}
\epsilon_2\epsilon_3 & \simeq 2\Mp^4\left[
\frac{V_{\phi \phi \phi}V_\phi}{V^2}-3\frac{V_{\phi \phi}}{V}
\left(\frac{V_\phi}{V}\right)^2
+2\left(\frac{V_\phi}{V}\right)^4\right].
\end{align} 
We see that the first Hubble flow parameter is also a measure of the
steepness of the potential and of its first derivative. The second
Hubble flow parameter is a measure of the second derivative of the
potential and so on. Therefore, if one can observationally constrain
the values of the Hubble flow parameters, we can say something about
the shape of the inflationary potential. The slow-roll approximation
also allows us to simplify the equations of motion and to analytically
integrate the inflaton trajectory. Indeed, in this regime,
Eqs.~(\ref{eq:friedman}) and~(\ref{eq:kg}), which control the
evolution of the system, can be approximated by $H^2\simeq V/(3\Mp^2)$
and ${\rm d}\phi/{\rm d}N\simeq -\Mp^2{\rm d}\ln V/{\rm d}\phi$, from
which one obtains
\begin{equation}
\label{eq:srtrajectory}
N-\Nini=-\frac{1}{\Mp^2}\int_{\phiini}^{\phi}\frac{V(\chi)}
{V_\chi(\chi)}\, \ud \chi \, ,
\end{equation} 
$\phiini$ being the initial value of the inflaton. If the above
integral can be performed, one gets $N=N(\phi)$ and if this last
equation can be inverted, one has the trajectory, $\phi=\phi(N)$.

Let us now describe the end of inflation. As already mentioned, this
is the phase during which the inflaton decays into the particles of
the standard model. During that phase, the interaction term is
obviously crucial. This means that, in principle, in order to have a
fair description of that process, one must specify all the interaction
terms of $\phi$ with the other scalars, the gauge bosons and the
fermions present in the universe together with the corresponding
coupling constants. Then, one must solve the (non linear) equations of
motion of all these fields. Clearly, this is a very complicated
task. However, in a cosmological context, one can proceed in a simpler
way. Indeed, the reheating phase can in fact be described by two
numbers, $\rho_{\rm reh}$, the energy density at which the radiation
dominated era starts (and, therefore, at which the reheating epochs
stops) and the mean equation of state $\wrehbar$. Of course, one
should also know at which energy density reheating starts but this is
not a new parameter since it is determined by the condition
$\epsilon_1=1$. In the following, we denote this quantity
$\rho_{\rm end}$. Let us notice that the knowledge of $\rho_{\rm reh}$
is equivalent to the knowledge of the reheating temperature since
\begin{align}
\rho_{\rm reh}=g_*\frac{\pi^2}{30}T_{\rm reh}^4,
\end{align}
where $g_*$ encodes the number of relativistic degrees of freedom.  On
the other hand, the mean equation of state controls the expansion rate
of the Universe during reheating. Let $\rho_{_{\rm T}} =\sum _i\rho_i$
and $p_{_{\rm T}}=\sum _i p_i$ be the total energy density and
pressure, where the sum is over all the species present during
reheating. Let us define the ``instantaneous'' equation of state by
$\wreh\equiv p_{_{\rm T}}/\rho_{_{\rm T}}$. Then the mean equation of
state parameter, $\wrehbar$, is given by
\begin{equation} 
\label{eq:wrehbar}
\wrehbar \equiv \frac{1}{\Delta N}\int_{\Nend}^{\Nreh} \wreh(n)\dd n,
\end{equation}
where $\Delta N \equiv \Nreh - \Nend$ is the total number of e-folds
during reheating. The quantity $\wrehbar$ allows us to determine the 
evolution of the total energy density since this quantity obeys
\begin{equation}
  \rho_{\rm reh}=\rho_{\rm end}\, e^{-3(1+\wrehbar)\Delta N},
\end{equation} 
where we recall that $\rho_{\rm end}$ can be determined once the model
of inflation is known.

In fact, as long as the CMB is concerned, only one parameter can be
constrained and this parameter is a combination of $\rho_{\rm reh}$
and $\wrehbar$. It is known as the reheating parameter and is defined
by
\begin{equation}
\label{eq:defRrad}
\Rrad\equiv 
\left(\frac{\rho_{\rm reh}}
{\rho_{\rm end}}\right)^{(1-3\wrehbar)/(12+12\wrehbar)}.
\end{equation}
The justification for this definition can be found in
Refs.~\cite{Martin:2006rs,Martin:2010kz,Martin:2014nya,Martin:2015dha,Martin:2016oyk}
but a simple argument shows that it makes sense. It is clear that one
cannot make the difference between a model of instantaneous reheating
where $\rho_{\rm end}=\rho_{\rm reh}$ and a model where reheating
proceeds with a mean equation of state of radiation, namely
$\wrehbar=1/3$, since in this last case reheating cannot be
distinguished from the subsequent radiation dominated era. We see on
the above definition that, in both cases, the reheating parameters has
the same numerical value, $\Rrad=1$, which is consistent.

It may come as a surprise that a very complicated phenomenon such as
reheating can be described by only one number. But one should keep in
mind that this is the case only if one tries to constrain reheating
from the CMB or, to put it differently, the reheating parameter is the
only quantity that can be measured if one uses CMB data. Moreover,
this is not a new situation. This is indeed very similar to what
happens for re-ionization~\cite{1965ApJ...142.1633G} for
instance. Clearly, re-ionization is, from a particle physics point of
view, a very complicated process. But despite this complexity, as long
as one considers CMB data only, it is described by one quantity, the
optical depth $\tau$~\cite{1965ApJ...142.1633G}.

\subsection{Inflationary Cosmological Perturbations}
\label{subsec:pert}

So far, we have described the background spacetime during
inflation. We now turn to the
perturbations~\cite{Mukhanov:1990me,Martin:2007bw,Martin:2004um,Martin:2003bt}. As
is well-known, this is a crucial part of the inflationary theory since
it gives a convincing explanation for the origin of the large scale
structures observed in our Universe. However, in order to deal with
this question, one must go beyond homogeneity and isotropy which is a
complicated task. But, we know that, in the early Universe, the
deviations from the cosmological principle were small as revealed, for
instance, by the magnitude of the CMB anisotropy
$\delta T/T\sim 10^{-5}$.  During inflation, we expect the
fluctuations to be even smaller since they grow with time according to
the mechanism of gravitational collapse. This means that we can treat
the inhomogeneities perturbatively and, in fact, restrict ourselves to
linear perturbations. Then, the idea is to write the metric tensor as
$g_{\mu \nu}(\eta, {\bm x})=g_{\mu \nu}^{_{\rm FLRW}}(\eta) +\delta
g_{\mu \nu}(\eta, {\bm x})+\cdots $,
where $g_{\mu \nu}^{_{\rm FLRW}}(\eta)$ represents the metric tensor
of the FLRW Universe, see Eq.~(\ref{eq:metric}), and where
$\delta g_{\mu \nu}(\eta, {\bm x})\ll g_{\mu \nu}^{_{\rm
    FLRW}}(\eta)$.
Here, $\eta$ is the conformal time, related to the cosmic time by
${\rm d}\eta=a{\rm d}t$. In the same way, the inflaton field is
expanded as
$\phi(\eta,{\bm x})=\phi^{_{\rm FLRW}}(\eta)+\delta \phi(\eta,{\bm
  x})$
with $\delta \phi(\eta,{\bm x})\ll \phi^{_{\rm FLRW}}(\eta)$. In fact,
$\delta g_{\mu \nu}(\eta, {\bm x})$ can be expressed in terms of three
types of perturbations, scalar, vector and tensor. In the context of
inflation, only scalar and tensor are important. Scalar perturbations
are directly coupled to the perturbed scalar field
$\delta \phi(\eta,{\bm x})$ while tensor fluctuations represent
primordial gravitational waves. The equations of motion of each type
of fluctuations are given by the perturbed Einstein equations, namely
$\delta G_{\mu \nu}=\delta T_{\mu \nu}/\Mp^2$. But we also need to
specify the initial conditions. A crucial assumption of inflation is
that the source of the perturbations are the unavoidable quantum
vacuum fluctuations of the gravitational and scalar fields. It is
clear that this has drastic implications: it means that the large
scale structures in the Universe are nothing but quantum fluctuations
made classical and stretched to cosmological scales.

Let us now turn to a quantitative characterization of the cosmological
fluctuations. The amplitude of scalar perturbations is described by
the curvature perturbations~\cite{Bardeen:1983qw,Martin:1997zd}
$\zeta(\eta, {\bm x}) \equiv \Phi+2({\cal H}^{-1}\Phi'+\Phi)/(3+3w)$,
with $w=p/\rho$ the equation of state during inflation and $\Phi$ the
Bardeen potential~\cite{Bardeen:1980kt} (not to be confused with the
scalar field $\phi$). The Bardeen potential is the quantity that
describes scalar perturbations as revealed by writing explicitly the
perturbed metric in longitudinal gauge,
${\rm d}s^2=a^2(\eta)[-(1-2\Phi){\rm d}\eta^2+(1-2\Phi)\delta_{ij}{\rm
  d}x^i{\rm d}x^j]$.
Since we deal with a linear theory, we can go to Fourier space and
follow the time evolution of the Fourier component
$\zeta_{\bm k}(\eta)$. Then, the properties of the fluctuations are
described by the power spectrum of scalar perturbations, which is
given by
\begin{equation}
\label{eq:meanzetasquare}
\calP_{\zeta}(k)=\frac{k^3}{2\pi^2}\vert \zeta_{\bm k}\vert^2.
  \end{equation}

The power spectrum depends on the model of inflation that is to say,
for the simple class of models discussed here, on the potential
$V(\phi)$. Unfortunately, there exists no exact analytic
calculation of $\calP_{\zeta}(k)$ for an arbitrary
$V(\phi)$. Therefore, one must either rely on numerical calculations
or on perturbative methods. Here again, the slow-roll approximation 
can be used and leads to the following result~\cite{Leach:2002ar}
\begin{equation} 
\label{spectrumsr}
\calP_\zeta(k)= \calP_{\zeta 0}(k_{_{\rm P}})
\left[a_0^{_{({\rm S})}} + 
a_1^{_{({\rm S})}} \ln \left(\dfrac{k}{k_{_{\rm P}}}\right) 
+ \frac{a_2^{_{({\rm S})}}}{2} \ln^2\left(\dfrac{k}{k_{_{\rm P}}}\right)
+ \cdots \right]\, ,
\end{equation}
where $k_{_{\rm P}}$ is a pivot scale and the
overall amplitude can be written as
\begin{equation}
\label{eq:scalaramp}
\calP_{\zeta \zero} =\frac{H_*^2}{8 \pi^2 \epsilon_{1*} \Mp^2}\,.
\end{equation}
In the above expression (and in the subsequent ones), a star means
that the corresponding quantity has been evaluated at the time at
which the pivot scale crossed out the Hubble radius during inflation,
namely $k_{_{\rm P}}\sim a_*H_*$. The amplitude of the spectrum
depends on (the square of) the strength of the gravitational field
during inflation which is described by the expansion rate $H_*$. It is
also inversely proportional to the first derivative of the potential
through the presence of $\epsilon_{1*}$ at the denominator. The main
property of $\calP_{\zeta \zero}$ is that it is does not depend on the
wave number, in other words it is scale independent. This result
represents one of the main success of inflation since a scale
invariant power spectrum was known for a long time to be in agreement
with the observations. But there is even more. We see that that the
scale invariant piece of the power spectrum receives scale dependent
logarithmic corrections the amplitudes of which are controlled by the
Hubble flow parameters and are given
by~\cite{Schwarz:2001vv,Casadio:2004ru,Casadio:2005xv,Casadio:2005em,Gong:2001he,Choe:2004zg,Leach:2002ar,Lorenz:2008et,Martin:2013uma},
\begin{eqnarray}
\label{eqn:as0}
a_{0}^{\usssPSP} &=& 1 - 2\left(C + 1\right)\epsilon_{1*} - C \epsilon_{2*}
+ \left(2C^2 + 2C + \frac{\pi^2}{2} - 5\right) \epsilon_{1*}^2 \nonumber
\\ & + & \left(C^2 - C + \frac{7\pi^2}{12} - 7\right)
\epsilon_{1*}\epsilon_{2*} + \left(\frac12 C^2 + \frac{\pi^2}{8} -
1\right)\epsilon_{2*}^2 \nonumber \\ & + & \left(-\frac12 C^2  +
\frac{\pi^2}{24}\right)  \epsilon_{2*}\epsilon_{3*} +\cdots \, , \\
\label{eqn:as1}
a_{1}^{\usssPSP} & = & - 2\epsilon_{1*} - \epsilon_{2*} 
+ 2(2C+1)\epsilon_{1*}^2
+ (2C - 1)\epsilon_{1*}\epsilon_{2*} + C\epsilon_{2*}^2 
- C\epsilon_{2*}\epsilon_{3*}
+\cdots \, ,\\ 
a_{2}^{\usssPSP} &=& 4\epsilon_{1*}^2 + 2\epsilon_{1*}\epsilon_{2*} +
\epsilon_{2*}^2 - \epsilon_{2*}\epsilon_{3*} +\cdots \, ,\\
a_{3}^{\usssPSP} &=& \calO(\epsilon_{n*}^3)\, ,
\label{eqn:as2}
\end{eqnarray}
where $C \equiv \gamma_{\usssE} + \ln 2 - 2 \approx -0.7296$,
$\gamma _\usssE$ being the Euler constant. Since the coefficients
$a_{1}^{\usssPSP}$, $a_{2}^{\usssPSP}$ etc \dots are small (being
proportional to the Hubble flow parameters), this means that the
inflationary power spectrum is not exactly scale-invariant but, in
fact, almost scale invariant. This is the main prediction of inflation
and it was confirmed recently by the CMB Planck data. We stress that
this is a prediction since it was made before it was measured. In
terms of spectral index, being defined as the logarithmic derivative
of $\ln \calP_{\zeta}(k)$, one has
\begin{equation}
\label{eq:specindices}
\nS=1-2\epsilon_{1*}-\epsilon_{2*},
\end{equation}
where $\nS=1$ corresponds to exact scale invariance. We see on the
above expression that the small deviations from exact scale invariance
carry information about the shape of the inflationary potential since
$\epsilon_1$ and $\epsilon_2$ respectively depend on the first and
second derivative of $V(\phi)$. Therefore, an accurate measurement of
the power spectrum can provide information about which version of
inflation was realized in the early universe.

\begin{figure}
\begin{center}
\includegraphics[width=15cm]{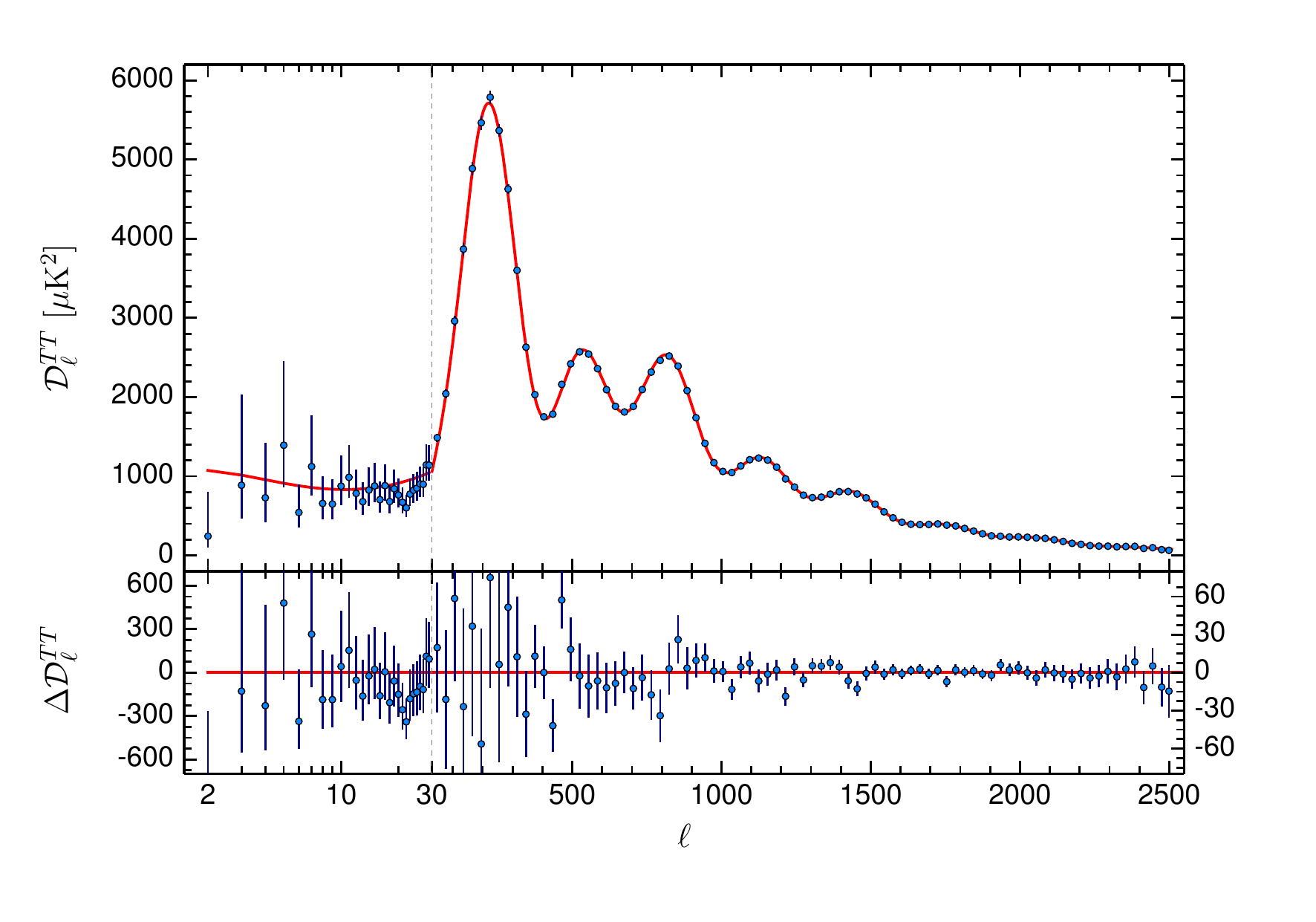}
\end{center}
\caption{Multipole moments versus angular scale obtained from the
  Planck $2015$ data. The multipole moments are defined from the
  following expression of the temperature fluctuation two-point
  correlation function:
  $\langle \delta T/T({\bm e}_1)\delta T/T({\bm
    e}_2)\rangle=(4\pi)^{-1}\sum _{\ell}(2\ell+1)C_{\ell}P_{\ell}(\cos
  \theta)$
  where $\theta$ is the angle between the two directions ${\bm e}_1$
  and ${\bm e}_2$. The multipole moments $C_{\ell}$ represent the
  power of the signal at a given spatial frequency $\ell$. Notice that
  the quantity ${\cal D}_{\ell}$ is defined by
  ${\cal D}_{\ell}=\ell(\ell+1)C_{\ell}/(2\pi)$. The red curve
  corresponds to the best fit in the parameter space of the
  $\Lambda$CDM model. This result is consistent with the predictions
  of inflation, for instance because of the presence of the Doppler
  peaks. Figure taken from Ref.~\cite{Ade:2015xua}.}
\label{fig:TTplanck2015}
\end{figure}

\begin{figure}
\begin{center}
\includegraphics[width=13cm]{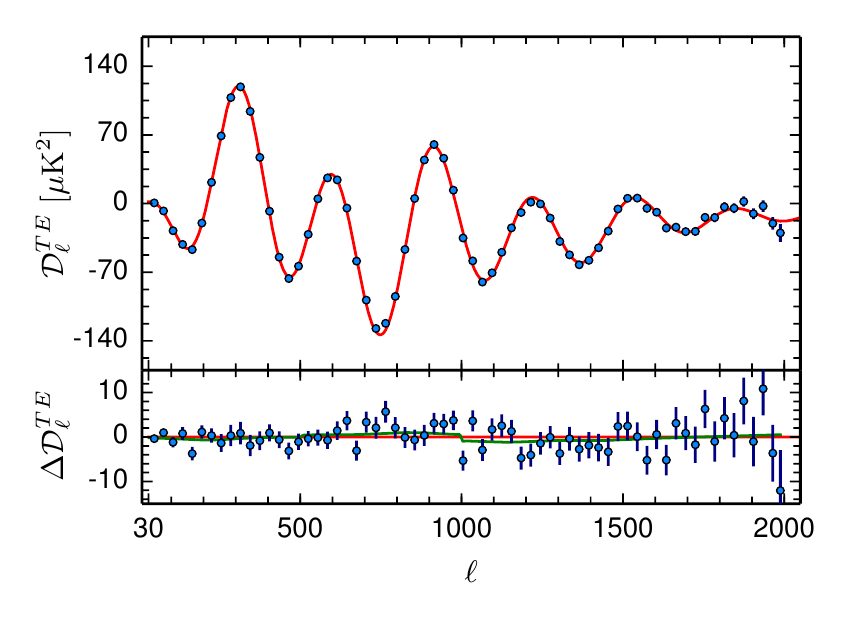}
\end{center}
\caption{Multipole moments corresponding to the correlation between
  temperature and so-called $E$-mode polarization anisotropies (we
  refer the reader to Ref.~\cite{Kosowsky:1994cy} for definitions of
  polarized CMB quantities) obtained from Planck $2015$. The red solid
  line corresponds to prediction of the $\Lambda$CDM model obtained
  from the best fit in Fig.~\ref{fig:TTplanck2015} (namely with
  temperature measurements only). The lower panel shows the residual
  with respect to this best fit. Figure taken from
  Ref.~\cite{Ade:2015xua}.}
\label{fig:TEplanck2015}
\end{figure}

\begin{figure}
\begin{center}
\includegraphics[width=13cm]{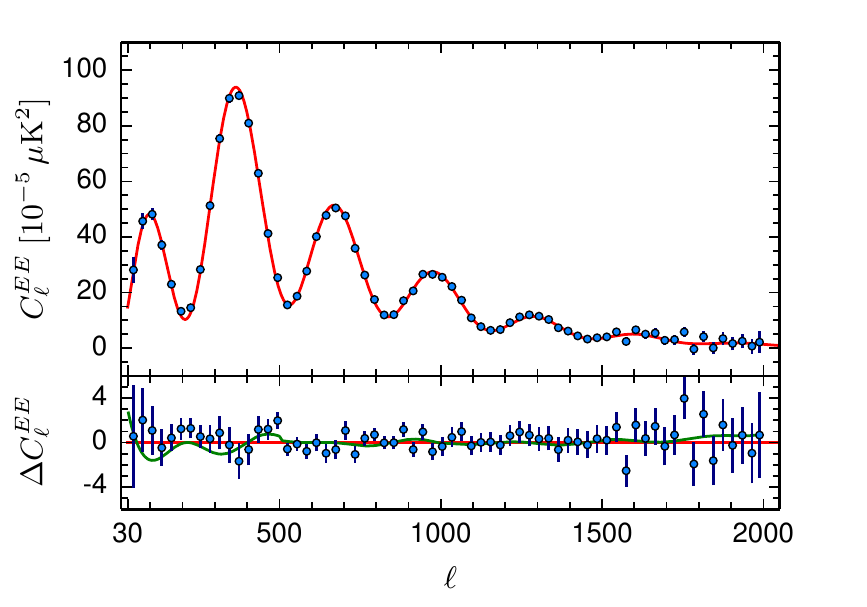}
\end{center}
\caption{Same as in Fig.~\ref{fig:TEplanck2015} but for the $E$-mode
  power spectrum obtained from Planck $2015$. Figure taken from
  Ref.~\cite{Ade:2015xua}.}
\label{fig:EEplanck2015}
\end{figure}

We have also mentioned that gravitational waves are produced during
inflation. The corresponding treatment is very similar to the one we
have just described. In particular, the tensor power spectrum
$\calP_h$ can be written in the same way as Eq.~(\ref{spectrumsr}),
namely
\begin{equation} 
\label{spectrumsr}
\calP_h(k)= \calP_{h 0}(k_{_{\rm P}})
\left[a_0^{_{({\rm T})}} + 
a_1^{_{({\rm T})}} \ln \left(\dfrac{k}{k_{_{\rm P}}}\right) 
+ \frac{a_2^{_{({\rm T})}}}{2} \ln^2\left(\dfrac{k}{k_{_{\rm P}}}\right)
+ \cdots \right]\, ,
\end{equation}
with a scale invariant overall amplitude that can be expressed as
\begin{equation}
\label{eq:tensoramp}
\calP_{h \zero} =\frac{2 H_*^2}{\pi^2 \Mp^2}\,.
\end{equation}
This time, and contrary to scalar perturbations, the amplitude only
depends on the Hubble parameter during inflation. This has a very
important implication: if one can measure the amplitude of tensor
power spectrum, then one immediately determines the expansion rate
during inflation or, in other words, the energy scale of
inflation. Unfortunately, the inflationary gravitational waves have
not yet been detected. As for scalar perturbations, the tensor power
spectrum has small scale dependent logarithmic corrections which can
be written as~\cite{Leach:2002ar}
\begin{eqnarray}
a_{0}^{\usssPTP} &=&
 1 - 2\left(C + 1\right)\epsilon_{1*} 
 + \left(2C^2 + 2C + \frac{\pi^2}{2} - 5\right) 
 \epsilon_{1*}^2 \nonumber \\ 
& + & \left(-C^2 - 2C + \frac{\pi^2}{12} - 2\right) 
 \epsilon_{1*}\epsilon_{2*} +\cdots \, ,\\
a_{1}^{\usssPTP} &=& 
 - 2\epsilon_{1*} + 2(2C + 1)\epsilon_{1*}^2 
 - 2(C + 1)\epsilon_{1*}\epsilon_{2*}+\cdots  \, ,
\label{eqn:at1} \\
a_{2}^{\usssPTP} &=& 4\epsilon_{1*}^2 - 2\epsilon_{1*}\epsilon_{2*}
+\cdots  \, , \\
a_{3}^{\usssPTP} &=& \calO(\epsilon_{n*}^3)\, ,
\label{eqn:at2}
\end{eqnarray}
corresponding to tensor spectral index given by
\begin{equation}
\label{eq:gwspecindices}
\nT=-2\epsilon_1,
\end{equation}
an exact scale invariance corresponding, with these conventions, to
$\nT=0$ (and not one as for the scalars). Since, by definition of what
inflation is, one has $\epsilon_1>0$, this means that $\nT<0$, i.e. we
say that inflation predicts a red power spectrum (that is to say more
power on large scales) for gravitational waves. It is also interesting
to measure the relative amplitude of the tensors compared to the 
scalars and this is done in terms of the parameter $r$ defined by
\begin{equation}
\label{eq:defr}
r \equiv \frac{\calP_h}{\calP_\zeta}=16\epsilon_{1*}.
\end{equation}
Clearly, since $\epsilon_{1*}\ll 1$, tensor are sub-dominant which is
compatible with the fact that they have not yet been
detected~\cite{Ade:2015tva,Martin:2014lra}.

\subsection{Constraints on Inflation}
\label{subsec:constraints}

After having discussed the main features and predictions of the
inflationary scenario, let us now review what the CMB Planck data
imply for inflation. The Planck data are represented in
Figs.~\ref{fig:TTplanck2015}, \ref{fig:TEplanck2015}
and~\ref{fig:EEplanck2015}. As already mentioned, the most important
discovery made by the Planck satellite is probably the measurement of
the scalar spectral index which is found to be~\cite{Ade:2015lrj}
\begin{align}
\nS &= 0.9645\pm 0.0049.
\end{align}
It is a crucial result since this is the first time that a deviation
from $\nS=1$ is measured at a statistical significant level (say, more
than $5\sigma$). It is clearly a strong point in favor of
inflation. As was discussed previously, inflation also predicts the
presence of a background of gravitational waves and, unfortunately, we
do not yet have a detection of those primordial gravity waves. This
means that we only have an upper bound on the parameter $r$, namely
\begin{align}
r\lesssim 0.07
\end{align}
obtained by combining the Planck data and the BICEP/Keck
data~\cite{Ade:2015tva}. As already mentioned, the Planck data are
also compatible with no Non-Gaussianity~\cite{Ade:2013ydc} and no
non-adiabatic perturbations~\cite{Ade:2015lrj} which is compatible
with the simplest model of inflation.

\begin{figure}
\begin{center}
\includegraphics[width=\wdblefig]{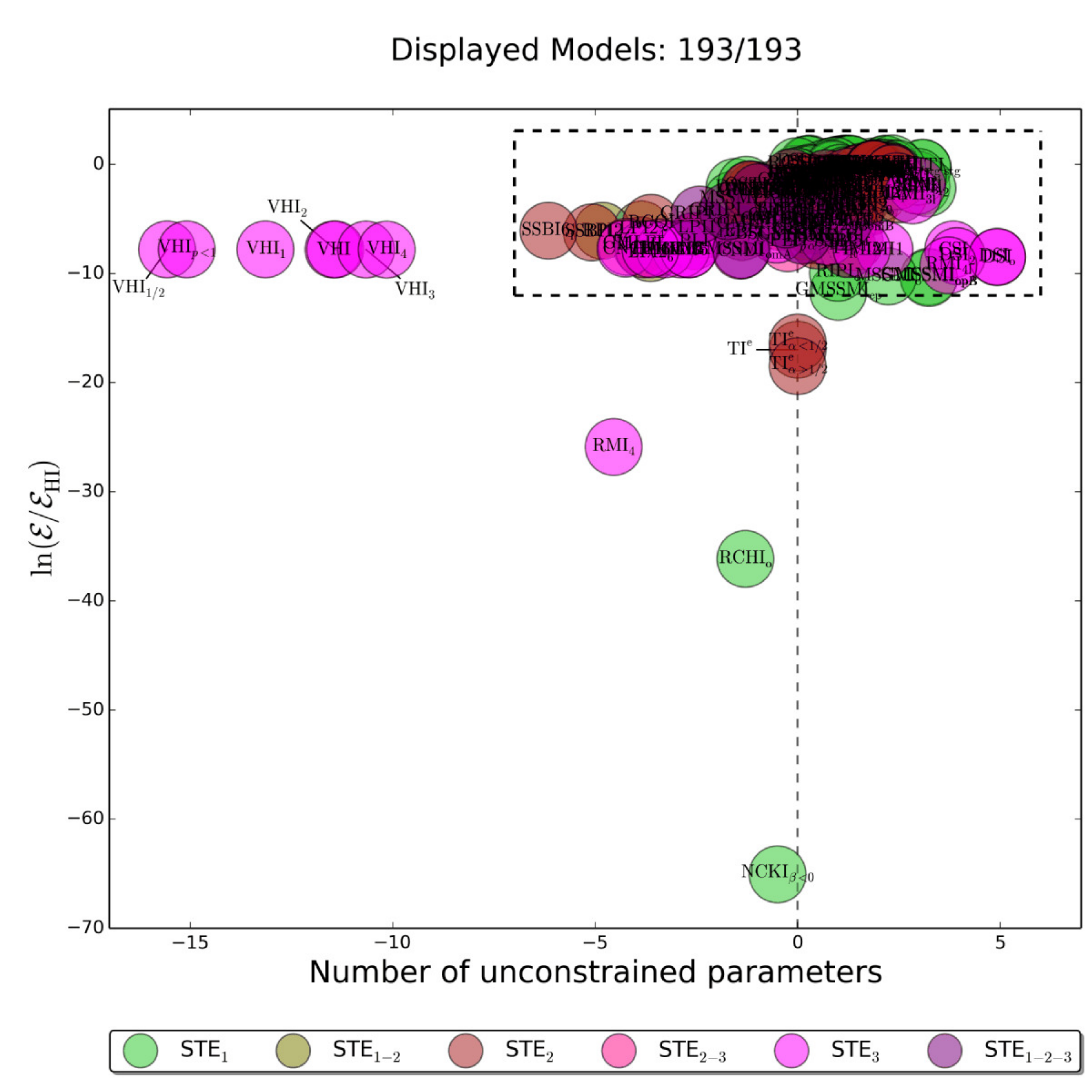}
\includegraphics[width=\wdblefig]{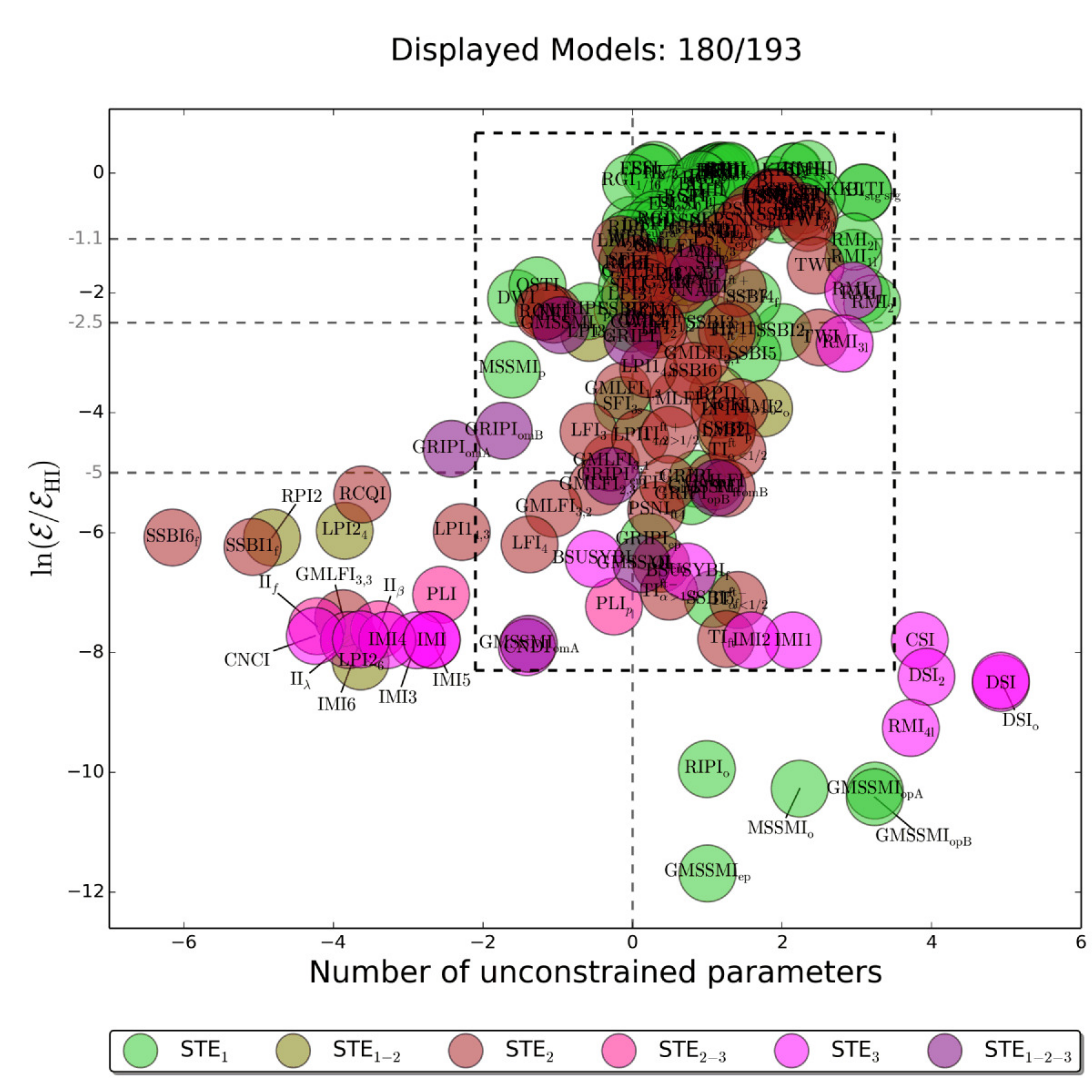}
\includegraphics[width=\wdblefig]{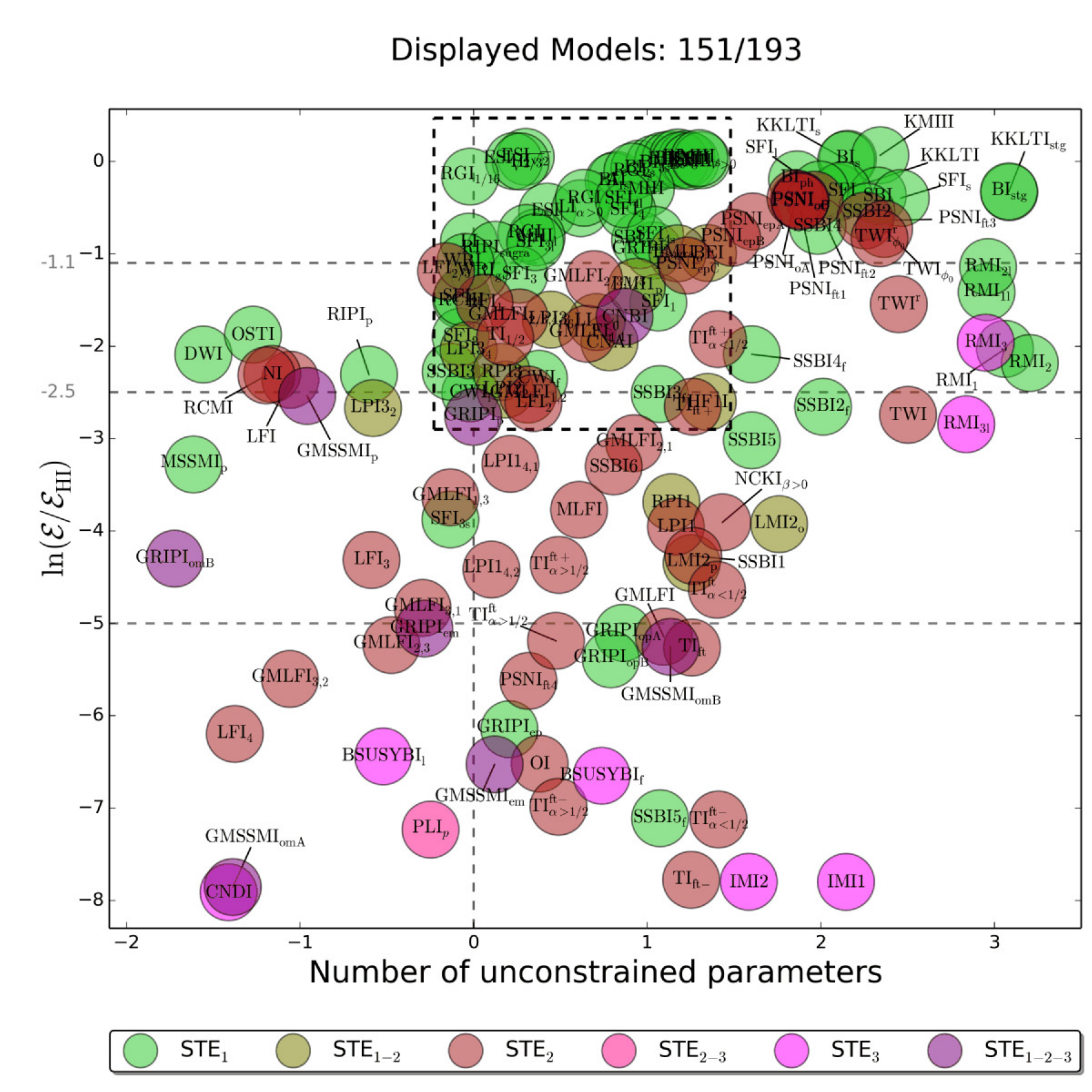}
\includegraphics[width=\wdblefig]{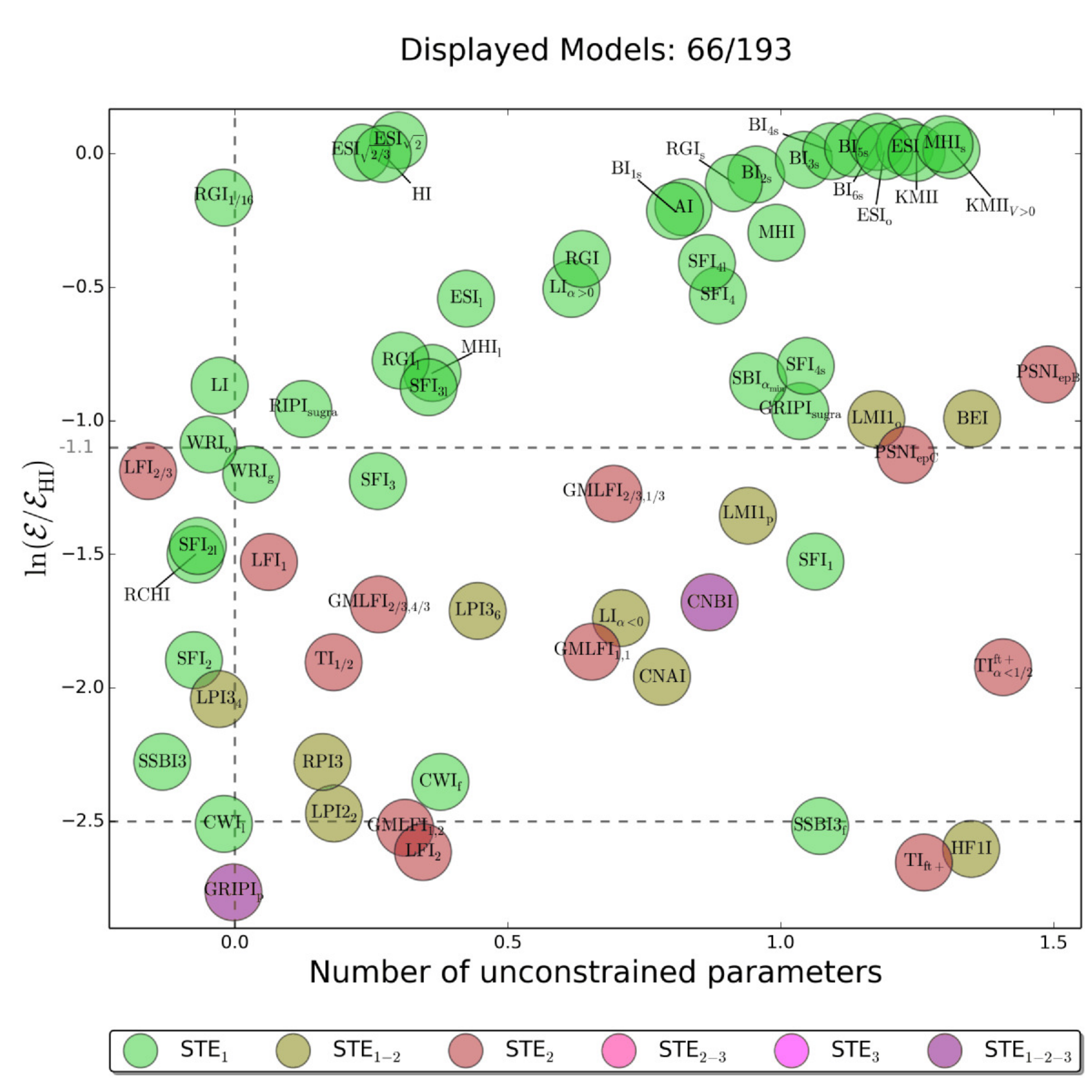}
\end{center}
\caption{Inflationary models in the space
  $\left(N_{\rm uc}, \ln \Bref{i} \right)$. $N_{\rm nuc}$ represents
  the number of unconstrained parameters of a given model while
  $\Bref{i}$ is the evidence of a given model ``$i$'' to evidence of a
  reference model ratio. Each model is represented by a circle (the
  radius of which has no meaning) with its acronym, taken from
  Ref.~\cite{Martin:2013nzq}, written inside. The four panels
  corresponds to successive zooms towards the best region (indicated
  by the dashed rectangles). Figures taken from
  Ref.~\cite{Martin:2013nzq}.}
\label{fig:compevid}
\end{figure}

\begin{figure}
\begin{center}
\includegraphics[width=13cm]{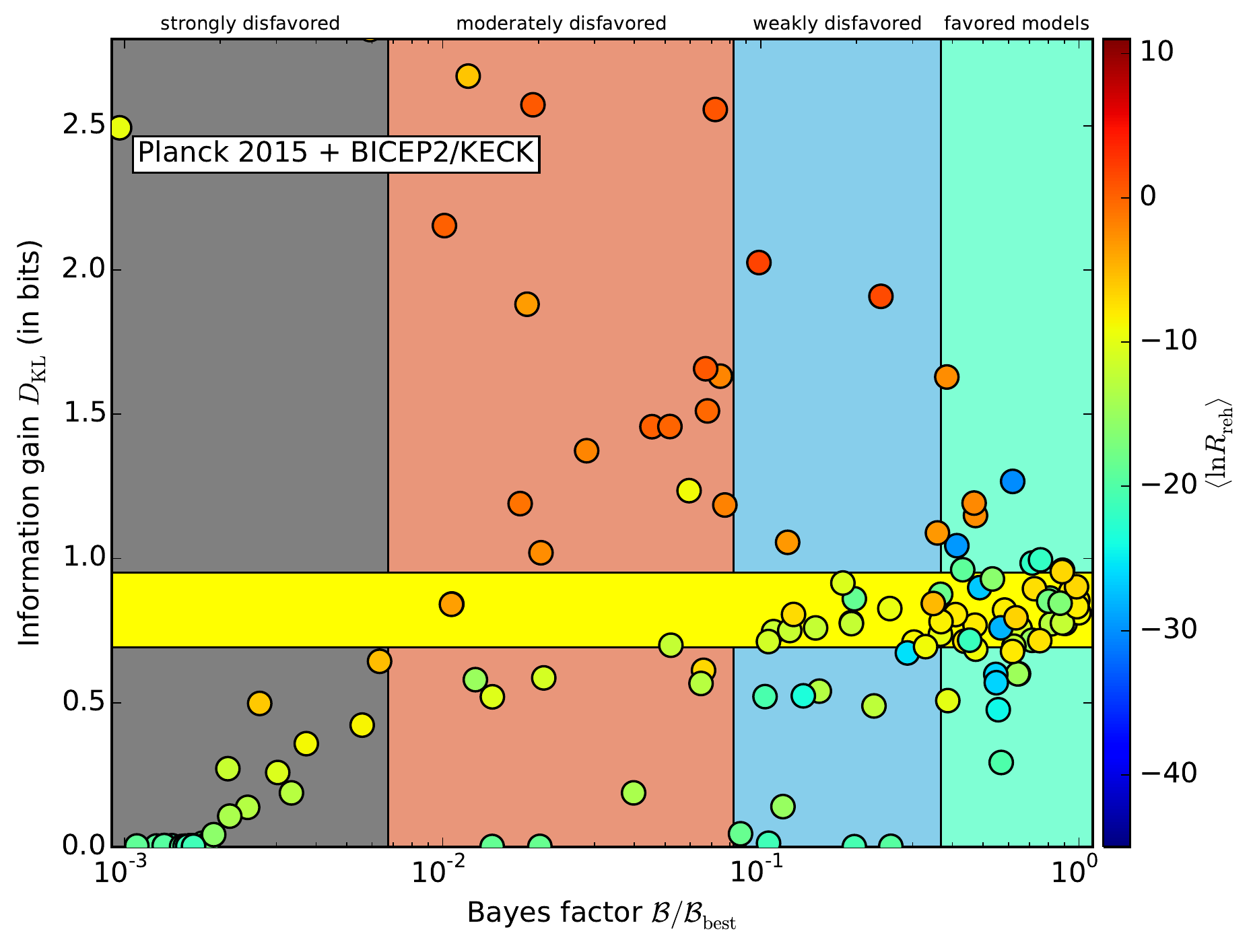}
\end{center}
\caption{Observational constraints on reheating from the Planck and
  BICEP2/KECK data. The vertical axis is a measure of how tight is the
  constraint on the reheating parameter while the horizontal axis
  represents the Bayesian evidence, namely the performance of a
  model. Each circle represent a model of inflation. The color code
  gives the best value of the reheating parameter.}
\label{fig:reheating}
\end{figure}

One can also use the Planck data to constrain the shape of the
inflationary potential. The performance of a model can be described by
two numbers: the Bayesian evidence~\cite{Trotta:2008qt,Trotta:2017wnx}
which characterizes the ability of the model to fit the data in a
simple way and the Bayesian complexity~\cite{Kunz:2006mc} which is
related to the number of unconstrained parameters (given a data
set). A good model is a model that has a large Bayesian evidence and
no unconstrained parameters. In Fig.~\ref{fig:compevid}, we have
represented the Bayesian evidence and complexity for nearly $200$
models of inflation, given the Planck
data~\cite{Martin:2006rs,Lorenz:2007ze,Lorenz:2008je,Martin:2010hh,Martin:2013tda,Martin:2013nzq,Martin:2014lra,Martin:2015dha}. Based
on this analysis, it is found that potentials with a plateau are
favored by the data, the prototypical example being the Starobinsky
model~\cite{Starobinsky:1980te} for which the potential is given by
\begin{align}
\label{eq:potstaro}
V(\phi )=M^4\left(1-e^{-\sqrt{2/3}\phi/\Mp}\right)^2.
\end{align}
We recall that this potential is represented in Fig.~\ref{fig:pot}.

Reheating can also be constrained by means of the Planck
data~\cite{Martin:2006rs,Martin:2010kz,Martin:2014nya,Martin:2015dha,Martin:2016oyk},
see Fig.~\ref{fig:reheating}. We have seen that the only piece of
information about the end of inflation that can be extracted from CMB
data is the posterior distribution of the reheating parameter. In
order to quantify whether the constraint is tight or not, one has then
to compare the posterior to the prior. In technical terms, this is
given by the Kullback-Leibler divergence, $D_{\rm KL}$, between the
prior and the posterior. In Fig.~\ref{fig:reheating}, we have
represented $D_{\rm KL}$ as a function of the Bayesian evidence for
the nearly $200$ models of inflation already studied in
Fig.~\ref{fig:compevid}. Each model is represented by a circle. The
yellow band corresponds to the one-sigma deviation around the mean
value, which is given by $\langle D_{\rm KL}\rangle =0.82\pm 0.13$.
This corresponds to an information of almost one bit, and, therefore,
this confirms that reheating is constrained by CMB data. Of course, it
is not straightforward to translate these constraints into constraints
on the reheating temperature unless one specifies $\wrehbar $
explicitly, in which case the reheating parameter and the reheating
temperature are in one-to-one correspondence.

\section{Is Inflation Fine-Tuned? Choosing the 
Free Parameters of the Inflationary Potential}
\label{sec:ft}

In this section, we turn to the question of whether inflation, which
was invented in order to solve fine-tuning problems, is itself fine
tuned. Let us discuss the first aspect of the problem, namely how the
parameters of the potential must be chosen and what their numerical
values are in order for the model to correctly account for the
data. Let us start with a particular model, namely Large Field Model
(LFI) for which the potential is given by
\begin{align}
V(\phi)=M^4\left(\frac{\phi}{\Mp}\right)^p,
\end{align}
where $M$ and $p$ are two free parameters. Using
Eq.~(\ref{eq:srtrajectory}), one can calculate the slow-roll
trajectory and one finds
\begin{align}
\label{eq:trajeclfi}
\phi(N)=\sqrt{\phi_{\rm ini}^2-2p\Mp^2(N-N_{\rm ini})}.
\end{align}
In order to calculate the spectral index and the scalar-to-tensor
ratio, one must calculate the Hubble flow parameters. Using the
expressions of $\epsilon_1$ and $\epsilon_2$ in the slow-roll
approximation, one obtains, see Eqs.~(\ref{eq:epsfirst})
and~(\ref{eq:eps2}),
\begin{equation}
\epsilon_1=\frac{p^2\Mp^2}{2\phi^2}, \quad 
\epsilon_2=\frac{2p\Mp^2}{\phi^2}.
\end{equation}
This immediately leads to the vacuum expectation value at which
inflation ends since the condition $\epsilon_1=1$ implies
$\phi_{\rm end}/\Mp=p/\sqrt{2}$. Then, we must evaluate the Hubble
flow parameters at the time that was previously denoted with a star,
namely the time at which the pivot scale crossed out the Hubble radius
during inflation. Using the slow-roll trajectory, it is easy to show
that $\phi_*^2/\Mp^2=p^2/2+2p\Delta N_*$, where
$\Delta N_*=N_{\rm end}-N_*$ with $N_{\rm end}$ the number of e-folds
at the end of inflation and $N_*$ the number of e-folds at Hubble
radius exit. In terms of $\Delta N_*$, the Hubble flow parameters read
\begin{align}
\epsilon_{1*}=\frac{p}{4(\Delta N_*+p/4)}, \quad 
\epsilon_{2*}=\frac{1}{\Delta N_*+p/4}.
\end{align}
As a consequence, one has 
\begin{align}
\nS-1=-\frac{p+2}{2\Delta N_*+p/2}, \quad r=\frac{4p}{\Delta N_*+p/4}.
\end{align}
The measurements of $\nS$ and the constraints on $r$ can therefore
allow us to put constraints on the parameter $p$. But we also see that
the spectral index and the tensor-to-scalar ratio do not depend on the
other free parameter, namely $M$. This one is in fact fixed by the
amplitude of the fluctuations (i.e. the ``COBE normalization''), that
is to say by the fact that $\delta T/T\sim 10^{-5}$. Using
Eq.~(\ref{eq:scalaramp}) and the slow-roll approximation for the
Friedmann equation, one obtains that
\begin{align}
\label{eq:cmbnormalization}
\frac{M^4}{\Mp^4}=12\pi^2 p^2
\left(\frac{\phi_*}{\Mp}\right)^{-p-2}\calP_{\zeta \zero}
=12\pi^2p^2\left(\frac{p^2}{2}+2p\Delta N_*\right)^{-p/2-1}
\calP_{\zeta \zero}.
\end{align}
The value of $\calP_{\zeta \zero}$ is provided by the Planck $2015$
data~\cite{Ade:2015tva,Ade:2015xua,Ade:2015lrj} 
\begin{align}
\label{eq:cmbnormaPlanck}
\ln \left(10^{10}\calP_{\zeta \zero}\right)
=3.094\pm 0.0049,
\end{align}
and one finds that $M/\Mp\simeq 1.3 \times 10^{-3}$ for $p=2$ and
$M/\Mp\simeq 3\times 10^{-4}$ for $p=4$. In order to obtain these
numbers we have assumed $\Delta N_*=55$ and a comment is in order at
this stage. In principle, one should not assume a value for
$\Delta N_*$ since it is determined once the reheating temperature and
the mean equation of state parameter during reheating have been
chosen~\cite{Martin:2006rs,Martin:2010kz,Martin:2014nya,Martin:2015dha,Martin:2016oyk}. It
can be quite dangerous to choose a ``reasonable'' value blindly
because, sometimes, it could imply a reheating energy density higher
that the energy density at the end of inflation which is clearly
meaningless. In fact, the dependence in $\Delta N_*$ of $\nS$ and $r$
is precisely the reason why one can use the CMB to put constraints on
the reheating epoch, as explained in the previous sections. Indeed,
$\Delta N_*$ cannot take arbitrary values otherwise the corresponding
spectral index and tensor to scalar ratio would be incompatible with
the data. But since $\Delta N_*$ depends on $T_{\rm reh}$ and
$\wrehbar$, this means that those quantities cannot take arbitrary
values as well or, to put it differently, are constrained by the CMB
data. Nevertheless, one can show that, for large-field inflation,
$\Delta N_*$ can vary in a quite small range around the value
$\Delta N_*=55$ and this is the reason why we choose this
value. Considering another value would not affect much our numerical
estimate and would change nothing to the present discussion.

The estimates of the mass scale $M$ derived above show that inflation
in this model takes place around the GUT scale. But let us consider
the case $p=4$ and write the potential as $V(\phi)=\lambda \phi^4$
where $\lambda$ is a dimensionless coupling constant. Clearly,
$\lambda =M^4/\Mp^4$ which implies that $\lambda \sim 10^{-13}$.  This
very small value can be viewed as a fine tuning, at least if one
adopts the standard lore that absence of fine tuning means that
dimensionless quantities should be ``naturally'' of order one. Let us
now consider the case $p=2$ and write the corresponding potential as
$V(\phi)=m^2\phi^2/2$ where $m$ is the mass of the inflaton field. In
that case one has $m=\sqrt{2}(M/\Mp)^2\Mp$ which leads to
$m\sim 2\times 10^{-6}\Mp$. Is this fine tuning? In absence of a
rigorous definition of fine tuning, this is hard to tell. But one can
notice that $m/H\sim \sqrt{6}(2+4\Delta N_*)^{-1/2}<1$, which may be
viewed as unnatural. Indeed, we expect the mass of the inflaton to be
corrected by high-energy physics according to
$m^2\rightarrow m^2+gM^2\ln(\Lambda/\mu)$, where $\mu$ is the
renormalization scale, $M>\Lambda$ the mass of a heavy field,
$\Lambda$ the cut-off already discussed in Sec.~\ref{subsec:realizing}
and $g$ the coupling constant. The presence of these corrections
implies $m/H\sim 1$ and keeping $m/H<1$ may be problematic. This
problem is also known as the $\eta$-problem of
inflation~\cite{Baumann:2014nda}. But, at least, this illustrates the
fact that the fine-tuning of the parameters (if any) can depend on the
potential. For this reason, it is worth studying the situation for the
Starobinsky potential~(\ref{eq:potstaro}) since this is the favored
model.

The Starobinsky model can be derived from different
assumptions. Historically, it was derived by considering $R^2$
corrections to the Einstein-Hilbert action. However, more recently, it
was realized that it can also be viewed as a scenario in which the
inflaton field is the Higgs field, this one being non-minimally
coupled to gravity. In technical terms, the action of the model reads
\begin{align}
S=\frac{\Mp^2}{2}\int {\rm d}^4 {\bm x}\sqrt{-g}
\left[\left(1+\xi h^2\right)R-g^{\mu \nu}\partial_{\mu}h
\partial_{\nu}h-2\Mp^2\frac{\lambda}{4}\left(h^2-\frac{v^2}{\Mp^2}
\right)^2\right],
\end{align}
where $v$ is the Higgs vacuum expectation value and $\lambda $ the
self-interacting coupling constant. The quantity $\xi$ is a
dimensionless constant which describes the non-minimal coupling. If
one defines the field $\phi$ by
${\rm d}[\phi/(\sqrt{2}\Mp)]/{\rm d}h
=\sqrt{1+\xi(1+6\xi)h^2}/[\sqrt{2}(1+\xi h^2)]$
then this field has a standard Lagrangian with a potential which is
exactly the potential of Eq.~(\ref{eq:potstaro}), the scale $M$ being
given by
\begin{equation}
M^4=\frac{\Mp^4 \lambda}{4\xi^2}.
\end{equation}
Then the COBE normalization, which constrains the value of $M$, leads 
to
\begin{equation}
\xi \sim 46000 \sqrt{\lambda},
\end{equation}
where $\lambda =m_{_{\rm H}}^2/v^2$ with $v\simeq 175\, \GeV$ and
$m_{_{\rm H}}\simeq 125\, \GeV$. We see that $\xi\gg 1$, which can
imply many issues as far as the consistency of the model is
concerned. 

\begin{figure}
\begin{center}
\includegraphics[width=7.cm,height=6cm]{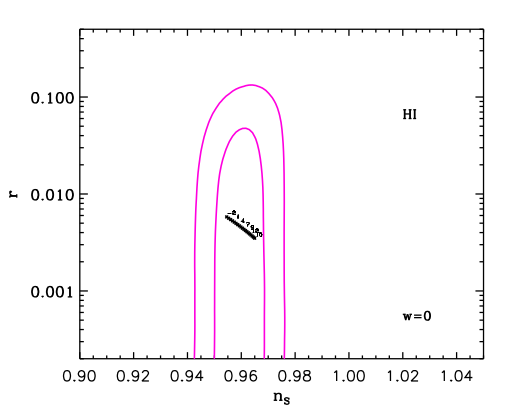}
\includegraphics[width=7cm,height=5.85cm]{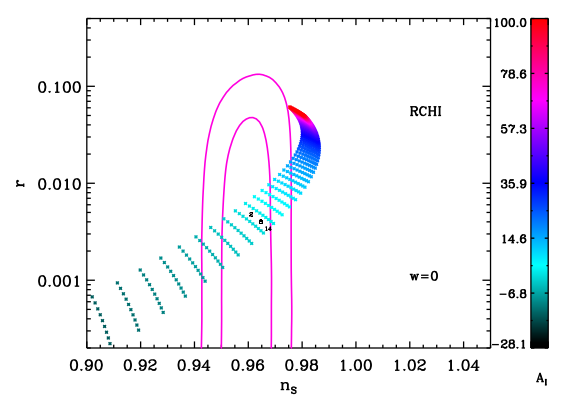}
\end{center}
\caption{Predictions in the $(\nS,r)$ space for two inflationary
  models, Higgs inflation (left panel) and Higgs inflation with
  quantum corrections (right panel), see
  Ref.~\cite{Martin:2013tda}. In both cases, a very good fit can be
  found but, in the case of Higgs inflation with quantum corrections,
  this requires a tuning of the free parameters characterizing the
  model. As a consequence, the Bayesian evidence is smaller than that
  of Higgs inflation and, given the data, the model is seen as ``less
  good''.}
\label{fig:compevidence}
\end{figure}

The overall picture that emerges from this section is that it is
difficult to say whether the parameters of the inflationary potential
are necessarily fine tuned if one wants to account for the data. It is
clear that this question is model dependent. For some potentials, the
fine-tuning seems to be present (at least if one adopts a naive
definition of fine-tuning) but for others, and in particular those
that fit the data well, it is unclear whether this is the case. The
situation of the Starobinsky model is particularly interesting. The
coupling between gravity and the Higgs is not small, or is not
perturbative, which may lead to technical difficulties but this strong
coupling problem is not necessarily associated with a fine-tuning
problem. Here, we are just missing an objective definition of what
fine tuning is.

In fact, one could argue that such a definition exists and is nothing
but the Bayesian evidence considered in
Sec.~\ref{subsec:constraints}. Technically, the Bayesian evidence is
the integral of the likelihood over prior space but its meaning can
easily be grasped intuitively. Let us consider a model depending on,
say, one free parameter. If, for all values of the parameter in the
prior range, one obtains a good fit, then the Bayesian evidence is
``good''. This is for instance the case of the model in
Fig.~\ref{fig:compevidence} (left panel). Different points correspond
to different values of the reheating temperature but all points are
within the $1\sigma$ Planck contour. On the contrary, if one needs to
tune the value of the free parameter in order to have a good fit, then
the Bayesian evidence will be ``bad''. This is the case for the model
in Fig.~\ref{fig:compevidence} (right panel). In order to have a good
compatibility with the data (\ie points within the $1\sigma$ contour),
one needs to tune the parameter $A_{_{\rm I}}$ (which controls the
amplitude of the quantum corrections) and the Bayesian evidence is
``bad''. In other words the wasted parameter space is
penalized. Obviously, the smaller the range of $A_{_{\rm I}}$ leading
to a good fit (compared to the prior), the smaller the evidence. We
conclude that the evidence is a good, objective, measure of
fine tuning. In this sense, the Starobinsky model is the best model
because it is the less fine-tuned one.

\section{Inflationary Initial Conditions}
\label{sec:ini}

\subsection{Homogeneous Initial Conditions}
\label{subsec:homoini}

\begin{figure}
\begin{center}
\includegraphics[width=11cm]{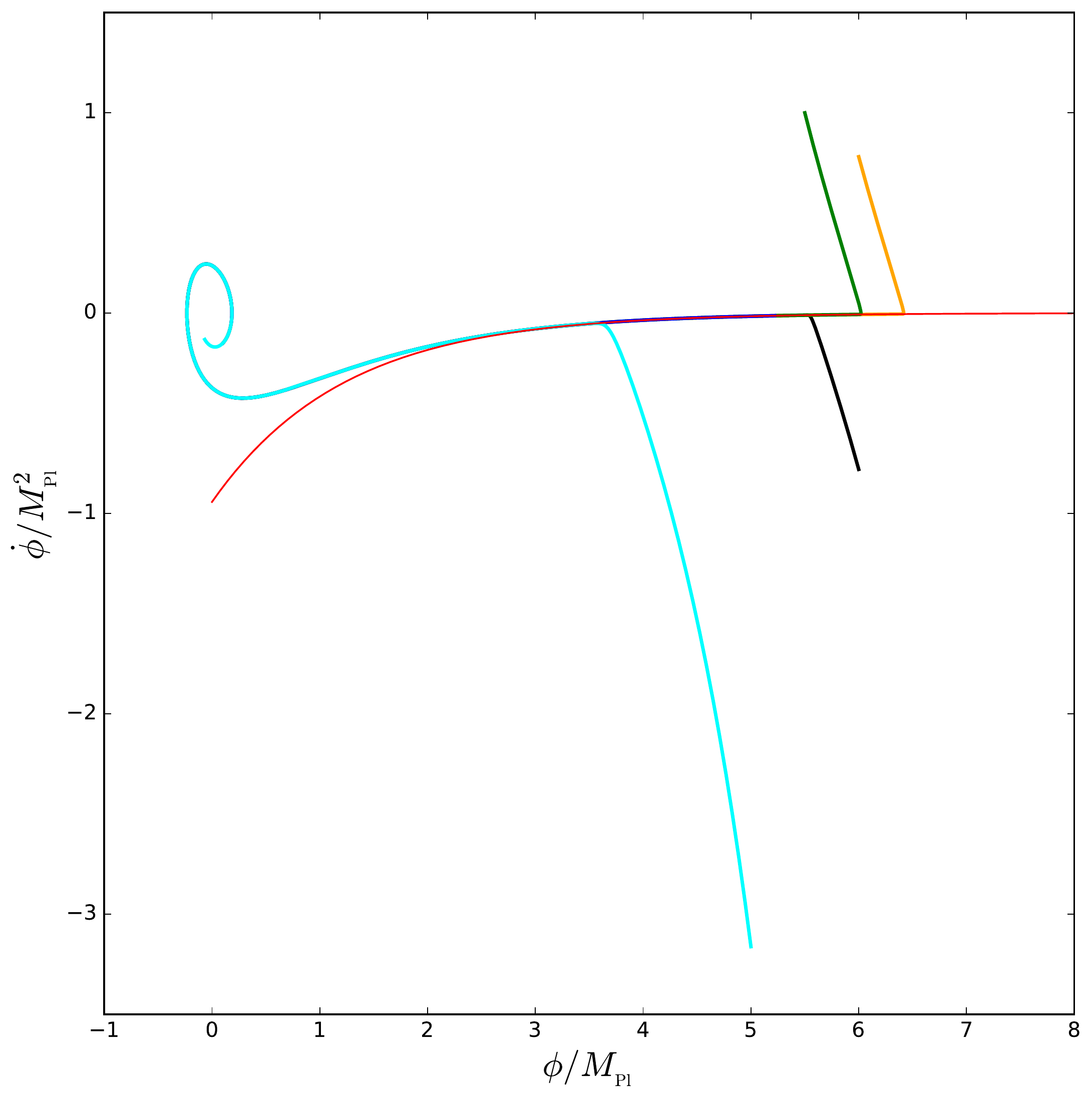}
\end{center}
\caption{Phase space for the Starobinsky model. The red line
  represents the slow roll trajectory while the other colored lines
  correspond to exact trajectories (numerically computed) with
  different initial conditions. It is evident from the plot that the
  slow-roll trajectory is an attractor in phase space. This question
  is treated in an exhaustive way in Ref.~\cite{Chowdhury:2019otk}.}
\label{fig:phasespacestaro}
\end{figure}

Let us now discuss another type of possible fine tuning, namely the
initial conditions (see also Ref.~\cite{Chowdhury:2019otk} for a
detailed discussion of this question). We have seen previously that
one of the main motivations for inflation is to avoid the fine tuning
of the initial conditions that is needed in order for the standard
model to work. If our solution to that issue were also fine tuned then
one could wonder whether something has been gained or not. In fact
this problem has different facets. If one restricts ourselves to an
homogeneous and isotropic solution, then the only question is how we
should choose $\phi_{\rm ini}$ and $\dot{\phi}_{\rm ini}$. The
slow-roll trajectory corresponds to
$\dot{\phi}_{\rm ini}\simeq -V_{\phi}(\phi_{\rm ini})/[3H(\phi_{\rm
  ini})]$
and, therefore, there could be the worry that we have to tune the
initial velocity to this value. However, this is not the case because
the slow-roll trajectory is an attractor as can be seen in
Fig.~\ref{fig:phasespacestaro}. It is true that, for some $V(\phi)$,
the corresponding basin of attraction is very small. This is for
instance the case for Small Field Inflation (SFI) if the size of the
hilltop part is sub-Planckian~\cite{Chowdhury:2019otk}. However, on
the contrary, it can be very large for other models, such as Large
Field Inflation (LFI) (let us also notice that the existence of an
attractor is immune to stochastic effects, see
Ref.~\cite{Grain:2017dqa}). The interesting point is that it is also
the case for the Starobinsky model and plateau
potentials~\cite{Chowdhury:2019otk}, namely the models favored by the
data. In this sense, in this restricted framework, there is no fine
tuning of the initial condition.

\subsection{Anisotropic Initial Conditions}
\label{subsec:aniini}

Obviously, however, the previous analysis is not entirely
satisfactory. Indeed, we start from a homogeneous and isotropic
situation while inflation is precisely supposed to explain why our
Universe is homogeneous and isotropic. The analysis can be improved by
considering that, initially, the Universe is not isotropic (but still
homogeneous)~\cite{Steigman:1983hr,Anninos:1991ma,Turner:1986gj}. For
this purpose let us consider the following metric (Bianchi I model):
\begin{align}
{\rm d}s^2=-{\rm d}t^2+a_i^2(t)\left({\rm d}x^i\right)^2,
\end{align}
that is to say we now have one scale factor for each space
direction. This metric can also be rewritten as
${\rm d}s^2=-{\rm d}t^2 +a^2(t)\gamma_{ij}{\rm d}x^i{\rm d}x^j$ with
\begin{align}
a(t)\equiv \left[a_1(t)a_2(t)a_3(t)\right]^{1/3}, 
\end{align}
and 
\begin{align}
\gamma_{ij}=
\begin{pmatrix}
e^{2\beta_1(t)} & 0 & 0 \\
0 & e^{2\beta_2(t)} & 0 \\
0 & 0 & e^{2\beta_3(t)} 
\end{pmatrix},
\end{align}
with $\sum_{i=1}^{i=3}\beta_i=0$. As usual, one can introduce the
conformal time $\eta$ in terms of which the metric can be expressed as
${\rm d}s^2=a^2(\eta)\left(-{\rm d}\eta^2+\gamma_{ij}{\rm d}x^i{\rm
    d}x^j\right)$.
Then, the next step is to introduce the shear $\sigma_{ij}$ which is
defined by (as usual a prime denotes a derivative with respect to
conformal time)
\begin{align}
\sigma_{ij}=\frac12 \gamma_{ij}'=
\begin{pmatrix}
\beta_1'e^{2\beta_1} & 0 & 0 \\
0 & \beta_2'e^{2\beta_2} & 0 \\
0 & 0 & \beta_3'e^{2\beta_3} 
\end{pmatrix}.
\end{align}
Assuming that matter is described by a scalar field, it is then easy to 
write the Einstein equations. They read
\begin{align}
3\frac{{\cal H}^2}{a^2}&=\frac{\rho}{\Mp^2}+\frac{\sigma^2}{2a^2}
=\frac{1}{\Mp^2}\left[\frac{\phi'^2}{2a^2}
+V(\phi)\right]+\frac{\sigma^2}{2a^2}, \\
-\frac{1}{a^2}\left({\cal H}^2+2{\cal H}'\right)&=
\frac{p}{\Mp^2}+\frac{\sigma ^2}{2a^2}
=\frac{1}{\Mp^2}\left[\frac{\phi'^2}{2a^2}
-V(\phi)\right]+\frac{\sigma^2}{2a^2},\\
\left(\sigma^i_j\right)'+2{\cal H}\sigma^i_j &=0,
\end{align}
where $\sigma ^2=\sigma_{ij}\sigma^{ij}=\sum_{i=1}^{i=3}\beta_i'^2$
and $\sigma ^i_j=\gamma ^{ik}\sigma_{kj}$, that is to say
\begin{align}
\sigma^i_j=
\begin{pmatrix}
\beta_1' & 0 & 0 \\
0 & \beta_2' & 0 \\
0 & 0 & \beta_3' 
\end{pmatrix}.
\end{align}
The solution for the shear can easily be found, namely
$\sigma ^i_j=S^i_j/a^2$, where $S^i_j$ is a constant tensor. This
implies that $\sigma^2=S^2/a^4$ where $S^2=S^i_jS^j_i$. As a
consequence, one sees that the shear is in fact equivalent to a stiff
fluid with an equation of state
$w_{\sigma}\equiv p_{\sigma}/\rho_{\sigma}=1$ and
$\rho_{\sigma}=\Mp^2S^2/(2a^6)$. Therefore, if initially the shear
dominates, $\rho_{\sigma}\gg \rho_{\phi}$, then the universe will
expand as $a\propto t^{1/3}$, see Eq.~(\ref{eq:scalefactorpower}), and
the expansion will not be accelerated. However, since
$\rho_{\sigma}\propto a^{-6}$ while $\rho_{\phi}$ is approximately
constant, the scalar field will eventually take over and inflation
will start. We conclude that, even if the Universe is not initially
isotropic, it will become so in the presence of a scalar field whose
energy density is dominated by its potential. In this sense, it is
legitimate to start from an isotropic situation as was done
previously. This is clearly not a fine tuning, but rather an attractor
of the dynamical evolution.

\subsection{Inhomogeneous Initial Conditions}
\label{subsec:inhomoini}

Despite the fact that taking into account the shear represents an
improvement, this still does not allow us to discuss the real
issue. For that, we need a framework where the initial state of the
Universe is neither isotropic nor homogeneous. Technically, this is
clearly very complicated since we have to solve the Einstein equations
in full generality. The only way to study these questions exactly is
therefore numerical relativity. However, some schemes of approximation
have also been developed and we now discuss them. Of course, the
perturbative approach described before, see Sec.~\ref{subsec:pert}, is
one way of taking into account the inhomogeneities. However, by
definition, these fluctuations must be small while we would like to
see whether inflation ``homogenizes'' the Universe even if it is
strongly inhomogeneous initially. Another method is the so-called
``effective-density
approximation''~\cite{Goldwirth:1989pr,Goldwirth:1991rj}, see also
Ref.~\cite{Chowdhury:2019otk}. The idea is to study an inhomogeneous
scalar field on a (isotropic and homogeneous) FLRW background and to
add to the Friedmann equation a term which describes the back-reaction
of the field gradient on the
geometry~\cite{Goldwirth:1989pr,Goldwirth:1991rj}. In practice, one
writes
\begin{align}
\phi(t,{\bm x})=\phi_0(t)+\Re\left[\delta \phi(t)
e^{i{\bm k}\cdot {\bm x}/a(t)}\right],
\end{align}
and assumes that the corresponding Klein-Gordon equation can be split
into two equations, namely
\begin{align}
\label{eq:kginhomo1}
\ddot{\phi}_0+3H\dot{\phi}_0+V_{\phi}(\phi_0) &=0, \\
\label{eq:kginhomo2}
\ddot{\delta \phi}+3H\dot{\delta \phi}+\frac{k^2}{a^2}\delta \phi &=0.
\end{align}
The Friedmann equation is then written as
\begin{equation}
\label{eq:friedinhomo}
H^2=\frac{1}{3\Mp^2}\left[\frac12 \dot{\phi}_0^2+V(\phi_0)
+\frac12 \dot{\delta \phi}^2+\frac12 \frac{k^2}{a^2}\delta \phi^2\right]
-\frac{{\cal K}}{a^2}.
\end{equation}
The wave number ${\bm k}$ should be chosen such that the wavelength of
the perturbations is much smaller than the Hubble radius, namely
$2\pi k/a\ll H^{-1}$. In the opposite limit, the contribution of
$\delta \phi$ should just be added to the background. The energy
density of the inhomogeneities
$\rho_{\delta \phi}=\rho_{\dot{\delta \phi}}+\rho_{\nabla}$, with
$\rho_{\dot{\delta \phi}}=\dot{\delta \phi}^2/2$ and
$\rho_{\nabla}=k^2\delta \phi^2/(2a^2)$ is supposed to dominate
initially (i.e. the Universe is inhomogeneous initially),
$\rho_{\delta \phi}\gg \rho_{\phi_0}$. The question is whether
$\rho_{\delta \phi}$ can decrease (i.e. the Universe becomes
homogeneous) such that, at some point, $\rho_{\phi_0}$ takes over and
inflation starts. 

\begin{figure}
\begin{center}
\includegraphics[width=12cm]{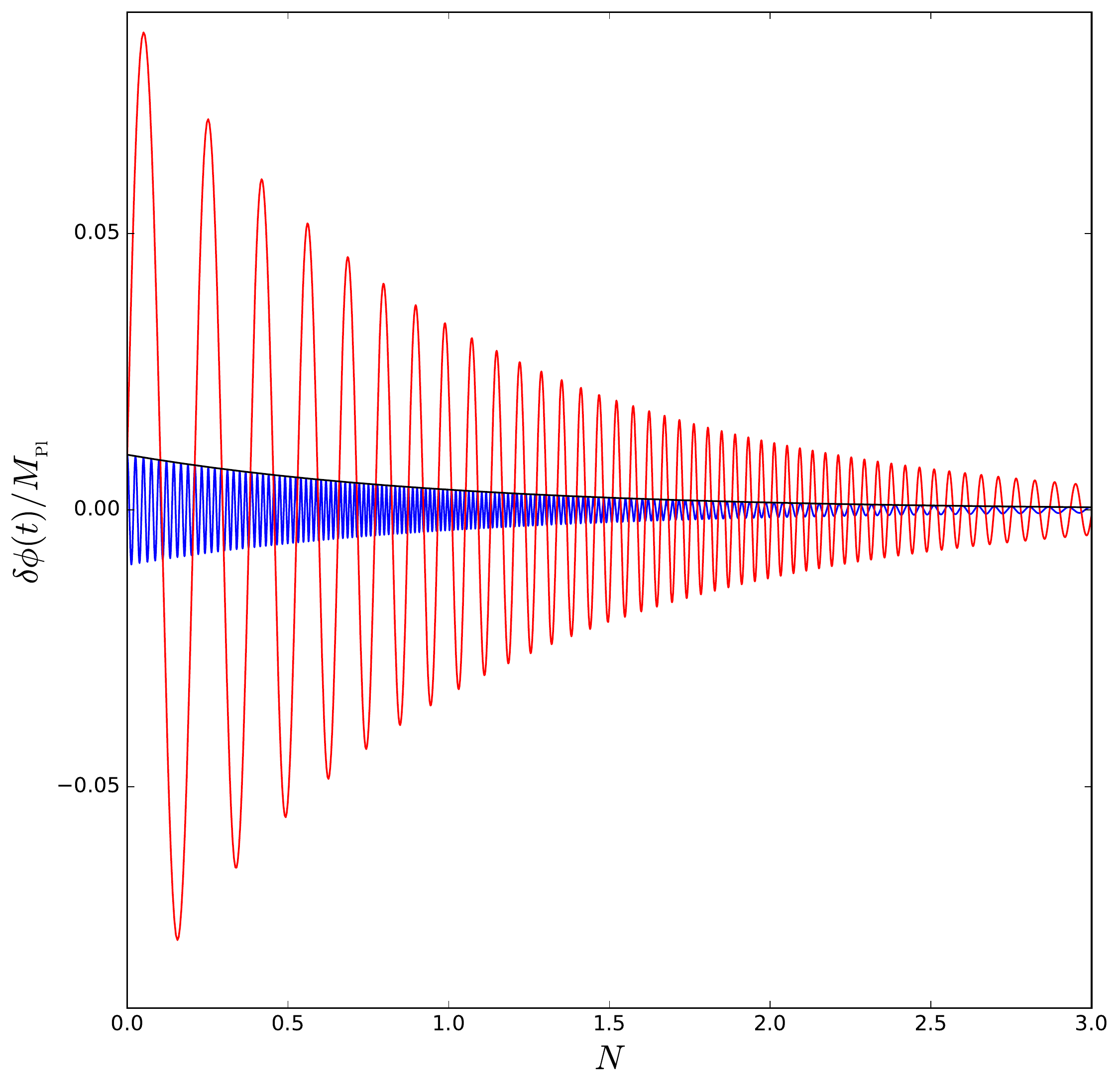}
\end{center}
\caption{Evolution of the scalar field $\delta \phi(t)$ obtained by
  numerical integration of
  Eqs.~(\ref{eq:kginhomo1}),~(\ref{eq:kginhomo2})
  and~(\ref{eq:friedinhomo}). The potential is chosen to be the
  Starobinsky one, see Eq.~(\ref{eq:potstaro}) with a scale
  $M=0.001\Mp$ which, roughly speaking, matches the CMB
  normalization. The initial value of the field $\phi_0$ is
  $\phi_0=4\Mp$ and $\dot{\phi}_0=-V_{\phi}(\phi_0)/[3V(\phi_0)]$
  (which is the slow-roll velocity). In absence of inhomogeneities,
  with these initial conditions, inflation would start and would lead
  to more than $60$ e-folds. The initial value of $\delta \phi$ is
  taken to be $0.01\Mp$ (and is therefore less than the Planck mass as
  required, see the main text) while the initial velocity of
  $\delta \phi(t)$ is given by $\dot{\delta \phi}_{\rm ini}=0$ (blue
  line) or $\dot{\delta \phi}_{\rm ini}=9\times 10^{-5}\Mp^2$ (red
  line). The scale $k$ is chosen to be $k/a_{\rm ini}=10^{-3}\Mp$ and
  the curvature is given by
  ${\cal K}/a_{\rm ini}^2\simeq 1.36\times 10^{-12}$. This implies the
  following Hubble parameter
  $H_{\rm ini}/\Mp\simeq 3.69\times 10^{-5}$ and
  $\rho_{\phi, {\rm ini}}\simeq 3.33\times 10^{-13}\Mp^4$,
  $\rho_{\delta \phi, {\rm ini}}\simeq 1.36\times 10^{-9}\Mp^4$. One
  easily checks that those initial conditions are such that
  $H_{\rm ini}^2a_{\rm ini}^2/k^2\simeq 1.36\times 10^{-3}<1$ and
  $\rho_{\phi, {\rm ini}}/\rho_{\delta \phi,{\rm ini}}\simeq
  2.44\times 10^{-4}$,
  namely the inhomogeneities largely dominate initially. Finally, the
  black line represents
  $\delta \phi_{\rm ini}/a=\delta \phi_{\rm ini}e^{-N}$ for the
  initial conditions corresponding to the blue line. We see that the
  envelope of the numerical solution indeed follows
  Eq.~(\ref{eq:deltaphisol}).}
\label{fig:deltaphi}
\end{figure}

\begin{figure}
\begin{center}
\includegraphics[width=12cm]{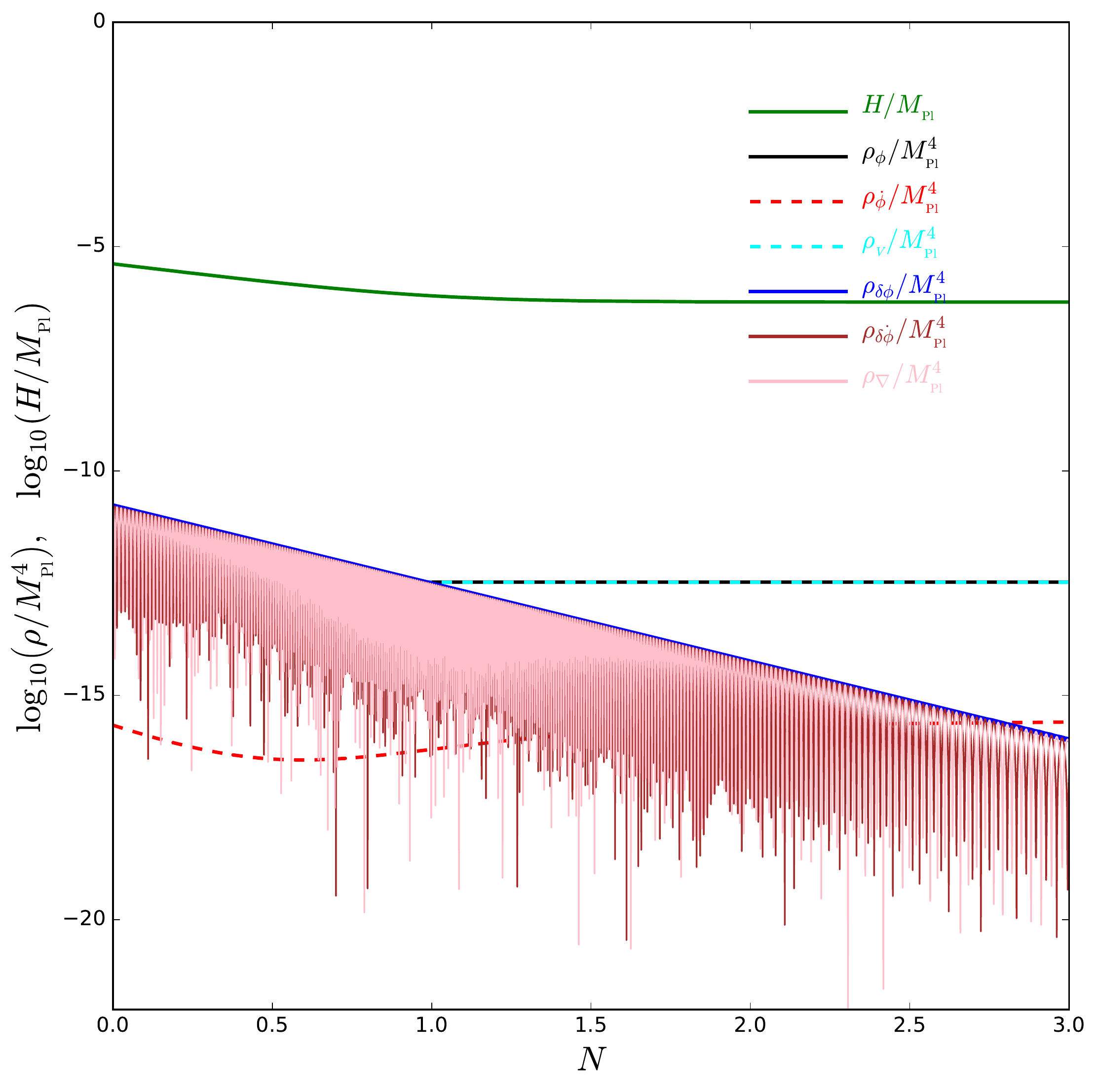}
\end{center}
\caption{Evolution of the Hubble parameter and of the various energy
  densities obtained by numerical integration of
  Eqs.~(\ref{eq:kginhomo1}),~(\ref{eq:kginhomo2})
  and~(\ref{eq:friedinhomo}). The initial conditions are those that
  lead to the red curve in Fig.~\ref{fig:deltaphi}. Initially the
  Universe is strongly inhomogeneous since
  $\rho_{\delta \phi}\gg \rho _{\phi}$. However, $\rho_{\phi}$ (black
  line) stays approximately constant while
  $\rho_{\delta \phi}\propto a^{-4}$ (blue line).  As a consequence,
  the expansion is first radiation dominated and then (at $N\simeq 1$
  in the above plot), $\rho_{\phi}$ takes over and inflation
  starts. Therefore, at least in this example, large inhomogeneities
  initially do not prevent the onset of inflation.}
\label{fig:energydensity}
\end{figure}

Let us now discuss the initial conditions. We take $\dot{\phi}_0$ and
$\phi_0$ such that, in absence of inhomogeneities, slow-roll inflation
starts. Initially, the Friedmann equations can be written as
\begin{align}
\label{eq:Friedinihomo}
\frac{3a^2H^2}{k^2}\simeq \frac12 \frac{\dot{\delta \phi}^2}{\Mp^2}
\frac{a^2}{k^2}+\frac 12\frac{\delta \phi^2}{\Mp^2},
\end{align}
since $\rho_{\delta \phi}\gg \rho_{\phi_0}$. For simplicity we have
taken ${\cal K}=0$ but it is straightforward to include the case where
curvature is not vanishing. We have already mentioned that the
effective density approximation is valid only if the wavelength of
$\delta \phi$ is smaller than the Hubble radius. This means that the
left hand side of Eq.~(\ref{eq:Friedinihomo}) must be small. This
immediately implies that $\delta \phi \ll \Mp$ and
$\dot{\delta \phi}^2/\Mp^2\ll k^2/a^2$ initially. Then, the
corresponding solution is easily determined: the field $\delta \phi$
oscillates and decays inversely proportional to the scale
factor~\cite{Chowdhury:2019otk}, namely
\begin{align}
\label{eq:deltaphisol}
\delta \phi(t)\simeq \Re\left[\frac{\delta 
\phi_{\rm ini}}{a(t)}e^{ikt/a(t)}\right].
\end{align}
This immediately implies that $\rho_{\delta \phi}$ behaves as
radiation, namely $\rho_{\delta \phi}\propto 1/a^4$. In
Fig.~\ref{fig:deltaphi},
Eqs.~(\ref{eq:kginhomo1}),~(\ref{eq:kginhomo2})
and~(\ref{eq:friedinhomo}) have been numerically integrated and the
evolution of $\delta \phi(t)$ is displayed and compared to
Eq.~(\ref{eq:deltaphisol}). We see that they match very well. In
Fig.~\ref{fig:energydensity}, we have represented the corresponding
energy densities. While $\rho_{\phi}$ remains constant,
$\rho_{\delta \phi}$ behaves as radiation and, as a consequence,
becomes very quickly sub-dominant. As a consequence, after a few
e-folds, the Universe becomes homogeneous and inflation starts as can
be seen on the evolution of the Hubble parameter (green line) which,
initially, decreases and, then, becomes almost constant.

The previous analysis seems to indicate that inflation does indeed
homogenize the Universe. However, one should be aware of its
limitations. Firstly, and obviously, there is the question of the
domain of validity of Eqs.~(\ref{eq:kginhomo1}),~(\ref{eq:kginhomo2})
and~(\ref{eq:friedinhomo}) and whether they can really represent a
strongly inhomogeneous situation. Clearly, if $2\pi k/a \simeq H^{-1}$
this is not the case and one has to rely on other
techniques. Basically, one has two possibilities: either one obtains
exact
solutions~\cite{SteinSchabes:1986sy,Calzetta:1992gv,Perez:2012pn} but
they are very hard to find in the inhomogeneous case or one uses
numerical
simulations~\cite{Goldwirth:1989pr,Goldwirth:1989vz,Goldwirth:1991rj,Goldwirth:1990pm,Albrecht:1985yf,Kung:1989xz,Brandenberger:1990wu,KurkiSuonio:1987pq,Laguna:1991zs,KurkiSuonio:1993fg}. These
ones are also complicated to study since they involve full numerical
relativity.

Historically, the first numerical
solutions~\cite{Goldwirth:1989pr,Goldwirth:1989vz,Goldwirth:1991rj,Goldwirth:1990pm}
were done under the assumption that spacetime is spherically
symmetric. This has the advantage to simplify the equations since they
only depend on time and $r$, the radial coordinate. Of course, in that
case one still has to numerically solve partial differential
equations. The metric considered in Ref.~\cite{Goldwirth:1991rj} reads
\begin{align}
{\rm d}s^2=-\left(N^2-R^2\beta ^2\right){\rm d}t^2
+2R^2\beta {\rm d}\chi {\rm d}t
+R^2\left({\rm d}\chi^2+\sin^2\chi{\rm d}\Omega^2\right),
\end{align}
where $0\le \chi\le \pi$ so that the spacelike sections are
closed. The lapse and shift functions $N$ and $\beta$ depend on $t$
and $r$ as well as the ``scale factor'' $R$. The matter content
assumed in Ref.~\cite{Goldwirth:1991rj} is a scalar field $\phi$,
which is the inflaton, and another scalar field $\psi$ without
potential and playing the role of an extra fluid. Some important
technical restrictions are also postulated on the initial
data. Firstly, it is assumed that the total energy density is
constant. Given an initial inhomogeneous distribution for the inflaton
$\phi(\chi)$, this is achieved by choosing the initial velocity of
$\psi$ to be such that the total energy density is constant. Secondly,
the initial momentum is taken to vanish. Based on the previous
calculations, see Eqs.~(\ref{eq:kginhomo1}),~(\ref{eq:kginhomo2})
and~(\ref{eq:friedinhomo}), it is argued in
Ref.~\cite{Goldwirth:1991rj} that, at least for large-field models,
this does not restrict the significance of the results. Thirdly, the
integration is performed for values of the inflaton self-coupling that
are larger than the ones needed to CMB normalize the model. Different
initial configurations for the inflaton field are considered. In
particular, the following Gaussian ansatz
\begin{align}
\phi_\uini(\chi)=\phi_0+\delta \phi \left[1-\exp\left(-\frac{\sin ^2\chi}
{\Delta ^2}\right)\right],
\end{align}
was studied in details. This initial profile depends on three
parameters: $\phi_0$, the value of the field at the origin $\chi=0$,
$\delta \phi$ which can be viewed as the value of the field on the
other side of the universe,
$\phi(\pi/2)=\phi_0+\delta \phi\left(1-e^{-1/\Delta^2}\right)$ and
$\Delta$ which represents the width of the Gaussian.

Let us now describe the results obtained for large-field inflation. If
$V(\chi=0)$ and $V(\chi=\pi/2)$, or $\phi_0$ and $\delta \phi$, are
such that, in a homogeneous situation, inflation would start, then it
also starts in the present case. If, on the contrary, $V(\chi=0)$ is
such that inflation would start in a homogeneous situation but not
$V(\chi=\pi/2)$ (therefore, the gradients are important), then
Ref.~\cite{Goldwirth:1991rj} has shown that the outcome crucially
depends on the width $\Delta$. More precisely, the numerical
simulations show that the crucial parameter is $R\Delta /H^{-1}$ which
has to be large enough in order for inflation to start. Moreover, the
larger the gradient, the shorter the duration of inflation. For small
field models, the sensitivity to the initial conditions is even
greater.

Few years later, the analysis was improved in a significant way and,
in particular, the assumption of spherical symmetry was
relaxed. Indeed,
Refs.~\cite{KurkiSuonio:1987pq,Laguna:1991zs,KurkiSuonio:1993fg} ran
simulations of strongly inhomogeneous inflation with a
three-dimensional numerical relativity code. These simulations are
such that the initial time slice has homogeneous total energy density
which means that $(\nabla \phi)^2/2<3\Mp^2H^2$ implying that 
\begin{align}
\nabla \phi<\sqrt{6}\frac{\Mp}{H^{-1}}.
\end{align}
Thus, inhomogeneities that have wavelengths smaller than the Hubble
radius must have a small amplitude or, to put it differently, large
inhomogeneities must necessarily extend over many Hubble patches. The
simulations were carried out for a quartic large field model with an
initial configuration given by 
\begin{align}
\phi_\uini(t_\uini,{\bm x})
&=\phi_0
\nonumber \\ & 
+\delta \phi \sum _{\ell, m,n=1}^2\frac{1}{\ell m n}
\sin\left(\frac{2\pi \ell x}{L}+\theta_{x\ell}\right)
\sin\left(\frac{2\pi \ell y}{L}+\theta_{ym}\right)
\sin\left(\frac{2\pi \ell z}{L}+\theta_{zn}\right),
\end{align}
where the $\theta$'s are random phases. Two runs have been carried out
in Ref.~\cite{KurkiSuonio:1993fg}, one with $L=H^{-1}$ and
$\delta \phi=0.0125\mpl$ and one with $L=32H^{-1}$ and
$\delta \phi=0.4\mpl$. In both cases, one has $\phi_0=5\mpl$ and
$H_0=0.1\mpl$. The simulations show that, in the first case, the
inhomogeneities oscillate and their amplitude is damped. At the end of
the run, the inflaton field is homogeneous. But, in the second case,
they do not oscillate (initially there are larger than the Hubble
radius) and are not damped.

In conclusion, it seems possible to start inflation with inhomogeneous
initial conditions and to homogenize the universe. However,
admittedly, the numerical simulations that have been carried out so
far all require some technical restrictions. The crucial question that
emerges from the simulations is the size of the initial homogeneous
patch. There is also a dependence in the model with large field
scenarios being the preferred class of scenarios. As a consequence,
the Starobinsky model is (again) among the good models. Let us also
notice that, even more recently, new simulations have been carried
out, see Refs.~\cite{East:2015ggf,Clough:2016ymm}. These new works
bring new insights into an issue that will probably be studied even
more in the future.

A last comment is that we have good reasons to believe the quantum
effects to play an important role at the beginning of inflation. For
this reason, studying the initial conditions at the classical level
only is maybe not sufficient and even more elaborated investigations
may be needed to settle this question.

\subsection{Initial Conditions for the Perturbations}
\label{subsec:inipert}

\begin{figure}
\includegraphics[width=1.\textwidth,height=.65\textwidth]{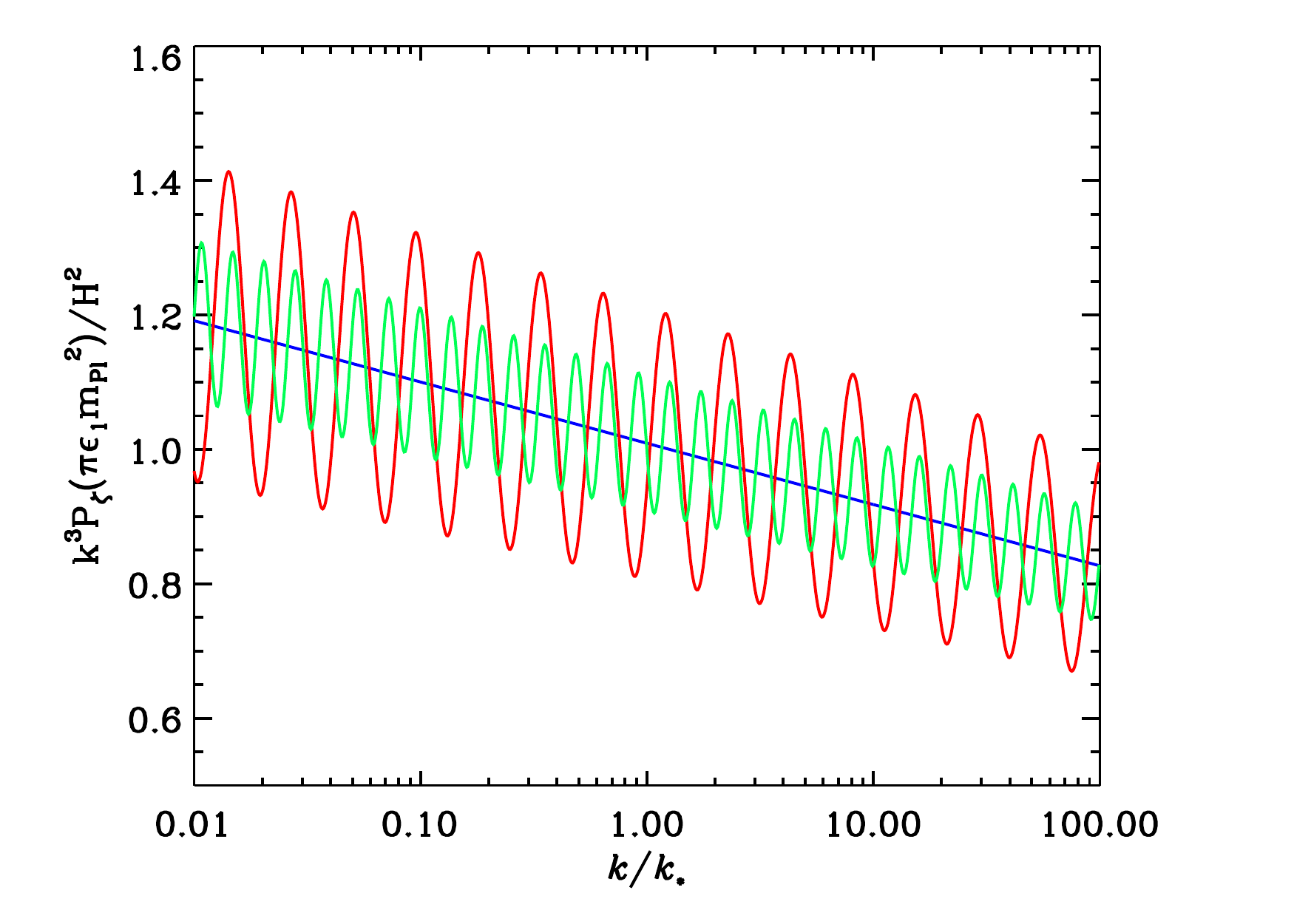}
\caption{Trans-Planckian power spectra given by
  Eq.~(\ref{eq:powerspectrum}).  The blue line corresponds to a
  vanilla model with $\vert x\vert =0$ and
  $\epsilon_1=1/(2\Delta N_*)$, $\epsilon_2=1/\Delta N_*$ with
  $\Delta N_*\simeq 50$ as predicted for the $m^2\phi^2$ inflationary
  model.  The red line corresponds to a model with the same values for
  the slow-roll parameters and $H/M_{_{\rm C}}\simeq 0.002$,
  $\vert x\vert \simeq 50$, $\varphi=3$.  Finally, the green line
  represents a model with $H/M_{_{\rm C}}\simeq 0.001$, $\varphi=2$
  and the same values for the other parameters.}
\label{fig:powerspectrum}
\end{figure}

So far, we have discussed the question of the fine tuning of the
initial conditions related to the background. Obviously, there is a
similar question for the perturbations. We have seen that they are
chosen such that the perturbations are initially placed in the vacuum
state. However, if one traces back the scale of astrophysical interest
today to the beginning of inflation, one notices that they correspond
to physical lengths smaller than the Planck length. Clearly, in this
regime, the framework used to derive the predictions of inflation,
namely quantum field theory in curved spacetime, is no longer
valid. This is the so-called trans-Planckian problem of
inflation~\cite{Martin:2000xs,Brandenberger:2000wr,Martin:2003kp,Brandenberger:2002hs,Brandenberger:2012aj,Brandenberger:2004kx,Lemoine:2001ar}. Notice
that, at the same time, one has $H\ll \Mp$ and, therefore, the concept
of classical background is perfectly well defined. So, a priori, one
could argue that the initial conditions for the perturbations are
tuned in an artificial way. Then, the next question is what happens if
one modifies those initial conditions: does it destroy the
inflationary predictions that are so successful?  To study the
robustness of inflation, one can introduce ad-hoc (since we do not
know the theory of quantum gravity which would control the behavior of
the perturbations on scales smaller than the Planck length), but
reasonable, modifications, then recompute the power spectrum of the
fluctuations and see whether we obtain a result which significantly
differs from the standard result. Various modifications have been
proposed, a modification of the dispersion relations of the
perturbations~\cite{Martin:2000xs,Brandenberger:2000wr}, a
modification of the commutation relations~\cite{Hassan:2002qk} etc
\dots However, the most general approach consists in parameterizing
the initial conditions of the perturbations when they emerge from the
quantum foam. Let $M_{_{\rm C}}$ be the energy scale at which the
regime of quantum field theory in curved spacetime breaks down
(possibly the Planck scale or the string
scale)~\cite{Martin:2003kp}. A Fourier mode emerges from the quantum
foam when its physical wavelength equals the length scale associated
to the scale $M_{_{\rm C}}$, namely
\begin{equation}
\label{eq:definitialtime}
\lambda(\eta)=\frac{2\pi}{k}a(\eta)=\ell_{_{\rm C}}
\equiv \frac{2\pi}{M_{_{\rm C}}},
\end{equation}
The initial time satisfying Eq.~(\ref{eq:definitialtime}) is, contrary
to what happens in the usual case, scale-dependent. As a consequence,
the corresponding power spectrum at the end of inflation is modified
and it now reads~\cite{Martin:2003kp}
\begin{align}
\label{eq:powerspectrum}
{\cal P}_{\zeta}(k) =& \frac{H^2}{\pi \epsilon_1\mpl^2}
\Biggl\{1-2\left(C+1\right)\epsilon_1-C\epsilon_2
-\left(2\epsilon_1+\epsilon_2\right)\ln \frac{k}{k_{_{\rm P}}} 
-2\vert x\vert \frac{H}{M_{_{\rm C}}}
\biggl[1-2(C+1)\epsilon_1
\nonumber \\ &
-C\epsilon_2
-\left(2\epsilon_1+\epsilon_2\right)\ln \frac{k}{k_{_{\rm P}}}\biggr]
\cos \left[\frac{2M_{_{\rm C}}}{H}
\left(1+\epsilon_1+\epsilon _1\ln \frac{k}{a_0M_{_{\rm C}}}
\right)+\varphi\right] \nonumber \\ & 
-\vert x\vert \frac{H}{M_{_{\rm C}}}\pi 
\left(2\epsilon_1+\epsilon_2\right)
\sin \left[\frac{2M_{_{\rm C}}}{H}
\left(1+\epsilon_1+\epsilon_1\ln \frac{k}{a_0M_{_{\rm C}}}
\right)+\varphi\right]\Biggr\}. 
\end{align}
This expression should be compared to Eq.~(\ref{spectrumsr}).  In this
expression, the scale $k_{_{\rm P}}$ is the pivot scale and $a_0$ is
the scale factor evaluated at the time where
$k_{_{\rm P}}/a_0=M_{_{\rm C}}$.  Finally, the initial quantum state
of the perturbations at the new scale-dependent initial time is
characterized by a complex number $x$ that can be written in polar
form $x\equiv \vert x\vert {\rm e}^{i\varphi}$, hence defining
$\vert x\vert $ and $\varphi$. This power spectrum is represented in
Fig.~\ref{fig:powerspectrum}.

\begin{figure*}
  \includegraphics[width=0.9\textwidth,height=.55\textwidth]
{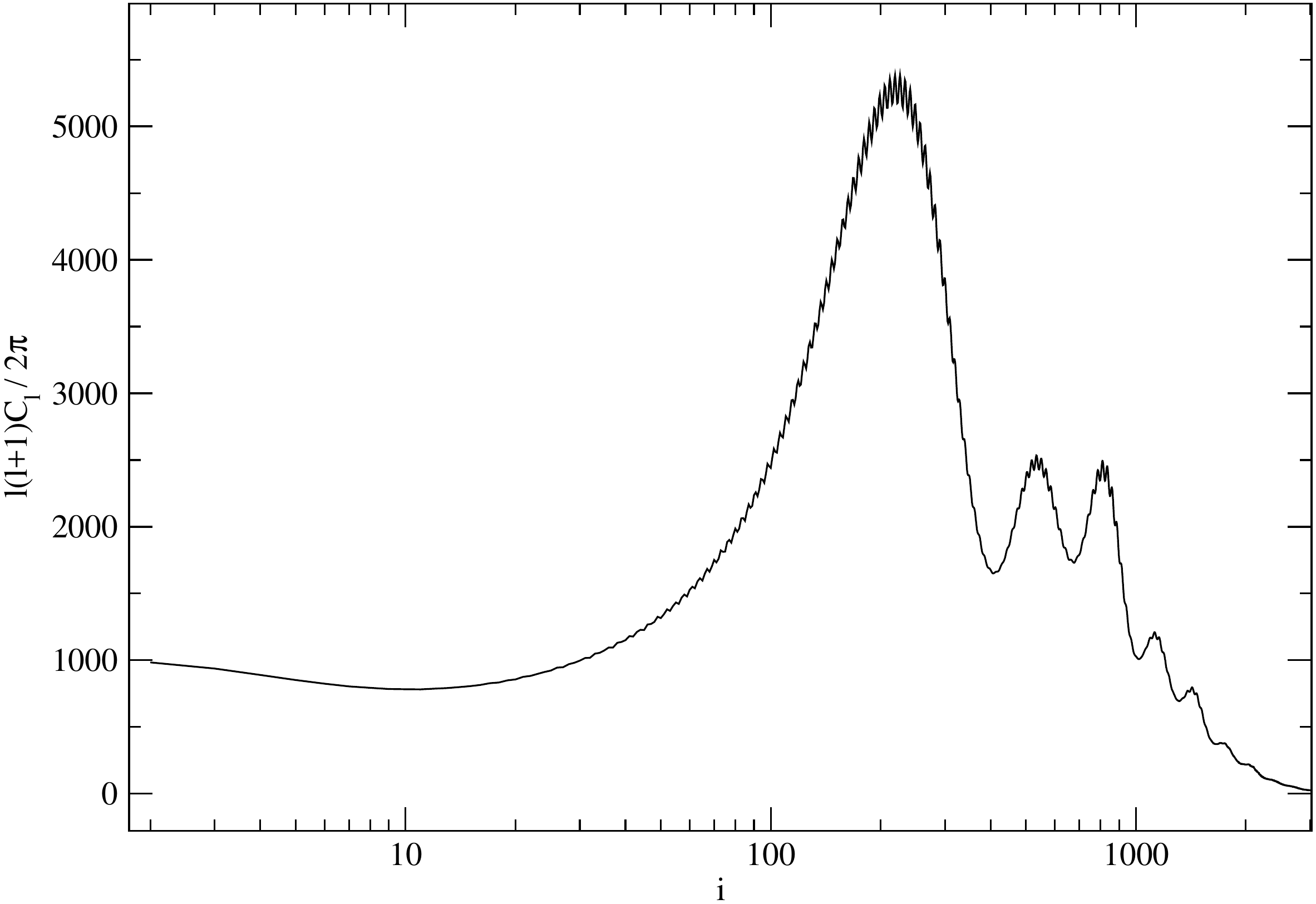}
\caption{Multipole moments in presence of super-imposed
  trans-Planckian oscillations. Figure taken from
  Ref.~\cite{Martin:2003sg}.}
\label{fig:tplcl}
\end{figure*}

Let us now comment on the power spectrum itself. The most obvious
remark is that it is modified by the presence of super-imposed
oscillations. These oscillations modify the CMB multipole moments as
shown in Fig.~\ref{fig:tplcl} and, therefore, have observational
consequences~\cite{Martin:2003sg,Martin:2004yi,Martin:2004iv}. The
amplitude of the oscillations is, roughly speaking, given by
$\vert x\vert H/M_{_{\rm C}}$, while the frequency is proportional to
$(H/M_{_{\rm C}})^{-1}$. On general grounds, we expect the ratio
$H/M_{_{\rm C}}$ to be a small number. Indeed, we know from the CMB
normalization that $H\lesssim 10^{-5}\Mp$. The scale $M_{_{\rm C}}$ is
not known but $M_{_{\rm C}}\in \left[10^{-1}\Mp,10^{-3}\Mp\right]$
seems reasonable and this implies that, at most,
$H/M_{_{\rm C}}\sim 0.01$. Therefore, unless the number $\vert x\vert$
is very large, the amplitude of the oscillations is small and one
could argue that inflation is robust against trans-Planckian
corrections. In this sense, assuming the vacuum state initially is not
a fine tuning. Of course, as already mentioned, $\vert x\vert$ could
be large and, in this case, the modification sizable. However, the
magnitude of $\vert x\vert $ is limited by the backreaction
problem~\cite{Brandenberger:2004kx}. Physically, this is due to the
fact that $\vert x\vert \neq 1$ corresponds to an excited state. But
the particles present in this quantum state carry energy density and
this energy density could prevent inflation to start. Therefore, it
has to be smaller than the inflationary energy density $H^2\Mp^2$. One
can show that this leads to an upper bound on the amplitude of the
oscillations given by~\cite{Martin:2003sg,Martin:2004yi,Martin:2004iv}
\begin{equation}
\label{eq:backreaction}
\vert x\vert \frac{H}{M_{_{\rm C}}}\lesssim \frac{10^4}{\sqrt{\epsilon_1}}
\left(\frac{H}{M_{_{\rm C}}}\right)^2\sim 4.3\times 10^{-4}\sqrt{r}\left(
\frac{\Mp}{M_{_{\rm C}}}\right)^2.
\end{equation}
This upper bound is not sufficient to exclude a possible detection of
the oscillations in the data (although for the moment nothing has been
seen). And, in this sense, one could argue that inflation is not
robust to a change of the initial conditions. However, detecting the
oscillations would mean opening a window on physics beyond the quantum
gravity scale, clearly a fascinating possibility.

\section{The Multiverse}
\label{sec:multi}

\subsection{Stochastic Inflation}
\label{subsec:stochainf}

The discussion of the previous section about the initial conditions
misses a crucial ingredient, namely the fact that the background field
is itself a quantum field. So far, the quantum effects have been taken
into account but only at the perturbative level. The question is now
whether they also play an important role in the evolution of the
background. Classically, the inflaton field evolves according to the
Klein-Gordon equation and, in the slow-roll regime, the typical
variation of $\phi$ is then given by
$\Delta \phi_{\rm cl}\simeq -V_{\phi}/(3H)\Delta t$. On the other
hand, the amplitude of the quantum kick received by $\phi$ during one
e-fold is, roughly speaking, of the order of the square root of the
power spectrum of $\delta \phi$, namely
$\Delta \phi_{\rm q}\simeq H/(2\pi)$. If
$\Delta \phi_{\rm q}\gg \Delta \phi_{\rm cl}$, then quantum effects
are likely to be dominant. In fact, it is easy to see that
\begin{align}
\label{eq:qclratio}
\frac{\Delta\phi_{\rm q}}{\Delta \phi_{\rm cl}}=\sqrt{{\cal P}_{\zeta \zero}},
\end{align}
where $\calP_{\zeta \zero} $ is the amplitude of scalar perturbations,
see Eq.~(\ref{eq:scalaramp}). This equation just tells us that, when
the fluctuations are of order one, quantum effects are relevant even
for the background. Notice that if we want to see whether stochastic
effects can modify the power spectrum of curvature perturbations, then
the criterion is different, see Ref.~\cite{Vennin:2015hra}.

If, for instance, we consider the model $V(\phi)=M^4(\phi/\Mp)^p$,
then the condition $\Delta \phi_{\rm q}>\Delta \phi_{\rm cl}$ is
equivalent to $\phi>\phi_{\rm s}$ with
\begin{equation}
\label{eq:stochalimit}
\frac{\phi_{\rm s}}{\Mp}=\left[\pi p\sqrt{6}\left(\frac{\Mp}{M}\right)^2
\right]^{\frac{2}{2+p}}.
\end{equation}
Then, if one uses the expression of $M$ given in
Eq.~(\ref{eq:cmbnormalization}), one arrives at
\begin{align}
\frac{\phi_{\rm s}}{\Mp}=2^{-\frac{1}{p+2}}
\left(\frac{p^2}{2}+2p\Delta N_*\right)
^{1/2}\left({\cal P}_{\zeta \zero}^{\rm Planck}\right)^{-\frac{1}{2+p}},
\end{align}
where ${\cal P}_{\zeta \zero}^{\rm Planck}$ is the amplitude of the
spectrum measured by the Planck satellite, see
Eq.~(\ref{eq:cmbnormaPlanck}). Using this result, namely
$\ln \left(10^{10}\calP_{\zeta \zero}\right) =3.094\pm 0.0049$, one
obtains $\phi_{\rm s}/\Mp\simeq 1743$ for the model $p=2$ (one has
taken $\Delta N_*\simeq 50$). It is also interesting to estimate the
Hubble parameter for this value of the field and one finds
\begin{align}
\frac{H_{\rm s}^2}{\Mp^2}=
4\pi^2p^2\left(\frac{p^2}{2}+2p\Delta N_*\right)^{-\frac{p}{2}-1}
{\cal P}_{\zeta \zero}^{\rm Planck}\left(\frac{\phi_{\rm s}}{\Mp}\right)^p.
\end{align}
For $p=2$, this gives $H_{\rm s}/\Mp\simeq 0.005$, the important point
being that we are in a regime where the quantum behavior of the
inflaton field must be taken into account but where, at the same time,
the concept of a background spacetime is still relevant
since $H_{\rm s}/\Mp\ll 1$.

After these qualitative considerations, let us now try to establish
more precisely the equations controlling the evolution of the system
in this
regime~\cite{Starobinsky:1986fx,Starobinsky:1994bd,Martin:2005hb,Martin:2005ir,Lorenz:2010vf,Martin:2011ib,Vennin:2015hra,Assadullahi:2016gkk,Grain:2017dqa}. Let
us first consider a quantum scalar field in a rigid, de Sitter,
background. This means that the backreaction of the quantum scalar
field is neglected or, in other words, that it is a test field living
in a de Sitter spacetime characterized by $H$. In this spacetime,
$H^{-1}$ is a preferred length and can be used to distinguish between
short and long wavelengths. Then one writes the scalar field according
to~\cite{Starobinsky:1986fx,Starobinsky:1994bd}
\begin{equation}
  \hat{\phi}(t,{\bm x})=\hat{\phi}_{_{\rm IR}}(t,{\bm x})
  +\frac{1}{(2\pi)^{3/2}}\int {\rm d}{\bm k}\,
  \Theta \left(k-\sigma aH\right)
  \left[\mu_{\bm k}(t)e^{i{\bm k}\cdot {\bm x}}\hat{c}_{\bm k}
    +\mu_{\bm k}^*(t)e^{-i{\bm k}\cdot {\bm x}}\hat{c}_{\bm k}^{\dagger}
  \right],
\end{equation}
where $\sigma \ll 1$ is a small constant. The quantity $\Theta $ is
the Heaviside function, $\mu_{\bm k}(t)$ is the field mode function
and $\hat{c}_{\bm k}$ and $\hat{c}_{\bm k}^{\dagger}$ are the
annihilation and creation operators satisfying the standard
commutation relations
$[\hat{c}_{\bm k},\hat{c}^{\dagger}_{\bm p}]=\delta({\bm k}-{\bm p})$.
One can then insert this expression into the Klein-Gordon equation to
find an equation of motion for the long-wavelength, infrared, part of
the field. In fact, it is possible to ignore that the infrared field
is a quantum field and see it as a stochastic quantity obeying a
Langevin equation given
by~\cite{Starobinsky:1986fx,Starobinsky:1994bd}
\begin{align}
\label{eq:langevin}
\frac{{\rm d}\phi_{_{\rm IR}}(N,{\bm x})}{{\rm d}N}=-
\frac{V_{\phi}(\phi_{_{\rm IR}})}{3H^2}
+\frac{H}{2\pi}\xi(N,{\bm x}),
\end{align}
where $\xi (N) $ is a white noise sourced by the ultraviolet part of
the field with correlation function
\begin{align}
\langle \xi(N,{\bm x})\xi(N',{\bm x}')\rangle =\delta(N-N')
j_0\left(\sigma a H\vert {\bm x}-{\bm x}'\vert\right).
\end{align}
Here, $j_0$ is a spherical Bessel function of order zero. By solving
the Langevin equation, one can calculate the various correlation
functions of the field and show that they coincide with the quantum
correlation functions (at least in some limit). This approach, called
stochastic inflation, is uncontroversial since it is a fact that the
two types of correlation function perfectly match. This is another
facet of the general fact that, on super-Hubble scales, the system can
be described by a classical stochastic
process~\cite{Polarski:1995jg,Martin:2015qta,Grain:2017dqa}.

\subsection{Eternal Inflation}
\label{subsec:eternal}

\begin{figure*}
  \includegraphics[width=0.9\textwidth,height=.9\textwidth]
{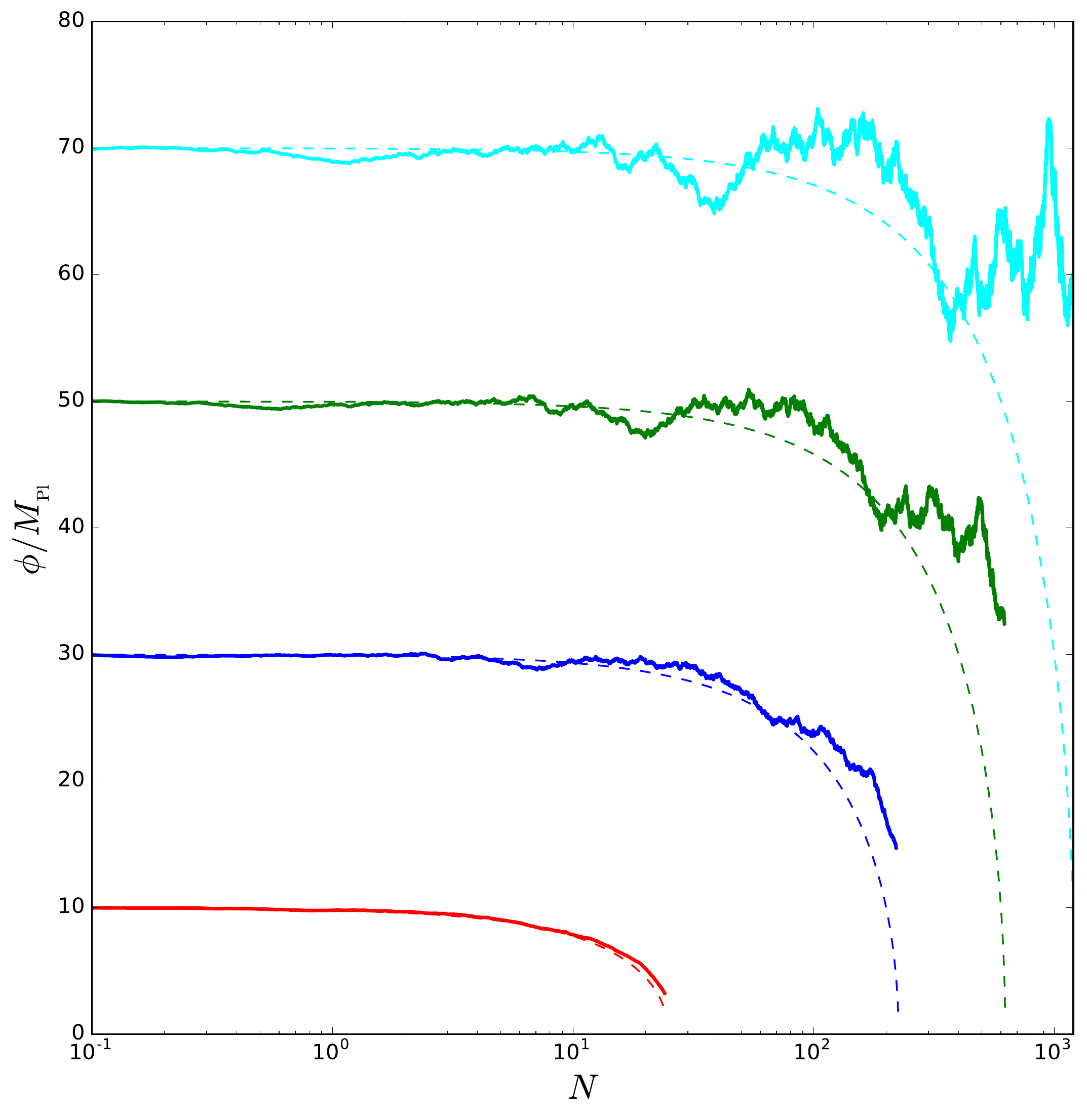}
\caption{Trajectories (vacuum expectation value of the inflaton field
  versus number of e-folds) for the inflationary model
  $V(\phi)=m^2\phi^2/2$ with $m=0.1\, \Mp$ and different initial
  conditions, $\phi_\uini=10\, \Mp$ (red line), $\phi_\uini=30\, \Mp$
  (blue line), $\phi_\uini=50\, \Mp$ (green line) and
  $\phi_\uini=70\, \Mp$ (cyan line). The solid lines represent the
  stochastic trajectories while the dashed ones correspond to the
  classical, slow-roll, ones.}
\label{fig:lfistocha}
\end{figure*}

Then, the next step is to relax the assumption that spacetime is rigid
and to take into account the back reaction of the scalar field on the
geometry~\cite{Linde:1982ur,Steinhardt:1982kg,Linde:1986fd,Linde:1993xx,Guth:2007ng,Guth:2000ka,Linde:2015edk}. This
is at this point that speculations enter the game. Since we study a
regime where the inflaton field is viewed as a quantum field, it seems
that there are two ways to take into account its backreaction. Either
we still view the background as classical, in which case, we need an
equation such as $G_{\mu \nu}=\langle \hat{T}_{\mu \nu}\rangle/\Mp^2$,
or the background spacetime becomes a quantum object, in which case we
need an equation similar to
$\hat{G}_{\mu \nu}=\hat{T}_{\mu \nu}/\Mp^2$. In the case of eternal
inflation, the second choice is made. Notice that one can even argue
that the first choice is inconsistent, see
Sec.~\ref{subsec:threat?}. However, since quantum objects are
represented by stochastic quantities, we are in fact led to the
concept of stochastic geometry (supposed to represent, in this
approach, the behavior of a quantum geometry). In this view, the
stochastic geometry is sourced by the stochastic scalar field. Then
comes the question of which equation controls the behavior of the
stochastic geometry. Here the common assumption consists in
postulating that
\begin{align}
\label{eq:friedstocha}
H^2=\frac{1}{3\Mp^2}V\left(\phi_{_{\rm IR}}\right), 
\end{align}
namely the classical equation promoted to an equation for the
stochastic quantities. Here, we really deal with an equation of the
type $\hat{G}_{\mu \nu}=\hat{T}_{\mu \nu}/\Mp^2$ since
$\phi_{_{\rm IR}}$ and $H$ are now considered as stochastic
quantities. We also notice that, obviously, the above equation is only
valid in a cosmological context. Then, the Langevin
equation~(\ref{eq:langevin}) becomes
\begin{align}
\label{eq:langevinback}
\frac{{\rm d}\phi_{_{\rm IR}}}{{\rm d}N}=-
\frac{V_{\phi}(\phi_{_{\rm IR}})}{3H^2(\phi_{_{\rm IR}})}
+\frac{H(\phi_{_{\rm IR}})}{2\pi}\xi(N).
\end{align}
Clearly this equation is not equivalent to Eq.~(\ref{eq:langevin}) and
can even be ambiguous because of the second term which is given by the
product of two stochastic quantities. In Fig.~\ref{fig:lfistocha}, we
present a numerical integration of this equation for the potential
$V=m^2\phi^2/2$ and for different initial conditions. It is easy to
see that, in that case, the criterion~(\ref{eq:stochalimit}) reads
$\phi_{\rm s}/\Mp\simeq \sqrt{4\pi\sqrt{6}}(m/\Mp)^{-1}$. For
numerical reasons, in order to clearly illustrate the effect, we
choose a value of $m$ much larger than the one implied by the CMB
normalization, namely $m=0.1 \Mp$. This leads to
$\phi_{\rm s}\simeq 55\, \Mp$. Then, we numerically integrate
Eq.~(\ref{eq:langevinback}) for four different initial conditions,
$\phi_\uini=10\, \Mp$, $\phi_\uini=30\, \Mp$, $\phi_\uini=50\, \Mp$
and $\phi_\uini=70\, \Mp$. Using the trajectory~(\ref{eq:trajeclfi})
and the fact that $\phi_{\rm end}/\Mp=p/\sqrt{2}$, classically, these
four initial conditions respectively correspond to a total of
$\sim 24.5$, $\sim 224.5$, $\sim 624.5$ and $\sim 1124.5$ e-folds of
inflation. This plot confirms the previous analysis. When
$\phi_\uini<\phi_{\rm s}$, we see that the stochastic trajectory
(solid line) is very close to the classical one (dashed line). On the
contrary, when $\phi_\uini\sim \phi_{\rm s}$ or
$\phi_\uini>\phi_{\rm s}$, the stochastic effects dominate, the
trajectory becomes ``chaotic'' and strongly differs from its classical
counterpart. In particular, we notice that, due to stochastic effects,
the value of the field can increase. This means that the field can in
fact climb its potential.

Let us now come back to Eq.~(\ref{eq:langevin}) where we assume that
the field is a test field living in a de Sitter spacetime. If
$V(\phi)=m^2\phi^2/2$, then this equation can be easily solved (since
it is a linear equation) and the solution reads
\begin{align}
\phi(N,{\bm x})=\phi_\uini (N,{\bm x})e^{-m^2(N-N_\uini)/(3H^2)}
+\frac{H}{2\pi}e^{-m^2N/(3H^2)}
\int_{N_\uini}^Ne^{m^2 n/(3H^2)}\xi(n,{\bm x}){\rm d}n.
\end{align}
Using this solution, one can then calculate the two-point correlation
function at equal time. One obtains
\begin{align}
\left \langle \phi\left(N,{\bm x}\right)
\phi \left(N,{\bm x}'\right) \right \rangle
&=\left[\phi_\uini(N,{\bm x})\phi_\uini(N,{\bm x}')
-\frac{3H^4}{8\pi^2 m^2}
j_0\left(\sigma a H\vert {\bm x}-{\bm x}'\vert\right)
\right]
e^{-\frac{2m^2}{3H^2}(N-N_\uini)}
\nonumber \\ &
+\frac{3H^4}{8\pi^2 m^2}j_0\left(\sigma a H\vert {\bm x}
-{\bm x}'\vert\right).
\end{align}
This expression is made of two pieces. The first one, which depends on
the initial conditions, decays away exponentially for
$N\gg N_{\rm ini}$ and quickly becomes sub-dominant. The second piece
shows that the ultra large scale structure of the field is made of a
collection of nearly homogeneous patches of size $H^{-1}$ (\ie the
Hubble radius) since this is the distance at which the correlation
function almost vanishes, thanks to the presence of the Bessel
function.

Then, since inflation is an almost de Sitter expansion, what we have
just described for a test field should also be true when the back
reaction is taken into account, namely for the field the behavior of
which is controlled by Eq.~(\ref{eq:langevinback})\footnote{For the
  potential $V(\phi)=m^2\phi^2/2$, this equation reads
\begin{align}
\frac{{\rm d}\phi_{_{\rm IR}}}{{\rm d}N}+\frac{2\Mp^2}{\phi_{_{\rm IR}}}
=\frac{m}{2\pi \Mp\sqrt{6}}\phi_{_{\rm IR}}\xi .
\end{align}
It is of the Bernouilli type and, therefore, can be solved explicitly. The 
solution takes the form
\begin{align}
\phi_{_{\rm IR}}^2=
e^{\frac{m}{\pi\Mp \sqrt{6}}\int _{N_\uini}^N\xi {\rm d}n}
\left[\phi_\uini^2-4\Mp^2 \int_{N_\uini}^N
e^{-\frac{m}{\pi\Mp \sqrt{6}}\int _{N_\uini}^n\xi(\bar{n}) {\rm d}\bar{n}}
{\rm d}n\right].
\end{align}
However, it is so complicated that it is not very useful. In
particular, it seems very difficult to calculate the two-point
correlation function of the field from this solution.}. Moreover, each
patch is isolated from the others as can be seen by computing the
event horizon in de Sitter spacetime. Let us indeed consider a
specific observer that we choose, for convenience, to be at the
origin. Then, its future horizon (the part of the Universe with which
the observer will be able to communicate in the future) is given by
\begin{equation}
d_{_{\rm E}}=a_0\int _{t_0}^{\infty}\frac{{\rm d}t}{a(t)}
=a_0\int _{t_0}^{\infty}{\rm d}t \frac{1}{a_0}e^{-H(t-t_0)}
=\frac{1}{H},
\end{equation}
namely the size of the patch itself. In other words, each patch is
causally disconnected from the others and this forever. These patches
are sometimes referred to as ``pocket universes''. The number of these
patches is growing with time. Indeed, in one e-fold, the ``size'' of
the Universe increases by a factor $e^3\sim 20$ while the ``size'' of
a patch is constant (since the Hubble parameter is constant). As a
consequence, each e-fold, one patch gives rise to about twenty new
patches, all causally disconnected.

There is also some kind of ergodic argument at play. When, see for
instance Fig.~\ref{fig:lfistocha}, we have solved the Langevin
equation, each realization of the solution of this equation was
supposed to represent a specific configuration of the field over the
entire homogeneous and isotropic spacetime. But one can also assume
that one realization corresponds to a specific value of the field in a
given patch since they are causally disconnected. And, as a
consequence, different realizations correspond to different values of
the field in different patches. So, in this interpretation, different
realizations do not represent an ensemble of different field
configurations over an homogeneous and isotropic spacetime but,
rather, the spatial distribution of $\phi_{_{\rm IR}}$ in different
patches.

The overall picture that emerges is that of an expanding spacetime
where the number of independent patches is increasing, the value of
the field in each pocket universe being a stochastic quantity
controlled by a Langevin equation. Since we have seen that, due to
stochastic effects, the field can climb up its potential, there are
patches where inflation will never stop. Obviously, the volume
occupied by those patches, compared to the volume occupied by the
patches where inflation stops, is growing which means that patches
where inflation is taking place occupy more and more regions of
spacetime. Globally, inflation will never stop meaning that there are
always regions of spacetime undergoing inflation. Of course, there
will also be regions of spacetime where inflation stops, those where,
by chance, the stochastic fluctuations do not push the field
upwards. This structure is referred to as ``eternal inflation''. The
stochastic effects are said to produce a ``multiverse''. Notice that
the word ``multiverse'' is especially awkward in the present context
since we do not produce many universes as in the many world
interpretation of quantum mechanics for instance but just a specific
spatial configuration of our single universe made of causally
independent regions, the pocket universes.

Before discussing the reliability and the implications of eternal
inflation, we would like to investigate the question of whether it is
unavoidable or not.

\subsection{Avoiding Self Replication}
\label{subsec:avoid}

\begin{figure*}
  \includegraphics[width=0.7\textwidth,height=.7\textwidth]
{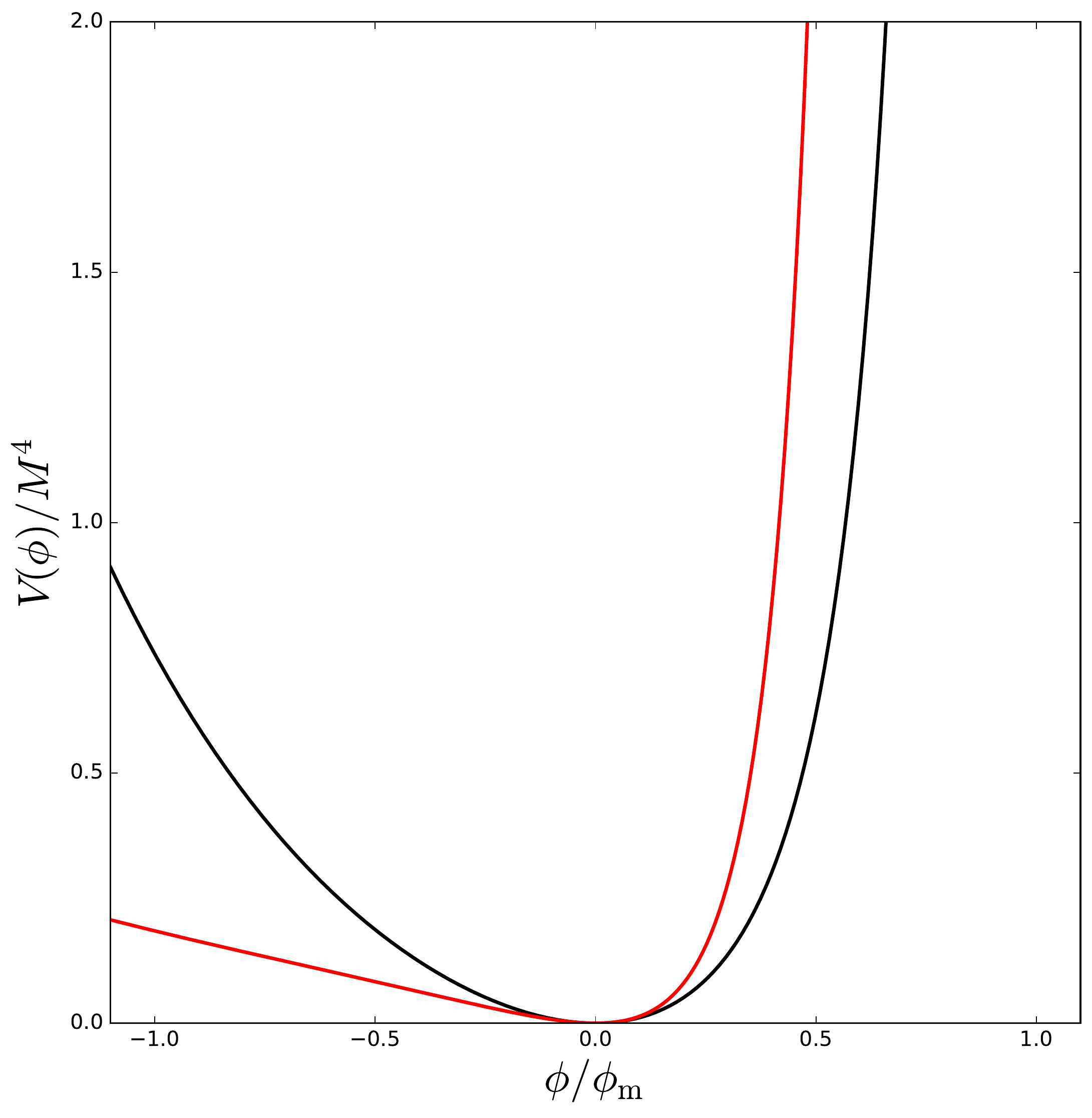}
\caption{Potential given by Eq.~(\ref{eq:potNSRI2}) for two values of
  $\alpha$, $\alpha=2$ (black line) and $\alpha=4$ (red line).}
\label{fig:potNSRI2}
\end{figure*}

\begin{figure*}
  \includegraphics[width=0.7\textwidth,height=.7\textwidth]
{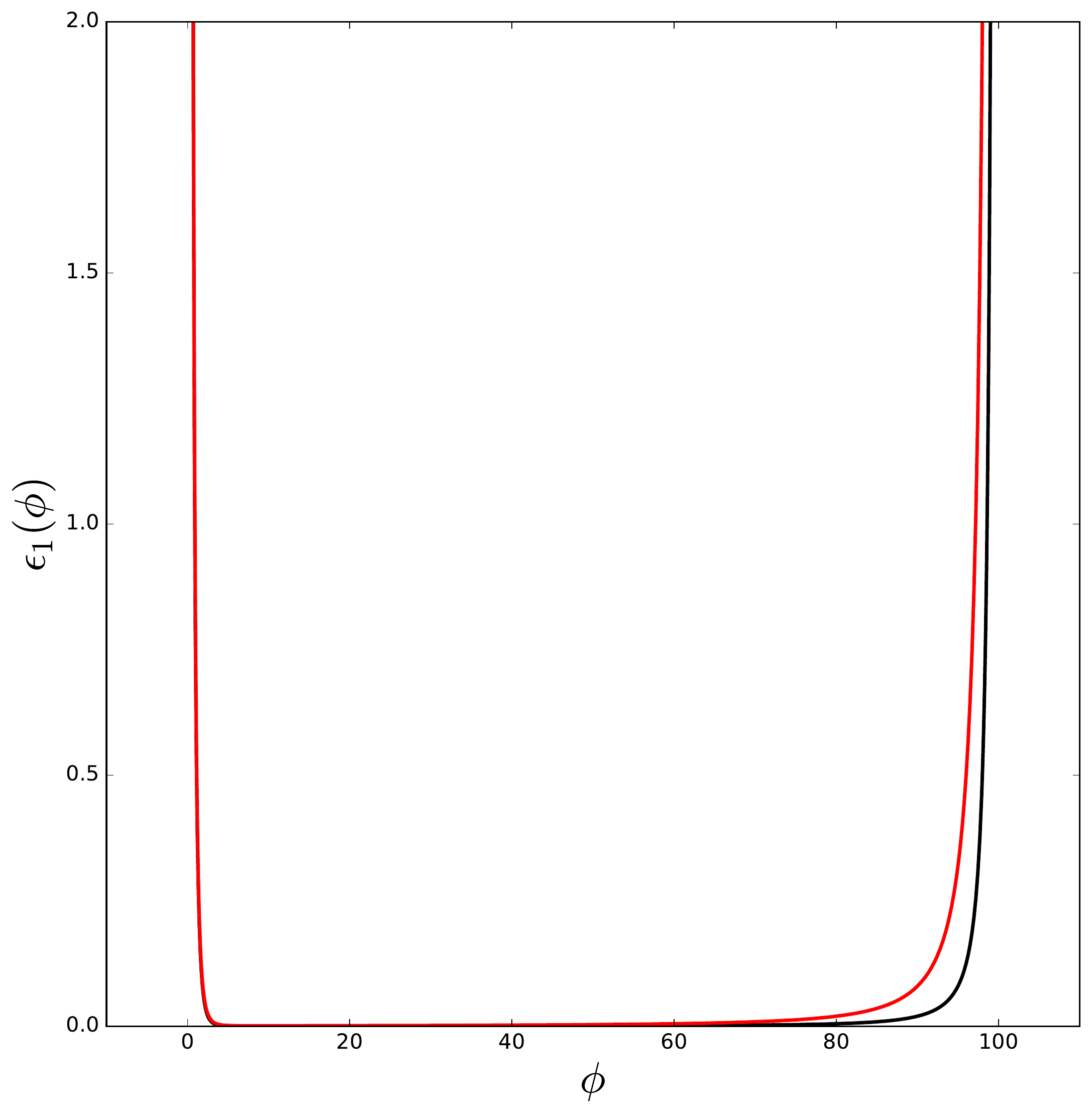}
\caption{First Hubble flow parameter $\epsilon_1(\phi)$ given by
  Eq.~(\ref{eq:eps1nsri2}) for two values of $\alpha$, $\alpha=2$
  (black line) and $\alpha=4$ (red line).}
\label{fig:eps1nsri2}
\end{figure*}

Before discussing the robustness of eternal inflation, it is
interesting to investigate whether this is an unavoidable consequence
of inflation. As recently discussed in Ref.~\cite{Mukhanov:2014uwa},
it turns out that this is not the case and, in this section, we
closely follow this paper although we also present some new
results. We have seen that the quantum-to-classical variation of the
field is given by the amplitude of the scalar power spectrum, see
Eq.~(\ref{eq:qclratio}). If there exists a field value for which this
amplitude
\begin{align}
{\cal P}_{\zeta \zero}(\phi)=\frac{H^2(\phi)}{8\pi \Mp^2\epsilon_1(\phi)},
\end{align}
is of order one, then this means that the quantum fluctuations of the
field are of order one and, if the considerations presented in the
previous section are correct, the regime of eternal inflation
starts. Usually, this happens in the regime where
$\epsilon_1(\phi)\rightarrow 0$ since $\epsilon_1(\phi)$ stands at the
denominator. But this also implies that, if the shape of the potential
is such that there is a field range such that $\epsilon_1\ll 1$ (in
order to have inflation!) but otherwise $\epsilon_1(\phi)$ is large,
then there could be no regime where ${\cal P}_{\zeta \zero}>1$. One
example was found by V.~Mukhanov in Ref.~\cite{Mukhanov:2014uwa}. The
corresponding potential is the following one
\begin{equation}
\label{eq:potNSRI2}
V(\phi)=M^4\left(1-e^{-\phi/\Mp}\right)^2
\left(1-\frac{\phi}{\phi_{\rm m}}\right)^{-\alpha},
\end{equation}
and is represented in Fig.~\ref{fig:potNSRI2}. Interestingly enough,
it looks like the Starobinsky model corrected by a term
$(1-\phi/\phi_{\rm m})^{-\alpha}$. The model depends on three
parameters: $M$, $\phi_{\rm m}$ and $\alpha$. As usual $M$ is fixed by
the CMB normalization.

The first two Hubble flow parameters are given by the following
expressions
\begin{align}
\label{eq:eps1nsri2}
\epsilon_1 &= \frac12 \left[2\frac{e^{-\phi/\Mp}}{1-e^{-\phi/\Mp}}+\alpha 
\frac{\Mp}{\phi_{\rm m}}\left(1-\frac{\phi}{\phi_{\rm m}}\right)^{-1}
\right]^2, \\
\epsilon_2 &=4\epsilon_1+4e^{-\phi/\Mp}\left(1-e^{-\phi/\Mp}\right)^{-1}
-4e^{-2\phi/\Mp}\left(1-e^{-\phi/\Mp}\right)^{-2}
\nonumber \\ & 
-8\alpha \frac{\Mp}{\phi_{\rm m}}
\left(1-e^{-\phi/\Mp}\right)^{-1}
\left(1-\frac{\phi}{\phi_{\rm m}}\right)^{-1}
-2\left(\alpha +\alpha^2\right)\frac{\Mp^2}{\phi_{\rm m}^2}
\left(1-\frac{\phi}{\phi_{\rm m}}\right)^{-2}.
\end{align}
The first Hubble flow parameter is represented in
Fig.~\ref{fig:eps1nsri2}. We see that it has exactly the expected
shape. There is a field range where $\epsilon_1$ is very small and
this is the regime during which inflation can take place. But, at
large-field values, the corrections play a crucial role and
$\epsilon_1\rightarrow +\infty $ as $\phi \rightarrow \phi_{\rm
  m}$.
As a consequence, the amplitude of the fluctuations is killed and we
never reach the regime of eternal inflation. 

\begin{figure*}
  \includegraphics[width=0.9\textwidth,height=.7\textwidth]
{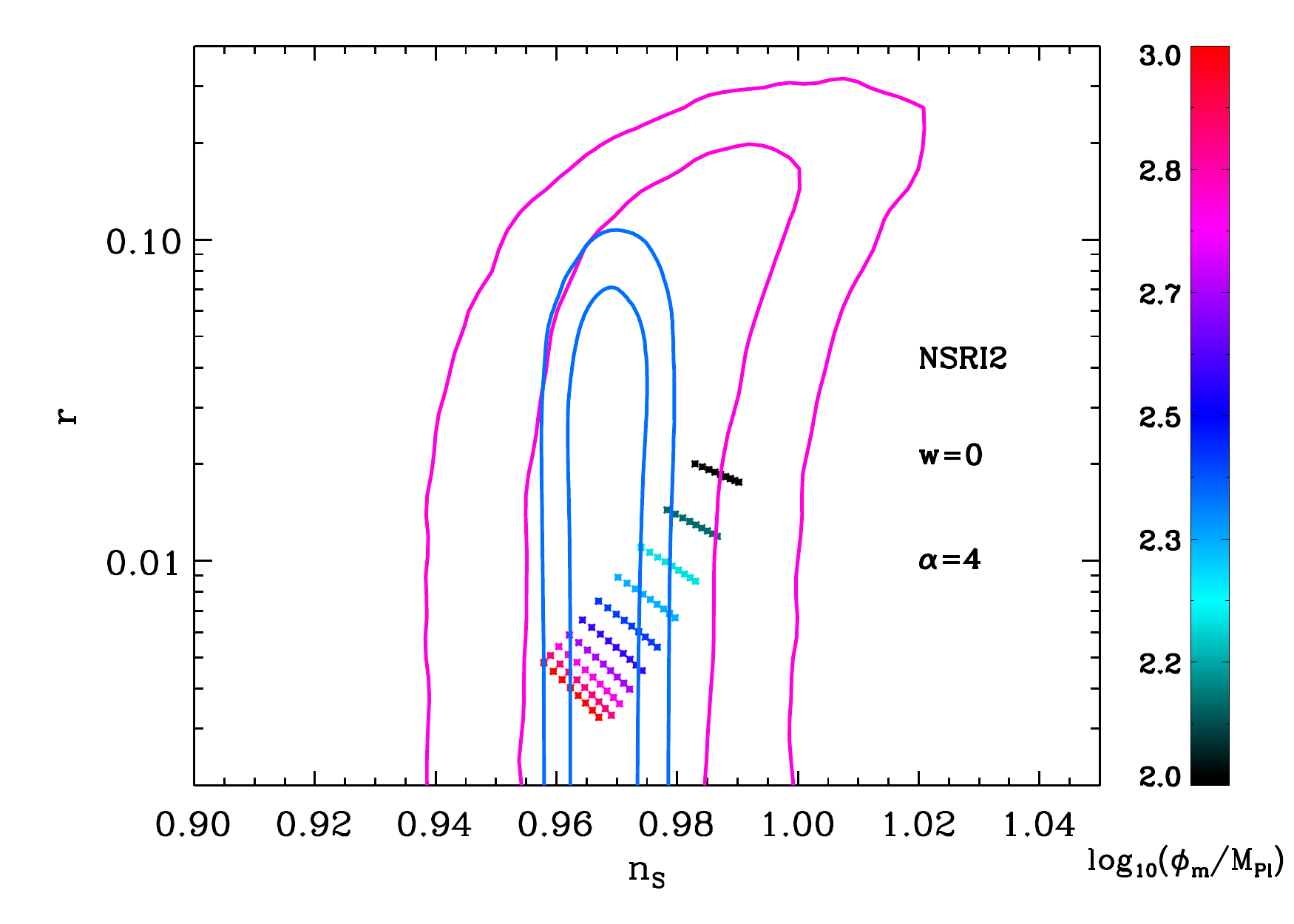}
\caption{Predictions in the $(r,\nS)$ space of the inflationary model
  with the potential given by Eq.~(\ref{eq:potNSRI2}). The scale $M$
  is CMB normalized, $\alpha=4$ and
  $\log_{10}\left(\phi_{\rm m}/\Mp\right)\in [2,3]$, its value being
  indicated by the color bar. Along the same interval, different
  points represent different reheating temperatures. The pink contours
  are the $1$ and $2\sigma$ WMAP7 contours while the blue ones are the
  $1$ and $2\sigma$ Planck contours.}
\label{fig:nsRNSRI2}
\end{figure*}

Moreover, this model is in perfect agreement with the observations. In
Fig.~\ref{fig:nsRNSRI2}, we have compared the predictions of the model
for $\alpha =4$ and different values of $\phi_{\rm m}$ (indicated by
the color bar) with the CMB data (the pink contours are the WMAP7
contours while the blue contours are the Planck contours). Evidently,
the model is in agreement with the data. 

\begin{figure*}
  \includegraphics[width=0.9\textwidth,height=.7\textwidth]
{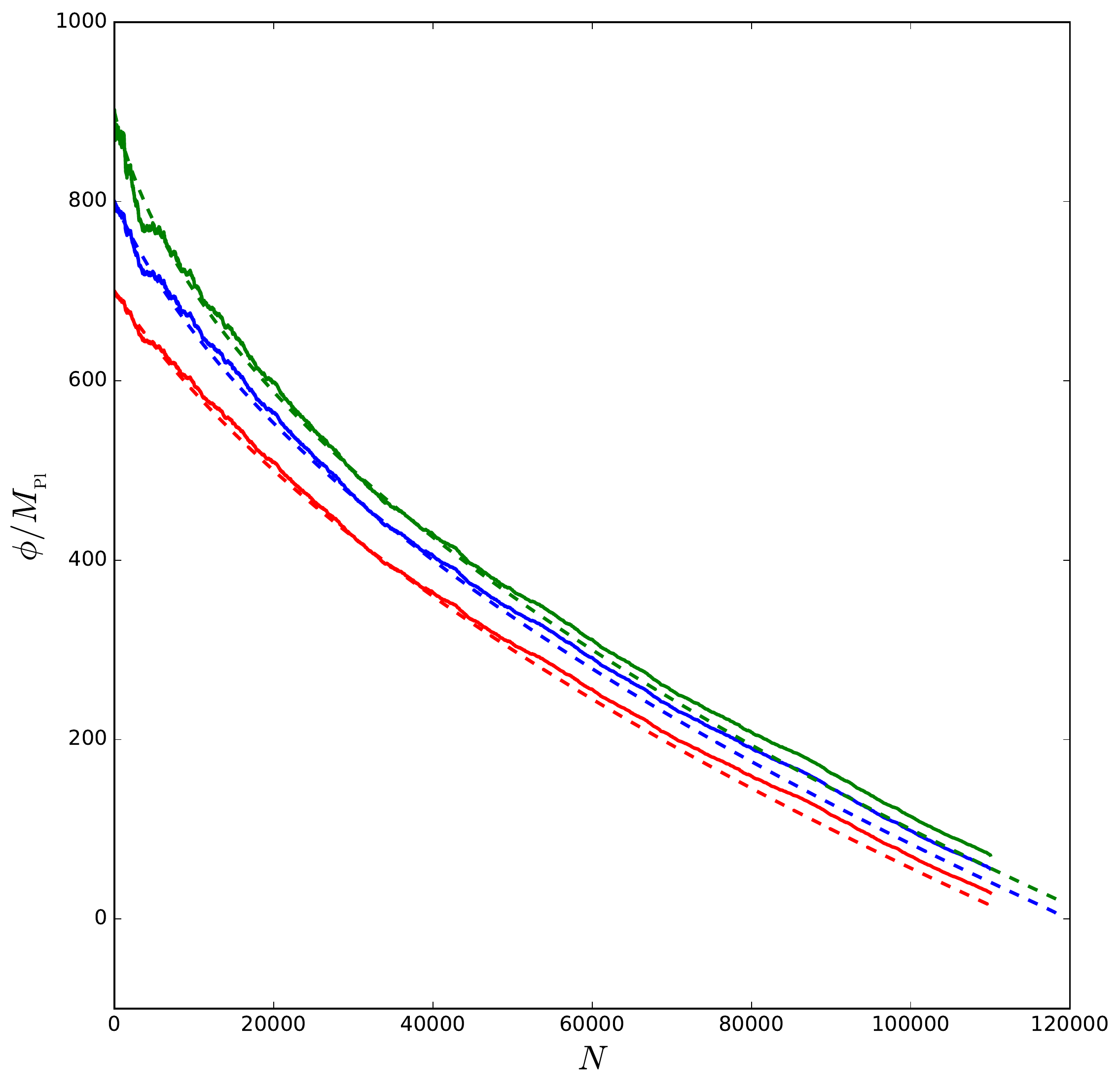}
\caption{The inflaton field vacuum expectation value versus number of
  e-folds, for the inflationary model given by Eq.~(\ref{eq:potNSRI2})
  with $\alpha=4$ and $\phi_{\rm m}=1000$, calculated by means of the
  Langevin equation (solid lines) and classically (dashed lines) for
  different initial conditions, $\phi_\uini=900$ (green lines),
  $\phi_\uini=800$ (blue lines) and $\phi_\uini=700$ (red lines).}
\label{fig:nsri2stocha}
\end{figure*}

From the previous considerations, as we have already discussed, it
should be obvious that the quantum fluctuations are suppressed. In
order to check this statement explicitly, we have integrated the
Langevin equation with the potential~(\ref{eq:potNSRI2}). The result
is represented in Fig.~\ref{fig:nsri2stocha} and should be compared to
Fig.~\ref{fig:lfistocha}. In both plots, the value of $M$ has been
artificially increased (compared to its CMB value) in order to see the
effects more clearly. It is evident that, for the
model~(\ref{eq:potNSRI2}), and contrary to what happens for large
field models, the quantum fluctuations never play an important
role. All the stochastic trajectories always remain close to the
classical one. 

Therefore, in conclusion, the results presented here clearly indicate
that eternal inflation is not mandatory at all and that it is
perfectly possible to build a model of inflation which is in perfect
agreement with the observations and where self replication never
starts. Moreover, from a physical point of view, this scenario seems
to make sense. In the slow-roll regime, the potential is flat and this
leads to predictions in agreement with CMB data. But, in the UV
regime, corrections kick in and modify the potential in such a way
that eternal inflation is avoided. The only limitation to the previous
argument is that, maybe, the field dependence of the corrections is
not such that self-replication is prevented. Indeed, for instance, one
has ${\cal P}_{\zeta}\sim V^3/V_{\phi}^2\sim \phi^{n+2}$ if
$V(\phi)\sim \phi^n$. For $n>0$, ${\cal P}_{\zeta} $ always grows with
$\phi$. So if the corrections take the form of monomials, quantum
corrections will unavoidably become of order one.

\subsection{Is the Multiverse a Threat for Inflation?}
\label{subsec:threat?}

In this sub-section, one would like to discuss the implications of the
previous considerations for inflation. The main point is that
inflation and eternal inflation should not be put on an equal
footing. The former provides a phenomenological description by means
of an effective model of the early universe which seems to be in good
agreement with the observations while the latter is, at this stage,
only a speculation although definitely an interesting one. The
arguments that support this point of view are the following.

Firstly, it is important to make the distinction between stochastic
inflation and eternal inflation. Stochastic inflation, which is not a
model of inflation, but a technique, appears to be very robust. It is
just a fact that the quantum correlation functions in an expanding
spacetime can be recovered by focusing on the long wavelength part of
the field and by requiring it to obey a Langevin equation. This has
been proven beyond any doubt, see for instance
Refs.~\cite{Starobinsky:1986fx,Starobinsky:1994bd}. Stochastic
inflation studies test quantum fields, namely neglects the back
reaction of the quantum field on the geometry. In stochastic
inflation, the geometry of spacetime is rigid and fixed once and for
all. 

On the contrary, in the case of eternal inflation, one takes into
account the backreaction which means that the geometry (\ie the
gravitational field) must be viewed as a quantum (or stochastic)
quantity. Clearly, this is reminiscent of quantum gravity. And, of
course, the big question is which theory controls the quantum behavior
of the geometry. The theory of eternal inflation just models the
coupling between the quantum field and the quantum geometry by
equation~(\ref{eq:friedstocha}), an equation that one could also write
as
\begin{equation}
\label{eq:quantumfried}
\hat{H^2}=\frac{1}{3\Mp^2}V\left(\hat{\phi}\right),
\end{equation}
where we have used hats to stress the fact that the geometry should
now be viewed as a stochastic quantity and that stochastic quantities
are in fact quantum quantities. If this equation happened to be too
simplistic, then the previous considerations about eternal inflation
could be drastically modified.

Let us now discuss the status of this equation in more detail (here,
we follow the treatment of
Refs.~\cite{Winitzki:2001fc,Vachaspati:2003de}). Classically, one has
$\dot{H}=-(\rho+p)/(2\Mp^2)$. If $H$ increases due to quantum jumps,
then $\rho+p<0$, which means that one must violate the Null Energy
Condition (NEC), namely $T_{\mu \nu}n^{\mu }n^{\nu}<0$, where
$n^{\mu}$ is a null vector. For a scalar field
$T_{\mu \nu}n^{\mu}n^{\nu}=(n^{\mu}\partial_{\mu}\phi)^2\ge 0$ and,
classically, the NEC cannot be violated. Quantum mechanically, a
natural way to describe the backreaction of quantum matter on the
geometry is to write the semi-classical Einstein equations,
$G_{\mu \nu}=\langle \hat{T}_{\mu \nu}\rangle/\Mp^2$. In this
approach, geometry remains classical. Then, let us introduce the NEC
operator
$\hat{O}\equiv \hat{T}_{\mu
  \nu}n^{\mu}n^{\nu}=\hat{P}^{\dagger}\hat{P}$,
where $\hat{P}\equiv n^{\mu}\partial_{\mu}\hat{\phi}$. Generically,
$\langle \hat{O}\rangle $ is infinite and must be renormalized. If
this is done in a quantum state compatible with the symmetry of de
Sitter, then, necessarily,
$\langle \hat{T}_{\mu \nu}^{\rm ren}\rangle \propto g_{\mu \nu}$ and,
therefore, $\langle \hat{O}_{\rm ren}\rangle =0$ and the NEC cannot be
violated. This means that it is necessary to go beyond semi-classical
gravity if we want to treat the eternal inflation case and allow for a
NEC. Notice that this is what is done in the theory of cosmological
perturbations where the equations controlling the evolution of the
system are $\delta \hat{G}_{\mu \nu}=\delta \hat{T}_{\mu \nu}/\Mp^2$,
\ie quantum operators on both sides. In the linear regime, this has
been shown to be consistent and is at the origin of the claim that
inflation implies an almost scale invariant power spectrum for
cosmological perturbations. Of course, eternal inflation corresponds
to a situation where the fluctuations are, by definition, not small. A
possible way out is to define a smeared NEC
operator~\cite{Winitzki:2001fc,Vachaspati:2003de},
\begin{equation}
\hat{O}_W^{\rm ren}\equiv \int {\rm d}^4x\sqrt{-g}W(x)\hat{O}^{\rm ren},
\end{equation}
where $W$ is a window function which has support on a finite part of
spacetime. This breaks de Sitter invariance and, as a consequence, one
can expect $\langle \hat{O}_W^{\rm ren}\rangle \neq 0$. Then, the next
step would be to calculate the effects of smeared fluctuations on the
metric, a framework which does not yet exist. Despite this, it is
usually assumed that this effect will be described by an equation
similar to Eq.~(\ref{eq:quantumfried}). As discussed in
Refs.~\cite{Winitzki:2001fc,Vachaspati:2003de}, the
equation~(\ref{eq:quantumfried}) may describe spacetime before and
after the fluctuation happens. But important issues are not addressed
as, for instance, the behavior of the metric through the fluctuation
or what role the conservation of energy plays in this picture. As
written in Ref.~\cite{GVW}, ``{\it An assumption is that
  Eq.~(28) is sufficient to describe this process}'' [where
``Eq.~(28)'' refers to Eq.~(\ref{eq:quantumfried}) and where ``{\it
  this process}'' refers to the response of quantum geometry to
stochastic fluctuations of the field], or ``{\it So the heuristic
  argument, while suggestive, is certainly not sufficient by itself to
  show that eternal inflation can occur}''. We conclude from the above
considerations that Eq.~(\ref{eq:quantumfried}), on which partially
rests eternal inflation, is an assumption.

To be completely fair, we should also mention an argument which is in
favor of Eq.~(\ref{eq:quantumfried}). Let us indeed consider the
Langevin equation~(\ref{eq:langevinback}) again. It can also be used
to write a Fokker-Planck equation for $P(\phi,N)$, the probability
density of having the field $\phi$ at time $N$. It reads
\begin{align}
\frac{\partial }{\partial N}P(\phi,N)
=\frac{\partial}{\partial \phi}
\left[\frac{V_\phi}{3H^2}P(\phi,N)\right]+
\frac{\partial^2}{\partial \phi^2}
\left[\frac{H^2}{8\pi^2}P(\phi,N)\right].
\end{align}
This equation can also be written as
$\partial P/\partial N=\partial J/\partial \phi$ where $J$ is a
current and a stationary solution $P_{\rm sta}(\phi)$ can be obtained
by requiring that $\partial P_{\rm sta}/\partial N=0$. Then, the
Fokker-Planck equation reduces to a first order differential equation
whose solution can be expressed as
\begin{align}
\label{eq:statiodistri}
P_{\rm sta}(\phi)\propto \exp\left[\frac{24\pi^2\Mp^4}{V(\phi)}\right],
\end{align}
where we have ignored the prefactor which does not play a crucial role
in our discussion. Notice that if one considers the Fokker-Planck
backward equation, then one obtains the same solution but, crucially,
with an overall minus sign in the argument of the exponential, namely 
\begin{align}
\label{eq:statiodistriback}
P_{\rm sta}(\phi)\propto \exp\left[-\frac{24\pi^2\Mp^4}{V(\phi)}\right].
\end{align}
Both equations~(\ref{eq:statiodistri}) and~(\ref{eq:statiodistriback})
are relevant for stochastic inflation. Notice that their derivation
implicitly assumes Eq.~(\ref{eq:quantumfried}).

Let us now consider the same situation but from a quantum cosmology
point of view~\cite{Coule:1999wg}. In quantum cosmology, both matter
and geometry are supposed to be quantized consistently. The
corresponding canonical Hamiltonian can be expressed as
\begin{align}
H_{\rm c} &=N\left[-\frac{\pi_a^2}{48\Mp^2v_{\cal K}a}
+\frac{\pi_\phi^2}{2v_{\cal K}a^3}
-12\Mp^2 kv_{\cal K}a+v_{\cal K}a^3V(\phi)\right],
\end{align}
where the quantity $v_{\cal K}$ represents the volume of the spacelike
hypersurfaces and $N$ is the lapse function. Carrying out Dirac
quantization leads to the Wheeler-De Witt equation for the
wave-function of the universe, $\Psi(a,\phi)$, namely
\begin{align}
\frac{\partial ^2}{\partial a^2}\Psi(a,\phi)
& +\frac{p}{a}\frac{\partial }{\partial a}\Psi(a,\phi)
-6\frac{\Mp^2}{a^2}\frac{\partial^2}{\partial \phi^2}\Psi(a,\phi)
\nonumber \\ &
-36v_{\cal K}^2\Mp^4a_0^2\left(\frac{a}{a_0}\right)^2
\left[{\cal K}-\left(\frac{a}{a_0}\right)^2\right]\Psi(a,\phi)=0.
\end{align}
Here the number $p$ takes into account the factor ordering ambiguity
and $a_0\equiv \left[V(\phi)/(3\Mp^2)\right]^{-1/2}$. If one neglects
the second derivative with respect to $\phi$ and chooses $p=-1$, then
the solution can be found explicitly and reads
\begin{align}
\Psi(a,\phi)=\frac{\alpha {\rm Ai}\left[z(a)\right]
+\beta {\rm Bi}\left[z(a)\right]}{\alpha {\rm Ai}\left[z(0)\right]
+\beta {\rm Bi}\left[z(0)\right]},
\end{align}
where ${\rm Ai}$ and ${\rm Bi}$ are Airy functions of first and second
kinds, respectively, and $z(0)=z(a=0)$. The quantity $z(a)$ is defined
by
$z(a)\equiv \left(3v_{\cal K}\Mp^2a_0^2\right)^{2/3}\left({\cal
    K}-a^2/a_0^2\right)$
and $\alpha$ and $\beta$ are complex numbers to be determined by
boundary conditions: the tunneling wave function corresponds to
$\alpha=1$ and $\beta=i$ and the no boundary one to $\alpha=1$ and
$\beta=0$. In order to make predictions, we need to calculate
probabilities but the Wheeler-De Witt equation does not lead to
positive-definite probabilities. Indeed, the associated current,
\begin{equation}
j=\frac{i}{2\Mp^2}a^p\left(\Psi^*\partial_a\Psi-\Psi\partial_a\Psi^*\right),
\end{equation}
is not positive-definite. However, in the limit
$a\gg \ell_{_{\rm Pl}}$, the Wentzel-Kramers-Brillouin (WKB)
approximation is valid and, in this regime, the probabilities are
positive. For the tunneling wave function, this gives
\begin{align}
j\simeq \frac{2}{\pi a_0^2\Mp^2\vert D\vert^2}
\left(3v_{\cal K}\Mp^2a_0^2\right)^{2/3}=6v_{\cal K}
e^{-12v_{\cal K}\Mp^4/V(\phi)}.
\end{align}
For the no-boundary wave function, one obtains the same result except
that there is no minus in the argument of the exponential. If, in
addition, the spacelike section are taken to be spheres, then
$v_{\cal K}=2\pi^2$ and the prediction of quantum cosmology reads
\begin{equation}
\label{eq:currentqm}
j\propto \exp\left[\pm \frac{24\pi^2 \Mp^2}{V(\phi)}\right],
\end{equation}
which is nothing but Eqs.~(\ref{eq:statiodistri})
and~(\ref{eq:statiodistriback}). We saw before that the use of an
equation $\hat{H}^2=V(\hat{\phi})$ is questionable. The previous
argument, however, seems to indicate that this could be
reasonable. Indeed, as already mentioned, the stationary distribution
of the Fokker-Planck equation was obtained by (implicitly) using this
equation. The fact that the Wheeler-De Witt equation, which is an
equation where the quantum effects of the geometry are taken into
account, leads to results consistent with those obtained from the
stochastic formalism retrospectively justifies the use of an equation
$\hat{H}^2=V(\hat{\phi})$. Of course, the argument is not completely
conclusive since the Wheeler-De Witt equation and the minisuperspace
approximation can also be questioned. We conclude that the tools used
in order to model backreaction in eternal inflation are, at least for
the moment, assumptions. These assumptions may be very reasonable (as
seems to be suggested by the above argument) but they remain
assumptions.

Let us now discuss a second argument. As is clearly illustrated on the
no self-reproduction potential of Sec.~\ref{subsec:avoid}, eternal
inflation also rests on an extrapolation of the potential $V(\phi)$
beyond the observable window. By observing the CMB anisotropy, we
probe only a limited part of $V(\phi)$ corresponding to about seven
e-folds. Eternal inflation depends on another region of the potential
which is not directly observed. Moreover, this part of the potential
is usually relevant at energies higher than the energy scale of
inflation (there are exceptions, for instance hybrid inflation, see
Ref.~\cite{Martin:2011ib}) where higher order operators can play a
crucial role. For instance, our calculation of eternal inflation in
large field models rests on the assumption that, even outside the
observational window, the potential is given by
$V(\phi)\propto \phi^p$. But nobody knows whether this is true since
this is not directly observable. The high-energy corrections could
maybe produce terms leading to the Mukhanov's potential of
Sec.~\ref{subsec:avoid}, in which case eternal inflation would be
irrelevant. Notice that, even if one considers a plateau model, these
corrections could play an important role. Indeed, it is true that, a
priori, corrections in $V/\Mp^4$ are, by construction, always
negligible for plateau models. But the potential itself will
generically receives corrections. For instance, if one adds a term
$\propto R^3$ to the Starobinsky model, then the effective potential
grows with $\phi$. As a consequence, when the field is pushed upwards
by the stochastic fluctuations, these corrections will be important.

Thirdly, eternal inflation suffers from a kind of ``trans-Planckian
problem''. Indeed, as discussed before, one expects the field to be
pushed upwards by stochastic fluctuations. Generically, this means
that the field will penetrate the region where $V(\phi)\gg \Mp^4$. In
this regime, even the notion of a background spacetime is lost.
Indeed, in Ref.~\cite{Linde:1993xx}, this problem was already
encountered and the potential made steeper by hand in order to prevent
the field to penetrate the trans-Planckian region. However, what
really happens in this regime remains a matter of debate.

Fourthly, the multiverse is in fact a combination of eternal inflation
with the string landscape. A priori, string theory only depends on one
parameter, the string tension. All the other parameters of high energy
physics, the masses of the particles, the coupling constant etc \dots
should be the vacuum expectation values of some fields appearing in
string theory. Since, according to eternal inflation, the fields
stochastically fluctuate from patch to patch, it should be the same
for the parameters. We are thus led to a picture where what we see as
fundamental parameters are in fact stochastic quantities fluctuating
from one patch (or one ``pocket universe'') to another. This is the
famous multiverse. As it turns out, the concept of string landscape is
not that obvious and has been discussed among string
theorists~\cite{Gross:2005qm}. At the moment, the best one could
conclude is that the multiverse may pose a question, possibly
justifying investigating alternatives to
inflation~\cite{Battefeld:2014uga}. So the multiverse problem is not
only based on an an extrapolation, it relies in fact on a combination
of extrapolations.

Based on the previous discussion, it seems therefore fair to call the
multiverse ``problem'' of inflation a wild speculation. Even if
eternal inflation happens, it is not completely obvious that a
multiverse will be present. Indeed, since the question of a stringy
landscape remains disputed among string experts, one could imagine a
situation where eternal inflation occurs but where there is no stringy
landscape. In this case, the inflaton vacuum expectation value would
still fluctuate from on patch to another but the fundamental constants
would be the same everywhere. This implies that the inflationary
predictions would also be the same everywhere (for instance, Doppler
peaks in the CMB would be present in each pocket universe), at least
in the patches where inflation came to an end. In any case, should we
reject single-field slow-roll inflation, a falsifiable, well tested,
effective approach to the early universe, in addition in perfect
agreement with observations because of the multiverse? To say the
least, it would be too hasty. It would be similar to rejecting the
standard model of particle physics because (at least for the moment)
it cannot be obtained from string theory.

\section{Conclusion}
\label{sec:conclusion}

In this article, we have discussed various aspects of inflation. The
picture that emerges is that inflation is a very successful model of
the early universe. It has all the criterions that a good scientific
theory should possess.

First, it is falsifiable. One can indeed quote two possible
observations that could potentially rule out inflation. All models of
inflation predict the presence of Doppler peaks in the CMB multipole
moments. Therefore, if instead of detecting them, we had obtained a
bump (as predicted, for instance, if the fluctuations entirely
originate from topological
defects~\cite{Gangui:1995vp,Durrer:1995ni,Ringeval:2005kr}), then
inflation would have been ruled out. Another observation that could
threaten the basic principles of inflation is the observation that
$\Omega_{\cal K}\neq 0$. It is true that an inflationary model with
$\Omega_{\cal K}\neq 0$ has been constructed in
Ref.~\cite{Linde:1995rv} but this model is so peculiar that it can be
viewed as a curiosity and cannot be considered as representative.
Some may argue that it shows the amount of arm-twisting that needs to
be done to inflation to make it predict $\Omega_{\cal K}\neq 0$. In
any case, it is the author's point of view that
$\Omega_{\cal K}\neq 0$ (beyond $10^{-5}$ since, of course, some
curvature is present in the perturbed universe) should be considered
as a fatal blow for inflation.

Second, inflation has been able to make predictions, most notably the
prediction that $\nS$ should be close to one but, and this is the
crucial point, excluding one (see, however, the
exception~\cite{Starobinsky:2005ab}). As discussed at length
previously, this prediction has been confirmed by the data. It is true
that a scale-invariant power spectrum, the so-called
Harrisson-Zeldovitch (HZ) power spectrum, was already considered
before inflation. But, precisely, the HZ power spectrum has $\nS=1$
while inflation has $\ns\sim 1 $ and, crucially, $\ns -1\neq 0$. The
prediction $\nS -1\neq 0$ was first made by inflation and its
observational confirmation is therefore a strong argument in favor of
inflation.

Third, the criticisms against inflation do not seem completely
compelling (see also Ref.~\cite{Chowdhury:2019otk} where the initial
conditions problem and the measure question are discussed in
detail). The initial condition problem does not seem to be very
severe, thanks to the presence of an attractor. It is true that the
attractor is not present for some models (for instance, small-field
inflation with sub-Planckian values) but, precisely, the Planck data
have singled out a model (namely the Starobinsky model) where it is
present.

The multiverse question is nowadays widely debated and there are
claims that its appearance implies that standard inflation makes no
prediction and, therefore, is not falsifiable. The argument is that if
everything happens, there could be patches in our universe where, for
instance, the Doppler peaks are present but there could be others
where it is not the case. Or there could be patches where $\nS$ is
close to one and others where it is far from one. All that is based on
the belief that the multiverse is unavoidable. However, it is, at the
moment, unreasonable to put the multiverse and standard inflation on
an equal footing. Indeed, at this stage, it is fair to say that the
multiverse is a speculation (if it is present at all since we have
seen that it can be avoided, see Sec.~\ref{subsec:avoid}) and one can
argue that it would be awkward to reject a good effective model
because of a mere speculation. As already mentioned, this would be
like rejecting the standard model of particle physics because, so far,
no one has been able to derive it from string theory. To be completely
fair with this analogy and the multiverse criticism, it is true that
the potential modifications of the standard model of particle physics
suggested by string theory are much less radical that what the
multiverse implies for standard inflation.

It is also true that we still do not know the physical nature of the
inflaton field even if the latest data raise the intriguing
possibility that it could the Higgs field itself. After all, we are
trying to develop a theory the typical energy scale of which could be
as high as the GUT scale. So, maybe this problem (if it is indeed one)
is not in the inflationary scenario but rather in our lack of
understanding of particle physics at $10^{15}\GeV$. In any case, with
the recent discovery of the Higgs boson, a common criticism against
inflation, namely that no scalar field has ever been seen, has fallen.

Of course, this does not mean that inflation has no drawback and
should not be criticized. Admittedly, the question of initial
conditions is clearly not completely settled. The question which is
left partially unanswered is what happens when one starts from
strongly inhomogeneous configurations in the most general situation:
impressive numerical simulations of fully inhomogeneous situations
have been performed but they do not yet cover all the
possibilities. This is technically complicated since this requires
numerical relativity. But it is fair to admit that this is a remaining
issue which is very important. On the other hand, it is not clear
whether this question can be treated classically. Most probably,
quantum effects also play an important role in this problem which
makes it even more complicated.

Another open issue is the Ultra-Violet (UV) sensitivity of
inflation. One example is of course eternal inflation itself. Indeed,
we have seen that it can happen or not depending on what we assume
about the shape of the potential at high energies, outside the
observational window.  Another example of UV dependence is the
trans-Planckian problem of inflation. If the fluctuations behave in a
non standard way when their physical wavelength becomes smaller than
the Planck length and if the trans-Planckian physics is non-adiabatic,
then the prediction of an almost scale-invariant power spectrum could
be modified. Let us nevertheless tone down this conclusion by
stressing out that the corresponding modification could be very small. As
was discussed earlier, we have indeed two scales in the problem, the
scale $M_{_{\rm C}}$ at which new physics pops up (typically the
Planck scale) and the Hubble parameter during inflation. If the effect
scales as the ratio $H/M_{_{\rm C}}$ to some power, then the
correction should be very small. Yet another example of UV dependence
is the importance of higher order operators for inflationary model
building, see Ref.~\cite{Silverstein:2017zfk}.

Therefore, it is true that inflation has some UV sensitivity. But,
after all, this is also the case of the standard model of particle
physics where the Higgs mass is not stable against quantum corrections
(the hierarchy problem). But no one would reject this model because of
this issue. Let us also add that it is inconsistent to claim at the
same time that inflation is UV dependent and that the multiverse is
unavoidable: if inflation is UV dependent, then one can modify it at
high energies to avoid the multiverse and this is exactly what the
calculation of Sec.~\ref{subsec:avoid} reveals. From a more general
perspective concerning the IR/UV connection, it is interesting that
inflation seems to provide an example in which the decoupling between
physics at different scales, which is at the basis of effective field
theory, does not work.

In conclusion, inflation appears to be a robust and reliable scenario
for the early universe, not completely free of open issues of course
but could it have been different for a theory which is trying to
describe the first instants of the universe, at energy scales as high
as $10^{15}\GeV$? At this stage, admittedly, one cannot yet trust it
as we trust, for example, the standard model of particle physics. The
situation, however, could change soon if, for instance, we could check
the consistency relation, $r=-\nT/8$. This is obviously a difficult
task and a first step would clearly be to detect primordial
gravitational waves. After all if the pieces of information that we
have gathered so far are correct, the next generation of experiments
should be able to see them. Indeed their target is $r\sim 10^{-4}$
while, our best model, the Starobinsky model, predicts
$r\sim 4\times 10^{-3}$. Then measuring $\nT$ will be even more
difficult but would be very important. The measurement of NG would
also be important. The expected level, $\fnl\simeq 10^{-2}$, is tiny
for our preferred class of models but people are already thinking
about experiments that could reach this level.

In brief, inflation continues to be an inspiration for many physicists
and continues to fuel new interesting works.  So, inflation, trick or
treat? Treat, definitively!

\acknowledgements{It is a pleasure to thank  P.~Peter and V.~Vennin for
  careful reading of the manuscript.}

\bibliographystyle{apsrev4-1}
\bibliography{ftinf}

\begin{thebibliography}{141}%
\makeatletter
\providecommand \@ifxundefined [1]{%
 \@ifx{#1\undefined}
}%
\providecommand \@ifnum [1]{%
 \ifnum #1\expandafter \@firstoftwo
 \else \expandafter \@secondoftwo
 \fi
}%
\providecommand \@ifx [1]{%
 \ifx #1\expandafter \@firstoftwo
 \else \expandafter \@secondoftwo
 \fi
}%
\providecommand \natexlab [1]{#1}%
\providecommand \enquote  [1]{``#1''}%
\providecommand \bibnamefont  [1]{#1}%
\providecommand \bibfnamefont [1]{#1}%
\providecommand \citenamefont [1]{#1}%
\providecommand \href@noop [0]{\@secondoftwo}%
\providecommand \href [0]{\begingroup \@sanitize@url \@href}%
\providecommand \@href[1]{\@@startlink{#1}\@@href}%
\providecommand \@@href[1]{\endgroup#1\@@endlink}%
\providecommand \@sanitize@url [0]{\catcode `\\12\catcode `\$12\catcode
  `\&12\catcode `\#12\catcode `\^12\catcode `\_12\catcode `\%12\relax}%
\providecommand \@@startlink[1]{}%
\providecommand \@@endlink[0]{}%
\providecommand \url  [0]{\begingroup\@sanitize@url \@url }%
\providecommand \@url [1]{\endgroup\@href {#1}{\urlprefix }}%
\providecommand \urlprefix  [0]{URL }%
\providecommand \Eprint [0]{\href }%
\providecommand \doibase [0]{http://dx.doi.org/}%
\providecommand \selectlanguage [0]{\@gobble}%
\providecommand \bibinfo  [0]{\@secondoftwo}%
\providecommand \bibfield  [0]{\@secondoftwo}%
\providecommand \translation [1]{[#1]}%
\providecommand \BibitemOpen [0]{}%
\providecommand \bibitemStop [0]{}%
\providecommand \bibitemNoStop [0]{.\EOS\space}%
\providecommand \EOS [0]{\spacefactor3000\relax}%
\providecommand \BibitemShut  [1]{\csname bibitem#1\endcsname}%
\let\auto@bib@innerbib\@empty
\bibitem [{\citenamefont {Starobinsky}(1979)}]{Starobinsky:1979ty}%
  \BibitemOpen
  \bibfield  {author} {\bibinfo {author} {\bibfnamefont {A.~A.}\ \bibnamefont
  {Starobinsky}},\ }\href@noop {} {\bibfield  {journal} {\bibinfo  {journal}
  {JETP Lett.}\ }\textbf {\bibinfo {volume} {30}},\ \bibinfo {pages} {682}
  (\bibinfo {year} {1979})}\BibitemShut {NoStop}%
\bibitem [{\citenamefont {Starobinsky}(1980)}]{Starobinsky:1980te}%
  \BibitemOpen
  \bibfield  {author} {\bibinfo {author} {\bibfnamefont {A.~A.}\ \bibnamefont
  {Starobinsky}},\ }\href {\doibase 10.1016/0370-2693(80)90670-X} {\bibfield
  {journal} {\bibinfo  {journal} {Phys. Lett.}\ }\textbf {\bibinfo {volume}
  {B91}},\ \bibinfo {pages} {99} (\bibinfo {year} {1980})}\BibitemShut
  {NoStop}%
\bibitem [{\citenamefont {Guth}(1981)}]{Guth:1980zm}%
  \BibitemOpen
  \bibfield  {author} {\bibinfo {author} {\bibfnamefont {A.~H.}\ \bibnamefont
  {Guth}},\ }\href {\doibase 10.1103/PhysRevD.23.347} {\bibfield  {journal}
  {\bibinfo  {journal} {Phys. Rev.}\ }\textbf {\bibinfo {volume} {D23}},\
  \bibinfo {pages} {347} (\bibinfo {year} {1981})}\BibitemShut {NoStop}%
\bibitem [{\citenamefont {Linde}(1982{\natexlab{a}})}]{Linde:1981mu}%
  \BibitemOpen
  \bibfield  {author} {\bibinfo {author} {\bibfnamefont {A.~D.}\ \bibnamefont
  {Linde}},\ }\href {\doibase 10.1016/0370-2693(82)91219-9} {\bibfield
  {journal} {\bibinfo  {journal} {Phys. Lett.}\ }\textbf {\bibinfo {volume}
  {B108}},\ \bibinfo {pages} {389} (\bibinfo {year}
  {1982}{\natexlab{a}})}\BibitemShut {NoStop}%
\bibitem [{\citenamefont {Mukhanov}\ and\ \citenamefont
  {Chibisov}(1981)}]{Mukhanov:1981xt}%
  \BibitemOpen
  \bibfield  {author} {\bibinfo {author} {\bibfnamefont {V.~F.}\ \bibnamefont
  {Mukhanov}}\ and\ \bibinfo {author} {\bibfnamefont {G.}~\bibnamefont
  {Chibisov}},\ }\href@noop {} {\bibfield  {journal} {\bibinfo  {journal} {JETP
  Lett.}\ }\textbf {\bibinfo {volume} {33}},\ \bibinfo {pages} {532} (\bibinfo
  {year} {1981})}\BibitemShut {NoStop}%
\bibitem [{\citenamefont {Mukhanov}\ and\ \citenamefont
  {Chibisov}(1982)}]{Mukhanov:1982nu}%
  \BibitemOpen
  \bibfield  {author} {\bibinfo {author} {\bibfnamefont {V.~F.}\ \bibnamefont
  {Mukhanov}}\ and\ \bibinfo {author} {\bibfnamefont {G.}~\bibnamefont
  {Chibisov}},\ }\href@noop {} {\bibfield  {journal} {\bibinfo  {journal}
  {Sov.Phys.JETP}\ }\textbf {\bibinfo {volume} {56}},\ \bibinfo {pages} {258}
  (\bibinfo {year} {1982})}\BibitemShut {NoStop}%
\bibitem [{\citenamefont {Starobinsky}(1982)}]{Starobinsky:1982ee}%
  \BibitemOpen
  \bibfield  {author} {\bibinfo {author} {\bibfnamefont {A.~A.}\ \bibnamefont
  {Starobinsky}},\ }\href {\doibase 10.1016/0370-2693(82)90541-X} {\bibfield
  {journal} {\bibinfo  {journal} {Phys. Lett.}\ }\textbf {\bibinfo {volume}
  {B117}},\ \bibinfo {pages} {175} (\bibinfo {year} {1982})}\BibitemShut
  {NoStop}%
\bibitem [{\citenamefont {Weinberg}(1972)}]{Weinberg:100595}%
  \BibitemOpen
  \bibfield  {author} {\bibinfo {author} {\bibfnamefont {S.}~\bibnamefont
  {Weinberg}},\ }\href {https://cds.cern.ch/record/100595} {\emph {\bibinfo
  {title} {{Gravitation and Cosmology: Principles and Applications of the
  General Theory of Relativity}}}}\ (\bibinfo  {publisher} {Wiley},\ \bibinfo
  {address} {New York, NY},\ \bibinfo {year} {1972})\BibitemShut {NoStop}%
\bibitem [{\citenamefont {Peter}\ and\ \citenamefont
  {Uzan}(2009)}]{Peter:1208401}%
  \BibitemOpen
  \bibfield  {author} {\bibinfo {author} {\bibfnamefont {P.}~\bibnamefont
  {Peter}}\ and\ \bibinfo {author} {\bibfnamefont {J.-P.}\ \bibnamefont
  {Uzan}},\ }\href {https://cds.cern.ch/record/1208401} {\emph {\bibinfo
  {title} {{Primordial cosmology}}}},\ Oxford Graduate Texts\ (\bibinfo
  {publisher} {Oxford Univ. Press},\ \bibinfo {address} {Oxford},\ \bibinfo
  {year} {2009})\BibitemShut {NoStop}%
\bibitem [{\citenamefont {Smoot}\ \emph {et~al.}(1992)\citenamefont {Smoot}
  \emph {et~al.}}]{Smoot:1992td}%
  \BibitemOpen
  \bibfield  {author} {\bibinfo {author} {\bibfnamefont {G.~F.}\ \bibnamefont
  {Smoot}} \emph {et~al.} (\bibinfo {collaboration} {COBE}),\ }\href {\doibase
  10.1086/186504} {\bibfield  {journal} {\bibinfo  {journal} {Astrophys. J.}\
  }\textbf {\bibinfo {volume} {396}},\ \bibinfo {pages} {L1} (\bibinfo {year}
  {1992})}\BibitemShut {NoStop}%
\bibitem [{\citenamefont {Bennett}\ \emph {et~al.}(1996)\citenamefont
  {Bennett}, \citenamefont {Banday}, \citenamefont {Gorski}, \citenamefont
  {Hinshaw}, \citenamefont {Jackson}, \citenamefont {Keegstra}, \citenamefont
  {Kogut}, \citenamefont {Smoot}, \citenamefont {Wilkinson},\ and\
  \citenamefont {Wright}}]{Bennett:1996ce}%
  \BibitemOpen
  \bibfield  {author} {\bibinfo {author} {\bibfnamefont {C.~L.}\ \bibnamefont
  {Bennett}}, \bibinfo {author} {\bibfnamefont {A.}~\bibnamefont {Banday}},
  \bibinfo {author} {\bibfnamefont {K.~M.}\ \bibnamefont {Gorski}}, \bibinfo
  {author} {\bibfnamefont {G.}~\bibnamefont {Hinshaw}}, \bibinfo {author}
  {\bibfnamefont {P.}~\bibnamefont {Jackson}}, \bibinfo {author} {\bibfnamefont
  {P.}~\bibnamefont {Keegstra}}, \bibinfo {author} {\bibfnamefont
  {A.}~\bibnamefont {Kogut}}, \bibinfo {author} {\bibfnamefont {G.~F.}\
  \bibnamefont {Smoot}}, \bibinfo {author} {\bibfnamefont {D.~T.}\ \bibnamefont
  {Wilkinson}}, \ and\ \bibinfo {author} {\bibfnamefont {E.~L.}\ \bibnamefont
  {Wright}},\ }\href {\doibase 10.1086/310075} {\bibfield  {journal} {\bibinfo
  {journal} {Astrophys. J.}\ }\textbf {\bibinfo {volume} {464}},\ \bibinfo
  {pages} {L1} (\bibinfo {year} {1996})},\ \Eprint
  {http://arxiv.org/abs/astro-ph/9601067} {arXiv:astro-ph/9601067 [astro-ph]}
  \BibitemShut {NoStop}%
\bibitem [{\citenamefont {Ade}\ \emph {et~al.}(2014{\natexlab{a}})\citenamefont
  {Ade} \emph {et~al.}}]{Ade:2013zuv}%
  \BibitemOpen
  \bibfield  {author} {\bibinfo {author} {\bibfnamefont {P.}~\bibnamefont
  {Ade}} \emph {et~al.} (\bibinfo {collaboration} {Planck}),\ }\href {\doibase
  10.1051/0004-6361/201321591} {\bibfield  {journal} {\bibinfo  {journal}
  {Astron.Astrophys.}\ }\textbf {\bibinfo {volume} {571}},\ \bibinfo {pages}
  {A16} (\bibinfo {year} {2014}{\natexlab{a}})},\ \Eprint
  {http://arxiv.org/abs/1303.5076} {arXiv:1303.5076 [astro-ph.CO]} \BibitemShut
  {NoStop}%
\bibitem [{\citenamefont {Ade}\ \emph {et~al.}(2014{\natexlab{b}})\citenamefont
  {Ade} \emph {et~al.}}]{Ade:2013sjv}%
  \BibitemOpen
  \bibfield  {author} {\bibinfo {author} {\bibfnamefont {P.~A.~R.}\
  \bibnamefont {Ade}} \emph {et~al.} (\bibinfo {collaboration} {Planck}),\
  }\href {\doibase 10.1051/0004-6361/201321529} {\bibfield  {journal} {\bibinfo
   {journal} {Astron. Astrophys.}\ }\textbf {\bibinfo {volume} {571}},\
  \bibinfo {pages} {A1} (\bibinfo {year} {2014}{\natexlab{b}})},\ \Eprint
  {http://arxiv.org/abs/1303.5062} {arXiv:1303.5062 [astro-ph.CO]} \BibitemShut
  {NoStop}%
\bibitem [{\citenamefont {Ade}\ \emph {et~al.}(2014{\natexlab{c}})\citenamefont
  {Ade} \emph {et~al.}}]{Planck:2013jfk}%
  \BibitemOpen
  \bibfield  {author} {\bibinfo {author} {\bibfnamefont {P.}~\bibnamefont
  {Ade}} \emph {et~al.} (\bibinfo {collaboration} {Planck}),\ }\href {\doibase
  10.1051/0004-6361/201321569} {\bibfield  {journal} {\bibinfo  {journal}
  {Astron.Astrophys.}\ }\textbf {\bibinfo {volume} {571}},\ \bibinfo {pages}
  {A22} (\bibinfo {year} {2014}{\natexlab{c}})},\ \Eprint
  {http://arxiv.org/abs/1303.5082} {arXiv:1303.5082 [astro-ph.CO]} \BibitemShut
  {NoStop}%
\bibitem [{\citenamefont {Ade}\ \emph {et~al.}(2014{\natexlab{d}})\citenamefont
  {Ade} \emph {et~al.}}]{Ade:2013ydc}%
  \BibitemOpen
  \bibfield  {author} {\bibinfo {author} {\bibfnamefont {P.}~\bibnamefont
  {Ade}} \emph {et~al.} (\bibinfo {collaboration} {Planck}),\ }\href {\doibase
  10.1051/0004-6361/201321554} {\bibfield  {journal} {\bibinfo  {journal}
  {Astron.Astrophys.}\ }\textbf {\bibinfo {volume} {571}},\ \bibinfo {pages}
  {A24} (\bibinfo {year} {2014}{\natexlab{d}})},\ \Eprint
  {http://arxiv.org/abs/1303.5084} {arXiv:1303.5084 [astro-ph.CO]} \BibitemShut
  {NoStop}%
\bibitem [{\citenamefont {Ade}\ \emph {et~al.}(2015)\citenamefont {Ade} \emph
  {et~al.}}]{Ade:2015tva}%
  \BibitemOpen
  \bibfield  {author} {\bibinfo {author} {\bibfnamefont {P.~A.~R.}\
  \bibnamefont {Ade}} \emph {et~al.} (\bibinfo {collaboration} {BICEP2,
  Planck}),\ }\href {\doibase 10.1103/PhysRevLett.114.101301} {\bibfield
  {journal} {\bibinfo  {journal} {Phys. Rev. Lett.}\ }\textbf {\bibinfo
  {volume} {114}},\ \bibinfo {pages} {101301} (\bibinfo {year} {2015})},\
  \Eprint {http://arxiv.org/abs/1502.00612} {arXiv:1502.00612 [astro-ph.CO]}
  \BibitemShut {NoStop}%
\bibitem [{\citenamefont {Ade}\ \emph {et~al.}(2016{\natexlab{a}})\citenamefont
  {Ade} \emph {et~al.}}]{Ade:2015xua}%
  \BibitemOpen
  \bibfield  {author} {\bibinfo {author} {\bibfnamefont {P.~A.~R.}\
  \bibnamefont {Ade}} \emph {et~al.} (\bibinfo {collaboration} {Planck}),\
  }\href {\doibase 10.1051/0004-6361/201525830} {\bibfield  {journal} {\bibinfo
   {journal} {Astron. Astrophys.}\ }\textbf {\bibinfo {volume} {594}},\
  \bibinfo {pages} {A13} (\bibinfo {year} {2016}{\natexlab{a}})},\ \Eprint
  {http://arxiv.org/abs/1502.01589} {arXiv:1502.01589 [astro-ph.CO]}
  \BibitemShut {NoStop}%
\bibitem [{\citenamefont {Ade}\ \emph {et~al.}(2016{\natexlab{b}})\citenamefont
  {Ade} \emph {et~al.}}]{Ade:2015lrj}%
  \BibitemOpen
  \bibfield  {author} {\bibinfo {author} {\bibfnamefont {P.~A.~R.}\
  \bibnamefont {Ade}} \emph {et~al.} (\bibinfo {collaboration} {Planck}),\
  }\href {\doibase 10.1051/0004-6361/201525898} {\bibfield  {journal} {\bibinfo
   {journal} {Astron. Astrophys.}\ }\textbf {\bibinfo {volume} {594}},\
  \bibinfo {pages} {A20} (\bibinfo {year} {2016}{\natexlab{b}})},\ \Eprint
  {http://arxiv.org/abs/1502.02114} {arXiv:1502.02114 [astro-ph.CO]}
  \BibitemShut {NoStop}%
\bibitem [{\citenamefont {Aghanim}\ \emph {et~al.}(2018)\citenamefont {Aghanim}
  \emph {et~al.}}]{Aghanim:2018eyx}%
  \BibitemOpen
  \bibfield  {author} {\bibinfo {author} {\bibfnamefont {N.}~\bibnamefont
  {Aghanim}} \emph {et~al.} (\bibinfo {collaboration} {Planck}),\ }\href@noop
  {} {\  (\bibinfo {year} {2018})},\ \Eprint {http://arxiv.org/abs/1807.06209}
  {arXiv:1807.06209 [astro-ph.CO]} \BibitemShut {NoStop}%
\bibitem [{\citenamefont {Akrami}\ \emph
  {et~al.}(2018{\natexlab{a}})\citenamefont {Akrami} \emph
  {et~al.}}]{Akrami:2018odb}%
  \BibitemOpen
  \bibfield  {author} {\bibinfo {author} {\bibfnamefont {Y.}~\bibnamefont
  {Akrami}} \emph {et~al.} (\bibinfo {collaboration} {Planck}),\ }\href@noop {}
  {\  (\bibinfo {year} {2018}{\natexlab{a}})},\ \Eprint
  {http://arxiv.org/abs/1807.06211} {arXiv:1807.06211 [astro-ph.CO]}
  \BibitemShut {NoStop}%
\bibitem [{\citenamefont {Akrami}\ \emph
  {et~al.}(2018{\natexlab{b}})\citenamefont {Akrami} \emph
  {et~al.}}]{Akrami:2018vks}%
  \BibitemOpen
  \bibfield  {author} {\bibinfo {author} {\bibfnamefont {Y.}~\bibnamefont
  {Akrami}} \emph {et~al.} (\bibinfo {collaboration} {Planck}),\ }\href@noop {}
  {\  (\bibinfo {year} {2018}{\natexlab{b}})},\ \Eprint
  {http://arxiv.org/abs/1807.06205} {arXiv:1807.06205 [astro-ph.CO]}
  \BibitemShut {NoStop}%
\bibitem [{\citenamefont {Penrose}(1989)}]{Penrose:1988mg}%
  \BibitemOpen
  \bibfield  {author} {\bibinfo {author} {\bibfnamefont {R.}~\bibnamefont
  {Penrose}},\ }\bibfield  {booktitle} {\emph {\bibinfo {booktitle} {{14th
  Texas Symposium on Relativistic Astrophysics Dallas, Texas, December 11-16,
  1988}}},\ }\href {\doibase 10.1111/j.1749-6632.1989.tb50513.x} {\bibfield
  {journal} {\bibinfo  {journal} {Annals N. Y. Acad. Sci.}\ }\textbf {\bibinfo
  {volume} {571}},\ \bibinfo {pages} {249} (\bibinfo {year}
  {1989})}\BibitemShut {NoStop}%
\bibitem [{\citenamefont {Peebles}(1999)}]{Peebles:1999xc}%
  \BibitemOpen
  \bibfield  {author} {\bibinfo {author} {\bibfnamefont {P.~J.~E.}\
  \bibnamefont {Peebles}}\ }(\bibinfo {year} {1999})\ \Eprint
  {http://arxiv.org/abs/astro-ph/9905390} {arXiv:astro-ph/9905390 [astro-ph]}
  \BibitemShut {NoStop}%
\bibitem [{\citenamefont {Ijjas}\ \emph {et~al.}(2013)\citenamefont {Ijjas},
  \citenamefont {Steinhardt},\ and\ \citenamefont {Loeb}}]{Ijjas:2013vea}%
  \BibitemOpen
  \bibfield  {author} {\bibinfo {author} {\bibfnamefont {A.}~\bibnamefont
  {Ijjas}}, \bibinfo {author} {\bibfnamefont {P.~J.}\ \bibnamefont
  {Steinhardt}}, \ and\ \bibinfo {author} {\bibfnamefont {A.}~\bibnamefont
  {Loeb}},\ }\href {\doibase 10.1016/j.physletb.2013.05.023} {\bibfield
  {journal} {\bibinfo  {journal} {Phys. Lett.}\ }\textbf {\bibinfo {volume}
  {B723}},\ \bibinfo {pages} {261} (\bibinfo {year} {2013})},\ \Eprint
  {http://arxiv.org/abs/1304.2785} {arXiv:1304.2785 [astro-ph.CO]} \BibitemShut
  {NoStop}%
\bibitem [{\citenamefont {Ijjas}\ \emph {et~al.}(2014)\citenamefont {Ijjas},
  \citenamefont {Steinhardt},\ and\ \citenamefont {Loeb}}]{Ijjas:2014nta}%
  \BibitemOpen
  \bibfield  {author} {\bibinfo {author} {\bibfnamefont {A.}~\bibnamefont
  {Ijjas}}, \bibinfo {author} {\bibfnamefont {P.~J.}\ \bibnamefont
  {Steinhardt}}, \ and\ \bibinfo {author} {\bibfnamefont {A.}~\bibnamefont
  {Loeb}},\ }\href {\doibase 10.1016/j.physletb.2014.07.012} {\bibfield
  {journal} {\bibinfo  {journal} {Phys. Lett.}\ }\textbf {\bibinfo {volume}
  {B736}},\ \bibinfo {pages} {142} (\bibinfo {year} {2014})},\ \Eprint
  {http://arxiv.org/abs/1402.6980} {arXiv:1402.6980 [astro-ph.CO]} \BibitemShut
  {NoStop}%
\bibitem [{\citenamefont {Turner}(1983)}]{Turner:1983he}%
  \BibitemOpen
  \bibfield  {author} {\bibinfo {author} {\bibfnamefont {M.~S.}\ \bibnamefont
  {Turner}},\ }\href {\doibase 10.1103/PhysRevD.28.1243} {\bibfield  {journal}
  {\bibinfo  {journal} {Phys. Rev.}\ }\textbf {\bibinfo {volume} {D28}},\
  \bibinfo {pages} {1243} (\bibinfo {year} {1983})}\BibitemShut {NoStop}%
\bibitem [{\citenamefont {Traschen}\ and\ \citenamefont
  {Brandenberger}(1990)}]{Traschen:1990sw}%
  \BibitemOpen
  \bibfield  {author} {\bibinfo {author} {\bibfnamefont {J.~H.}\ \bibnamefont
  {Traschen}}\ and\ \bibinfo {author} {\bibfnamefont {R.~H.}\ \bibnamefont
  {Brandenberger}},\ }\href {\doibase 10.1103/PhysRevD.42.2491} {\bibfield
  {journal} {\bibinfo  {journal} {Phys. Rev.}\ }\textbf {\bibinfo {volume}
  {D42}},\ \bibinfo {pages} {2491} (\bibinfo {year} {1990})}\BibitemShut
  {NoStop}%
\bibitem [{\citenamefont {Kofman}\ \emph {et~al.}(1997)\citenamefont {Kofman},
  \citenamefont {Linde},\ and\ \citenamefont {Starobinsky}}]{Kofman:1997yn}%
  \BibitemOpen
  \bibfield  {author} {\bibinfo {author} {\bibfnamefont {L.}~\bibnamefont
  {Kofman}}, \bibinfo {author} {\bibfnamefont {A.~D.}\ \bibnamefont {Linde}}, \
  and\ \bibinfo {author} {\bibfnamefont {A.~A.}\ \bibnamefont {Starobinsky}},\
  }\href {\doibase 10.1103/PhysRevD.56.3258} {\bibfield  {journal} {\bibinfo
  {journal} {Phys. Rev.}\ }\textbf {\bibinfo {volume} {D56}},\ \bibinfo {pages}
  {3258} (\bibinfo {year} {1997})},\ \Eprint
  {http://arxiv.org/abs/hep-ph/9704452} {arXiv:hep-ph/9704452 [hep-ph]}
  \BibitemShut {NoStop}%
\bibitem [{\citenamefont {Amin}\ \emph {et~al.}(2014)\citenamefont {Amin},
  \citenamefont {Hertzberg}, \citenamefont {Kaiser},\ and\ \citenamefont
  {Karouby}}]{Amin:2014eta}%
  \BibitemOpen
  \bibfield  {author} {\bibinfo {author} {\bibfnamefont {M.~A.}\ \bibnamefont
  {Amin}}, \bibinfo {author} {\bibfnamefont {M.~P.}\ \bibnamefont {Hertzberg}},
  \bibinfo {author} {\bibfnamefont {D.~I.}\ \bibnamefont {Kaiser}}, \ and\
  \bibinfo {author} {\bibfnamefont {J.}~\bibnamefont {Karouby}},\ }\href
  {\doibase 10.1142/S0218271815300037} {\bibfield  {journal} {\bibinfo
  {journal} {Int. J. Mod. Phys.}\ }\textbf {\bibinfo {volume} {D24}},\ \bibinfo
  {pages} {1530003} (\bibinfo {year} {2014})},\ \Eprint
  {http://arxiv.org/abs/1410.3808} {arXiv:1410.3808 [hep-ph]} \BibitemShut
  {NoStop}%
\bibitem [{\citenamefont {Starobinsky}(1992)}]{Starobinsky:1992ts}%
  \BibitemOpen
  \bibfield  {author} {\bibinfo {author} {\bibfnamefont {A.~A.}\ \bibnamefont
  {Starobinsky}},\ }\href@noop {} {\bibfield  {journal} {\bibinfo  {journal}
  {JETP Lett.}\ }\textbf {\bibinfo {volume} {55}},\ \bibinfo {pages} {489}
  (\bibinfo {year} {1992})},\ \bibinfo {note} {[Pisma Zh. Eksp. Teor.
  Fiz.55,477(1992)]}\BibitemShut {NoStop}%
\bibitem [{\citenamefont {Armendariz-Picon}\ \emph {et~al.}(2001)\citenamefont
  {Armendariz-Picon}, \citenamefont {Mukhanov},\ and\ \citenamefont
  {Steinhardt}}]{ArmendarizPicon:2000ah}%
  \BibitemOpen
  \bibfield  {author} {\bibinfo {author} {\bibfnamefont {C.}~\bibnamefont
  {Armendariz-Picon}}, \bibinfo {author} {\bibfnamefont {V.~F.}\ \bibnamefont
  {Mukhanov}}, \ and\ \bibinfo {author} {\bibfnamefont {P.~J.}\ \bibnamefont
  {Steinhardt}},\ }\href {\doibase 10.1103/PhysRevD.63.103510} {\bibfield
  {journal} {\bibinfo  {journal} {Phys. Rev.}\ }\textbf {\bibinfo {volume}
  {D63}},\ \bibinfo {pages} {103510} (\bibinfo {year} {2001})},\ \Eprint
  {http://arxiv.org/abs/astro-ph/0006373} {arXiv:astro-ph/0006373 [astro-ph]}
  \BibitemShut {NoStop}%
\bibitem [{\citenamefont {Wands}(2008)}]{Wands:2007bd}%
  \BibitemOpen
  \bibfield  {author} {\bibinfo {author} {\bibfnamefont {D.}~\bibnamefont
  {Wands}},\ }\bibfield  {booktitle} {\emph {\bibinfo {booktitle} {{22nd IAP
  Colloquium on Inflation + 25: The First 25 Years of Inflationary Cosmology
  Paris, France, June 26-30, 2006}}},\ }\href {\doibase
  10.1007/978-3-540-74353-8_8} {\bibfield  {journal} {\bibinfo  {journal}
  {Lect. Notes Phys.}\ }\textbf {\bibinfo {volume} {738}},\ \bibinfo {pages}
  {275} (\bibinfo {year} {2008})},\ \Eprint
  {http://arxiv.org/abs/astro-ph/0702187} {arXiv:astro-ph/0702187 [ASTRO-PH]}
  \BibitemShut {NoStop}%
\bibitem [{\citenamefont {Ávila}\ \emph {et~al.}(2014)\citenamefont {Ávila},
  \citenamefont {Martin},\ and\ \citenamefont {Steer}}]{Avila:2013ela}%
  \BibitemOpen
  \bibfield  {author} {\bibinfo {author} {\bibfnamefont {S.}~\bibnamefont
  {Ávila}}, \bibinfo {author} {\bibfnamefont {J.}~\bibnamefont {Martin}}, \
  and\ \bibinfo {author} {\bibfnamefont {D.}~\bibnamefont {Steer}},\ }\href
  {\doibase 10.1088/1475-7516/2014/08/032} {\bibfield  {journal} {\bibinfo
  {journal} {JCAP}\ }\textbf {\bibinfo {volume} {1408}},\ \bibinfo {pages}
  {032} (\bibinfo {year} {2014})},\ \Eprint {http://arxiv.org/abs/1304.3262}
  {arXiv:1304.3262 [hep-th]} \BibitemShut {NoStop}%
\bibitem [{\citenamefont {Martin}\ \emph
  {et~al.}(2014{\natexlab{a}})\citenamefont {Martin}, \citenamefont
  {Ringeval},\ and\ \citenamefont {Vennin}}]{Martin:2013tda}%
  \BibitemOpen
  \bibfield  {author} {\bibinfo {author} {\bibfnamefont {J.}~\bibnamefont
  {Martin}}, \bibinfo {author} {\bibfnamefont {C.}~\bibnamefont {Ringeval}}, \
  and\ \bibinfo {author} {\bibfnamefont {V.}~\bibnamefont {Vennin}},\ }\href
  {\doibase 10.1016/j.dark.2014.01.003} {\bibfield  {journal} {\bibinfo
  {journal} {Phys. Dark Univ.}\ }\textbf {\bibinfo {volume} {5-6}},\ \bibinfo
  {pages} {75–235} (\bibinfo {year} {2014}{\natexlab{a}})},\ \Eprint
  {http://arxiv.org/abs/1303.3787} {arXiv:1303.3787 [astro-ph.CO]} \BibitemShut
  {NoStop}%
\bibitem [{\citenamefont {Linde}(1982{\natexlab{b}})}]{Linde:1982ur}%
  \BibitemOpen
  \bibfield  {author} {\bibinfo {author} {\bibfnamefont {A.~D.}\ \bibnamefont
  {Linde}},\ }\href@noop {} {\  (\bibinfo {year}
  {1982}{\natexlab{b}})}\BibitemShut {NoStop}%
\bibitem [{\citenamefont {Steinhardt}(1982)}]{Steinhardt:1982kg}%
  \BibitemOpen
  \bibfield  {author} {\bibinfo {author} {\bibfnamefont {P.~J.}\ \bibnamefont
  {Steinhardt}},\ }in\ \href@noop {} {\emph {\bibinfo {booktitle} {{Nuffield
  Workshop on the Very Early Universe Cambridge, England, June 21-July 9,
  1982}}}}\ (\bibinfo {year} {1982})\ pp.\ \bibinfo {pages}
  {251--266}\BibitemShut {NoStop}%
\bibitem [{\citenamefont {Linde}(1986)}]{Linde:1986fd}%
  \BibitemOpen
  \bibfield  {author} {\bibinfo {author} {\bibfnamefont {A.~D.}\ \bibnamefont
  {Linde}},\ }\href {\doibase 10.1016/0370-2693(86)90611-8} {\bibfield
  {journal} {\bibinfo  {journal} {Phys. Lett.}\ }\textbf {\bibinfo {volume}
  {B175}},\ \bibinfo {pages} {395} (\bibinfo {year} {1986})}\BibitemShut
  {NoStop}%
\bibitem [{\citenamefont {Linde}\ \emph {et~al.}(1994)\citenamefont {Linde},
  \citenamefont {Linde},\ and\ \citenamefont {Mezhlumian}}]{Linde:1993xx}%
  \BibitemOpen
  \bibfield  {author} {\bibinfo {author} {\bibfnamefont {A.~D.}\ \bibnamefont
  {Linde}}, \bibinfo {author} {\bibfnamefont {D.~A.}\ \bibnamefont {Linde}}, \
  and\ \bibinfo {author} {\bibfnamefont {A.}~\bibnamefont {Mezhlumian}},\
  }\href {\doibase 10.1103/PhysRevD.49.1783} {\bibfield  {journal} {\bibinfo
  {journal} {Phys. Rev.}\ }\textbf {\bibinfo {volume} {D49}},\ \bibinfo {pages}
  {1783} (\bibinfo {year} {1994})},\ \Eprint
  {http://arxiv.org/abs/gr-qc/9306035} {arXiv:gr-qc/9306035 [gr-qc]}
  \BibitemShut {NoStop}%
\bibitem [{\citenamefont {Guth}(2007)}]{Guth:2007ng}%
  \BibitemOpen
  \bibfield  {author} {\bibinfo {author} {\bibfnamefont {A.~H.}\ \bibnamefont
  {Guth}},\ }\bibfield  {booktitle} {\emph {\bibinfo {booktitle} {{Quantum
  theories and renormalization group in gravity and cosmology: Proceedings, 2nd
  International Conference, IRGAC 2006, Barcelona, Spain, July 11-15, 2006}}},\
  }\href {\doibase 10.1088/1751-8113/40/25/S25} {\bibfield  {journal} {\bibinfo
   {journal} {J. Phys.}\ }\textbf {\bibinfo {volume} {A40}},\ \bibinfo {pages}
  {6811} (\bibinfo {year} {2007})},\ \Eprint
  {http://arxiv.org/abs/hep-th/0702178} {arXiv:hep-th/0702178 [HEP-TH]}
  \BibitemShut {NoStop}%
\bibitem [{\citenamefont {Guth}(2000)}]{Guth:2000ka}%
  \BibitemOpen
  \bibfield  {author} {\bibinfo {author} {\bibfnamefont {A.~H.}\ \bibnamefont
  {Guth}},\ }\href {\doibase 10.1016/S0370-1573(00)00037-5} {\bibfield
  {journal} {\bibinfo  {journal} {Phys. Rept.}\ }\textbf {\bibinfo {volume}
  {333}},\ \bibinfo {pages} {555} (\bibinfo {year} {2000})},\ \Eprint
  {http://arxiv.org/abs/astro-ph/0002156} {arXiv:astro-ph/0002156 [astro-ph]}
  \BibitemShut {NoStop}%
\bibitem [{\citenamefont {Linde}(2017)}]{Linde:2015edk}%
  \BibitemOpen
  \bibfield  {author} {\bibinfo {author} {\bibfnamefont {A.}~\bibnamefont
  {Linde}},\ }\href {\doibase 10.1088/1361-6633/aa50e4} {\bibfield  {journal}
  {\bibinfo  {journal} {Rept. Prog. Phys.}\ }\textbf {\bibinfo {volume} {80}},\
  \bibinfo {pages} {022001} (\bibinfo {year} {2017})},\ \Eprint
  {http://arxiv.org/abs/1512.01203} {arXiv:1512.01203 [hep-th]} \BibitemShut
  {NoStop}%
\bibitem [{\citenamefont {Chowdhury}\ \emph {et~al.}(2019)\citenamefont
  {Chowdhury}, \citenamefont {Martin}, \citenamefont {Ringeval},\ and\
  \citenamefont {Vennin}}]{Chowdhury:2019otk}%
  \BibitemOpen
  \bibfield  {author} {\bibinfo {author} {\bibfnamefont {D.}~\bibnamefont
  {Chowdhury}}, \bibinfo {author} {\bibfnamefont {J.}~\bibnamefont {Martin}},
  \bibinfo {author} {\bibfnamefont {C.}~\bibnamefont {Ringeval}}, \ and\
  \bibinfo {author} {\bibfnamefont {V.}~\bibnamefont {Vennin}},\ }\href@noop {}
  {\  (\bibinfo {year} {2019})},\ \Eprint {http://arxiv.org/abs/1902.03951}
  {arXiv:1902.03951 [astro-ph.CO]} \BibitemShut {NoStop}%
\bibitem [{\citenamefont {Maartens}(2011)}]{Maartens:2011yx}%
  \BibitemOpen
  \bibfield  {author} {\bibinfo {author} {\bibfnamefont {R.}~\bibnamefont
  {Maartens}},\ }\href {\doibase 10.1098/rsta.2011.0289} {\bibfield  {journal}
  {\bibinfo  {journal} {Phil. Trans. Roy. Soc. Lond.}\ }\textbf {\bibinfo
  {volume} {A369}},\ \bibinfo {pages} {5115} (\bibinfo {year} {2011})},\
  \Eprint {http://arxiv.org/abs/1104.1300} {arXiv:1104.1300 [astro-ph.CO]}
  \BibitemShut {NoStop}%
\bibitem [{\citenamefont {Battefeld}\ and\ \citenamefont
  {Peter}(2015)}]{Battefeld:2014uga}%
  \BibitemOpen
  \bibfield  {author} {\bibinfo {author} {\bibfnamefont {D.}~\bibnamefont
  {Battefeld}}\ and\ \bibinfo {author} {\bibfnamefont {P.}~\bibnamefont
  {Peter}},\ }\href {\doibase 10.1016/j.physrep.2014.12.004} {\bibfield
  {journal} {\bibinfo  {journal} {Phys. Rept.}\ }\textbf {\bibinfo {volume}
  {571}},\ \bibinfo {pages} {1} (\bibinfo {year} {2015})},\ \Eprint
  {http://arxiv.org/abs/1406.2790} {arXiv:1406.2790 [astro-ph.CO]} \BibitemShut
  {NoStop}%
\bibitem [{\citenamefont {Brandenberger}\ and\ \citenamefont
  {Peter}(2017)}]{Brandenberger:2016vhg}%
  \BibitemOpen
  \bibfield  {author} {\bibinfo {author} {\bibfnamefont {R.}~\bibnamefont
  {Brandenberger}}\ and\ \bibinfo {author} {\bibfnamefont {P.}~\bibnamefont
  {Peter}},\ }\href {\doibase 10.1007/s10701-016-0057-0} {\bibfield  {journal}
  {\bibinfo  {journal} {Found. Phys.}\ }\textbf {\bibinfo {volume} {47}},\
  \bibinfo {pages} {797} (\bibinfo {year} {2017})},\ \Eprint
  {http://arxiv.org/abs/1603.05834} {arXiv:1603.05834 [hep-th]} \BibitemShut
  {NoStop}%
\bibitem [{\citenamefont {Hartle}\ and\ \citenamefont
  {Hawking}(1983)}]{Hartle:1983ai}%
  \BibitemOpen
  \bibfield  {author} {\bibinfo {author} {\bibfnamefont {J.~B.}\ \bibnamefont
  {Hartle}}\ and\ \bibinfo {author} {\bibfnamefont {S.~W.}\ \bibnamefont
  {Hawking}},\ }\href {\doibase 10.1103/PhysRevD.28.2960} {\bibfield  {journal}
  {\bibinfo  {journal} {Phys. Rev.}\ }\textbf {\bibinfo {volume} {D28}},\
  \bibinfo {pages} {2960} (\bibinfo {year} {1983})}\BibitemShut {NoStop}%
\bibitem [{\citenamefont {Perlmutter}\ \emph {et~al.}(1997)\citenamefont
  {Perlmutter} \emph {et~al.}}]{Perlmutter:1996ds}%
  \BibitemOpen
  \bibfield  {author} {\bibinfo {author} {\bibfnamefont {S.}~\bibnamefont
  {Perlmutter}} \emph {et~al.} (\bibinfo {collaboration} {Supernova Cosmology
  Project}),\ }\href {\doibase 10.1086/304265} {\bibfield  {journal} {\bibinfo
  {journal} {Astrophys. J.}\ }\textbf {\bibinfo {volume} {483}},\ \bibinfo
  {pages} {565} (\bibinfo {year} {1997})},\ \Eprint
  {http://arxiv.org/abs/astro-ph/9608192} {arXiv:astro-ph/9608192 [astro-ph]}
  \BibitemShut {NoStop}%
\bibitem [{\citenamefont {Perlmutter}\ \emph {et~al.}(1998)\citenamefont
  {Perlmutter} \emph {et~al.}}]{Perlmutter:1997zf}%
  \BibitemOpen
  \bibfield  {author} {\bibinfo {author} {\bibfnamefont {S.}~\bibnamefont
  {Perlmutter}} \emph {et~al.} (\bibinfo {collaboration} {Supernova Cosmology
  Project}),\ }\href {\doibase 10.1038/34124} {\bibfield  {journal} {\bibinfo
  {journal} {Nature}\ }\textbf {\bibinfo {volume} {391}},\ \bibinfo {pages}
  {51} (\bibinfo {year} {1998})},\ \Eprint
  {http://arxiv.org/abs/astro-ph/9712212} {arXiv:astro-ph/9712212 [astro-ph]}
  \BibitemShut {NoStop}%
\bibitem [{\citenamefont {Kowalski}\ \emph {et~al.}(2008)\citenamefont
  {Kowalski} \emph {et~al.}}]{Kowalski:2008ez}%
  \BibitemOpen
  \bibfield  {author} {\bibinfo {author} {\bibfnamefont {M.}~\bibnamefont
  {Kowalski}} \emph {et~al.} (\bibinfo {collaboration} {Supernova Cosmology
  Project}),\ }\href {\doibase 10.1086/589937} {\bibfield  {journal} {\bibinfo
  {journal} {Astrophys. J.}\ }\textbf {\bibinfo {volume} {686}},\ \bibinfo
  {pages} {749} (\bibinfo {year} {2008})},\ \Eprint
  {http://arxiv.org/abs/0804.4142} {arXiv:0804.4142 [astro-ph]} \BibitemShut
  {NoStop}%
\bibitem [{\citenamefont {Astier}\ \emph {et~al.}(2006)\citenamefont {Astier}
  \emph {et~al.}}]{Astier:2005qq}%
  \BibitemOpen
  \bibfield  {author} {\bibinfo {author} {\bibfnamefont {P.}~\bibnamefont
  {Astier}} \emph {et~al.} (\bibinfo {collaboration} {SNLS}),\ }\href {\doibase
  10.1051/0004-6361:20054185} {\bibfield  {journal} {\bibinfo  {journal}
  {Astron. Astrophys.}\ }\textbf {\bibinfo {volume} {447}},\ \bibinfo {pages}
  {31} (\bibinfo {year} {2006})},\ \Eprint
  {http://arxiv.org/abs/astro-ph/0510447} {arXiv:astro-ph/0510447 [astro-ph]}
  \BibitemShut {NoStop}%
\bibitem [{\citenamefont {{Fixsen}}(2009)}]{2009ApJ...707..916F}%
  \BibitemOpen
  \bibfield  {author} {\bibinfo {author} {\bibfnamefont {D.~J.}\ \bibnamefont
  {{Fixsen}}},\ }\href {\doibase 10.1088/0004-637X/707/2/916} {\bibfield
  {journal} {\bibinfo  {journal} {\apj}\ }\textbf {\bibinfo {volume} {707}},\
  \bibinfo {pages} {916} (\bibinfo {year} {2009})},\ \Eprint
  {http://arxiv.org/abs/0911.1955} {arXiv:0911.1955} \BibitemShut {NoStop}%
\bibitem [{\citenamefont {{Gunn}}\ and\ \citenamefont
  {{Peterson}}(1965)}]{1965ApJ...142.1633G}%
  \BibitemOpen
  \bibfield  {author} {\bibinfo {author} {\bibfnamefont {J.~E.}\ \bibnamefont
  {{Gunn}}}\ and\ \bibinfo {author} {\bibfnamefont {B.~A.}\ \bibnamefont
  {{Peterson}}},\ }\href {\doibase 10.1086/148444} {\bibfield  {journal}
  {\bibinfo  {journal} {\apj}\ }\textbf {\bibinfo {volume} {142}},\ \bibinfo
  {pages} {1633} (\bibinfo {year} {1965})}\BibitemShut {NoStop}%
\bibitem [{\citenamefont {{Cyburt}}\ \emph {et~al.}(2016)\citenamefont
  {{Cyburt}}, \citenamefont {{Fields}}, \citenamefont {{Olive}},\ and\
  \citenamefont {{Yeh}}}]{2016RvMP...88a5004C}%
  \BibitemOpen
  \bibfield  {author} {\bibinfo {author} {\bibfnamefont {R.~H.}\ \bibnamefont
  {{Cyburt}}}, \bibinfo {author} {\bibfnamefont {B.~D.}\ \bibnamefont
  {{Fields}}}, \bibinfo {author} {\bibfnamefont {K.~A.}\ \bibnamefont
  {{Olive}}}, \ and\ \bibinfo {author} {\bibfnamefont {T.-H.}\ \bibnamefont
  {{Yeh}}},\ }\href {\doibase 10.1103/RevModPhys.88.015004} {\bibfield
  {journal} {\bibinfo  {journal} {Reviews of Modern Physics}\ }\textbf
  {\bibinfo {volume} {88}},\ \bibinfo {eid} {015004} (\bibinfo {year}
  {2016})},\ \Eprint {http://arxiv.org/abs/1505.01076} {arXiv:1505.01076}
  \BibitemShut {NoStop}%
\bibitem [{\citenamefont {{Peebles}}(1968)}]{1968ApJ...153....1P}%
  \BibitemOpen
  \bibfield  {author} {\bibinfo {author} {\bibfnamefont {P.~J.~E.}\
  \bibnamefont {{Peebles}}},\ }\href {\doibase 10.1086/149628} {\bibfield
  {journal} {\bibinfo  {journal} {\apj}\ }\textbf {\bibinfo {volume} {153}},\
  \bibinfo {pages} {1} (\bibinfo {year} {1968})}\BibitemShut {NoStop}%
\bibitem [{\citenamefont {Kolb}\ and\ \citenamefont
  {Turner}(1990)}]{Kolb:1990vq}%
  \BibitemOpen
  \bibfield  {author} {\bibinfo {author} {\bibfnamefont {E.~W.}\ \bibnamefont
  {Kolb}}\ and\ \bibinfo {author} {\bibfnamefont {M.~S.}\ \bibnamefont
  {Turner}},\ }\href@noop {} {\bibfield  {journal} {\bibinfo  {journal} {Front.
  Phys.}\ }\textbf {\bibinfo {volume} {69}},\ \bibinfo {pages} {1} (\bibinfo
  {year} {1990})}\BibitemShut {NoStop}%
\bibitem [{\citenamefont {{Lynch}}(2014)}]{2014arXiv1406.1125L}%
  \BibitemOpen
  \bibfield  {author} {\bibinfo {author} {\bibfnamefont {P.}~\bibnamefont
  {{Lynch}}},\ }\href@noop {} {\bibfield  {journal} {\bibinfo  {journal} {arXiv
  e-prints}\ } (\bibinfo {year} {2014})},\ \Eprint
  {http://arxiv.org/abs/1406.1125} {arXiv:1406.1125 [nlin.SI]} \BibitemShut
  {NoStop}%
\bibitem [{\citenamefont {Chialva}\ and\ \citenamefont
  {Mazumdar}(2015)}]{Chialva:2014rla}%
  \BibitemOpen
  \bibfield  {author} {\bibinfo {author} {\bibfnamefont {D.}~\bibnamefont
  {Chialva}}\ and\ \bibinfo {author} {\bibfnamefont {A.}~\bibnamefont
  {Mazumdar}},\ }\href {\doibase 10.1142/S0217732315400088} {\bibfield
  {journal} {\bibinfo  {journal} {Mod. Phys. Lett.}\ }\textbf {\bibinfo
  {volume} {A30}},\ \bibinfo {pages} {1540008} (\bibinfo {year} {2015})},\
  \Eprint {http://arxiv.org/abs/1405.0513} {arXiv:1405.0513 [hep-th]}
  \BibitemShut {NoStop}%
\bibitem [{\citenamefont {Schwarz}\ \emph {et~al.}(2001)\citenamefont
  {Schwarz}, \citenamefont {Terrero-Escalante},\ and\ \citenamefont
  {Garcia}}]{Schwarz:2001vv}%
  \BibitemOpen
  \bibfield  {author} {\bibinfo {author} {\bibfnamefont {D.~J.}\ \bibnamefont
  {Schwarz}}, \bibinfo {author} {\bibfnamefont {C.~A.}\ \bibnamefont
  {Terrero-Escalante}}, \ and\ \bibinfo {author} {\bibfnamefont {A.~A.}\
  \bibnamefont {Garcia}},\ }\href {\doibase 10.1016/S0370-2693(01)01036-X}
  {\bibfield  {journal} {\bibinfo  {journal} {Phys. Lett.}\ }\textbf {\bibinfo
  {volume} {B517}},\ \bibinfo {pages} {243} (\bibinfo {year} {2001})},\ \Eprint
  {http://arxiv.org/abs/astro-ph/0106020} {arXiv:astro-ph/0106020 [astro-ph]}
  \BibitemShut {NoStop}%
\bibitem [{\citenamefont {Leach}\ \emph {et~al.}(2002)\citenamefont {Leach},
  \citenamefont {Liddle}, \citenamefont {Martin},\ and\ \citenamefont
  {Schwarz}}]{Leach:2002ar}%
  \BibitemOpen
  \bibfield  {author} {\bibinfo {author} {\bibfnamefont {S.~M.}\ \bibnamefont
  {Leach}}, \bibinfo {author} {\bibfnamefont {A.~R.}\ \bibnamefont {Liddle}},
  \bibinfo {author} {\bibfnamefont {J.}~\bibnamefont {Martin}}, \ and\ \bibinfo
  {author} {\bibfnamefont {D.~J.}\ \bibnamefont {Schwarz}},\ }\href {\doibase
  10.1103/PhysRevD.66.023515} {\bibfield  {journal} {\bibinfo  {journal} {Phys.
  Rev.}\ }\textbf {\bibinfo {volume} {D66}},\ \bibinfo {pages} {023515}
  (\bibinfo {year} {2002})},\ \Eprint {http://arxiv.org/abs/astro-ph/0202094}
  {arXiv:astro-ph/0202094 [astro-ph]} \BibitemShut {NoStop}%
\bibitem [{\citenamefont {Liddle}\ \emph {et~al.}(1994)\citenamefont {Liddle},
  \citenamefont {Parsons},\ and\ \citenamefont {Barrow}}]{Liddle:1994dx}%
  \BibitemOpen
  \bibfield  {author} {\bibinfo {author} {\bibfnamefont {A.~R.}\ \bibnamefont
  {Liddle}}, \bibinfo {author} {\bibfnamefont {P.}~\bibnamefont {Parsons}}, \
  and\ \bibinfo {author} {\bibfnamefont {J.~D.}\ \bibnamefont {Barrow}},\
  }\href {\doibase 10.1103/PhysRevD.50.7222} {\bibfield  {journal} {\bibinfo
  {journal} {Phys. Rev.}\ }\textbf {\bibinfo {volume} {D50}},\ \bibinfo {pages}
  {7222} (\bibinfo {year} {1994})},\ \Eprint
  {http://arxiv.org/abs/astro-ph/9408015} {arXiv:astro-ph/9408015 [astro-ph]}
  \BibitemShut {NoStop}%
\bibitem [{\citenamefont {Martin}\ and\ \citenamefont
  {Ringeval}(2006)}]{Martin:2006rs}%
  \BibitemOpen
  \bibfield  {author} {\bibinfo {author} {\bibfnamefont {J.}~\bibnamefont
  {Martin}}\ and\ \bibinfo {author} {\bibfnamefont {C.}~\bibnamefont
  {Ringeval}},\ }\href {\doibase 10.1088/1475-7516/2006/08/009} {\bibfield
  {journal} {\bibinfo  {journal} {JCAP}\ }\textbf {\bibinfo {volume} {0608}},\
  \bibinfo {pages} {009} (\bibinfo {year} {2006})},\ \Eprint
  {http://arxiv.org/abs/astro-ph/0605367} {arXiv:astro-ph/0605367 [astro-ph]}
  \BibitemShut {NoStop}%
\bibitem [{\citenamefont {Martin}\ and\ \citenamefont
  {Ringeval}(2010)}]{Martin:2010kz}%
  \BibitemOpen
  \bibfield  {author} {\bibinfo {author} {\bibfnamefont {J.}~\bibnamefont
  {Martin}}\ and\ \bibinfo {author} {\bibfnamefont {C.}~\bibnamefont
  {Ringeval}},\ }\href {\doibase 10.1103/PhysRevD.82.023511} {\bibfield
  {journal} {\bibinfo  {journal} {Phys. Rev.}\ }\textbf {\bibinfo {volume}
  {D82}},\ \bibinfo {pages} {023511} (\bibinfo {year} {2010})},\ \Eprint
  {http://arxiv.org/abs/1004.5525} {arXiv:1004.5525 [astro-ph.CO]} \BibitemShut
  {NoStop}%
\bibitem [{\citenamefont {Martin}\ \emph {et~al.}(2015)\citenamefont {Martin},
  \citenamefont {Ringeval},\ and\ \citenamefont {Vennin}}]{Martin:2014nya}%
  \BibitemOpen
  \bibfield  {author} {\bibinfo {author} {\bibfnamefont {J.}~\bibnamefont
  {Martin}}, \bibinfo {author} {\bibfnamefont {C.}~\bibnamefont {Ringeval}}, \
  and\ \bibinfo {author} {\bibfnamefont {V.}~\bibnamefont {Vennin}},\ }\href
  {\doibase 10.1103/PhysRevLett.114.081303} {\bibfield  {journal} {\bibinfo
  {journal} {Phys. Rev. Lett.}\ }\textbf {\bibinfo {volume} {114}},\ \bibinfo
  {pages} {081303} (\bibinfo {year} {2015})},\ \Eprint
  {http://arxiv.org/abs/1410.7958} {arXiv:1410.7958 [astro-ph.CO]} \BibitemShut
  {NoStop}%
\bibitem [{\citenamefont {Martin}(2016)}]{Martin:2015dha}%
  \BibitemOpen
  \bibfield  {author} {\bibinfo {author} {\bibfnamefont {J.}~\bibnamefont
  {Martin}},\ }\bibfield  {booktitle} {\emph {\bibinfo {booktitle} {{The Cosmic
  Microwave Background}}},\ }\href {\doibase 10.1007/978-3-319-44769-8_2}
  {\bibfield  {journal} {\bibinfo  {journal} {Astrophys. Space Sci. Proc.}\
  }\textbf {\bibinfo {volume} {45}},\ \bibinfo {pages} {41} (\bibinfo {year}
  {2016})},\ \Eprint {http://arxiv.org/abs/1502.05733} {arXiv:1502.05733
  [astro-ph.CO]} \BibitemShut {NoStop}%
\bibitem [{\citenamefont {Martin}\ \emph {et~al.}(2016)\citenamefont {Martin},
  \citenamefont {Ringeval},\ and\ \citenamefont {Vennin}}]{Martin:2016oyk}%
  \BibitemOpen
  \bibfield  {author} {\bibinfo {author} {\bibfnamefont {J.}~\bibnamefont
  {Martin}}, \bibinfo {author} {\bibfnamefont {C.}~\bibnamefont {Ringeval}}, \
  and\ \bibinfo {author} {\bibfnamefont {V.}~\bibnamefont {Vennin}},\ }\href
  {\doibase 10.1103/PhysRevD.93.103532} {\bibfield  {journal} {\bibinfo
  {journal} {Phys. Rev.}\ }\textbf {\bibinfo {volume} {D93}},\ \bibinfo {pages}
  {103532} (\bibinfo {year} {2016})},\ \Eprint
  {http://arxiv.org/abs/1603.02606} {arXiv:1603.02606 [astro-ph.CO]}
  \BibitemShut {NoStop}%
\bibitem [{\citenamefont {Mukhanov}\ \emph {et~al.}(1992)\citenamefont
  {Mukhanov}, \citenamefont {Feldman},\ and\ \citenamefont
  {Brandenberger}}]{Mukhanov:1990me}%
  \BibitemOpen
  \bibfield  {author} {\bibinfo {author} {\bibfnamefont {V.~F.}\ \bibnamefont
  {Mukhanov}}, \bibinfo {author} {\bibfnamefont {H.}~\bibnamefont {Feldman}}, \
  and\ \bibinfo {author} {\bibfnamefont {R.~H.}\ \bibnamefont
  {Brandenberger}},\ }\href {\doibase 10.1016/0370-1573(92)90044-Z} {\bibfield
  {journal} {\bibinfo  {journal} {Phys. Rept.}\ }\textbf {\bibinfo {volume}
  {215}},\ \bibinfo {pages} {203} (\bibinfo {year} {1992})}\BibitemShut
  {NoStop}%
\bibitem [{\citenamefont {Martin}(2008)}]{Martin:2007bw}%
  \BibitemOpen
  \bibfield  {author} {\bibinfo {author} {\bibfnamefont {J.}~\bibnamefont
  {Martin}},\ }\href {\doibase 10.1007/978-3-540-74353-8_6} {\bibfield
  {journal} {\bibinfo  {journal} {Lect. Notes Phys.}\ }\textbf {\bibinfo
  {volume} {738}},\ \bibinfo {pages} {193} (\bibinfo {year} {2008})},\ \Eprint
  {http://arxiv.org/abs/0704.3540} {arXiv:0704.3540 [hep-th]} \BibitemShut
  {NoStop}%
\bibitem [{\citenamefont {Martin}(2005)}]{Martin:2004um}%
  \BibitemOpen
  \bibfield  {author} {\bibinfo {author} {\bibfnamefont {J.}~\bibnamefont
  {Martin}},\ }\bibfield  {booktitle} {\emph {\bibinfo {booktitle} {{Planck
  scale effects in astrophysics and cosmology. Proceedings, 40th Karpacs Winter
  School, Ladek Zdroj, Poland, February 4-14, 2004}}},\ }\href {\doibase
  10.1007/11377306_7} {\bibfield  {journal} {\bibinfo  {journal} {Lect. Notes
  Phys.}\ }\textbf {\bibinfo {volume} {669}},\ \bibinfo {pages} {199} (\bibinfo
  {year} {2005})},\ \bibinfo {note} {[,199(2004)]},\ \Eprint
  {http://arxiv.org/abs/hep-th/0406011} {arXiv:hep-th/0406011 [hep-th]}
  \BibitemShut {NoStop}%
\bibitem [{\citenamefont {Martin}(2004)}]{Martin:2003bt}%
  \BibitemOpen
  \bibfield  {author} {\bibinfo {author} {\bibfnamefont {J.}~\bibnamefont
  {Martin}},\ }\bibfield  {booktitle} {\emph {\bibinfo {booktitle} {{Particles
  and fields. Proceedings, 24th National Meeting, ENFPC 24, Caxambu, Brazil,
  September 30-October 4, 2003}}},\ }\href {\doibase
  10.1590/S0103-97332004000700005} {\bibfield  {journal} {\bibinfo  {journal}
  {Braz. J. Phys.}\ }\textbf {\bibinfo {volume} {34}},\ \bibinfo {pages} {1307}
  (\bibinfo {year} {2004})},\ \Eprint {http://arxiv.org/abs/astro-ph/0312492}
  {arXiv:astro-ph/0312492 [astro-ph]} \BibitemShut {NoStop}%
\bibitem [{\citenamefont {Bardeen}\ \emph {et~al.}(1983)\citenamefont
  {Bardeen}, \citenamefont {Steinhardt},\ and\ \citenamefont
  {Turner}}]{Bardeen:1983qw}%
  \BibitemOpen
  \bibfield  {author} {\bibinfo {author} {\bibfnamefont {J.~M.}\ \bibnamefont
  {Bardeen}}, \bibinfo {author} {\bibfnamefont {P.~J.}\ \bibnamefont
  {Steinhardt}}, \ and\ \bibinfo {author} {\bibfnamefont {M.~S.}\ \bibnamefont
  {Turner}},\ }\href {\doibase 10.1103/PhysRevD.28.679} {\bibfield  {journal}
  {\bibinfo  {journal} {Phys. Rev.}\ }\textbf {\bibinfo {volume} {D28}},\
  \bibinfo {pages} {679} (\bibinfo {year} {1983})}\BibitemShut {NoStop}%
\bibitem [{\citenamefont {Martin}\ and\ \citenamefont
  {Schwarz}(1998)}]{Martin:1997zd}%
  \BibitemOpen
  \bibfield  {author} {\bibinfo {author} {\bibfnamefont {J.}~\bibnamefont
  {Martin}}\ and\ \bibinfo {author} {\bibfnamefont {D.~J.}\ \bibnamefont
  {Schwarz}},\ }\href {\doibase 10.1103/PhysRevD.57.3302} {\bibfield  {journal}
  {\bibinfo  {journal} {Phys. Rev.}\ }\textbf {\bibinfo {volume} {D57}},\
  \bibinfo {pages} {3302} (\bibinfo {year} {1998})},\ \Eprint
  {http://arxiv.org/abs/gr-qc/9704049} {arXiv:gr-qc/9704049 [gr-qc]}
  \BibitemShut {NoStop}%
\bibitem [{\citenamefont {Bardeen}(1980)}]{Bardeen:1980kt}%
  \BibitemOpen
  \bibfield  {author} {\bibinfo {author} {\bibfnamefont {J.~M.}\ \bibnamefont
  {Bardeen}},\ }\href {\doibase 10.1103/PhysRevD.22.1882} {\bibfield  {journal}
  {\bibinfo  {journal} {Phys. Rev.}\ }\textbf {\bibinfo {volume} {D22}},\
  \bibinfo {pages} {1882} (\bibinfo {year} {1980})}\BibitemShut {NoStop}%
\bibitem [{\citenamefont {Casadio}\ \emph
  {et~al.}(2005{\natexlab{a}})\citenamefont {Casadio}, \citenamefont {Finelli},
  \citenamefont {Luzzi},\ and\ \citenamefont {Venturi}}]{Casadio:2004ru}%
  \BibitemOpen
  \bibfield  {author} {\bibinfo {author} {\bibfnamefont {R.}~\bibnamefont
  {Casadio}}, \bibinfo {author} {\bibfnamefont {F.}~\bibnamefont {Finelli}},
  \bibinfo {author} {\bibfnamefont {M.}~\bibnamefont {Luzzi}}, \ and\ \bibinfo
  {author} {\bibfnamefont {G.}~\bibnamefont {Venturi}},\ }\href {\doibase
  10.1103/PhysRevD.71.043517} {\bibfield  {journal} {\bibinfo  {journal} {Phys.
  Rev.}\ }\textbf {\bibinfo {volume} {D71}},\ \bibinfo {pages} {043517}
  (\bibinfo {year} {2005}{\natexlab{a}})},\ \Eprint
  {http://arxiv.org/abs/gr-qc/0410092} {arXiv:gr-qc/0410092 [gr-qc]}
  \BibitemShut {NoStop}%
\bibitem [{\citenamefont {Casadio}\ \emph
  {et~al.}(2005{\natexlab{b}})\citenamefont {Casadio}, \citenamefont {Finelli},
  \citenamefont {Luzzi},\ and\ \citenamefont {Venturi}}]{Casadio:2005xv}%
  \BibitemOpen
  \bibfield  {author} {\bibinfo {author} {\bibfnamefont {R.}~\bibnamefont
  {Casadio}}, \bibinfo {author} {\bibfnamefont {F.}~\bibnamefont {Finelli}},
  \bibinfo {author} {\bibfnamefont {M.}~\bibnamefont {Luzzi}}, \ and\ \bibinfo
  {author} {\bibfnamefont {G.}~\bibnamefont {Venturi}},\ }\href {\doibase
  10.1016/j.physletb.2005.08.056} {\bibfield  {journal} {\bibinfo  {journal}
  {Phys. Lett.}\ }\textbf {\bibinfo {volume} {B625}},\ \bibinfo {pages} {1}
  (\bibinfo {year} {2005}{\natexlab{b}})},\ \Eprint
  {http://arxiv.org/abs/gr-qc/0506043} {arXiv:gr-qc/0506043 [gr-qc]}
  \BibitemShut {NoStop}%
\bibitem [{\citenamefont {Casadio}\ \emph
  {et~al.}(2005{\natexlab{c}})\citenamefont {Casadio}, \citenamefont {Finelli},
  \citenamefont {Luzzi},\ and\ \citenamefont {Venturi}}]{Casadio:2005em}%
  \BibitemOpen
  \bibfield  {author} {\bibinfo {author} {\bibfnamefont {R.}~\bibnamefont
  {Casadio}}, \bibinfo {author} {\bibfnamefont {F.}~\bibnamefont {Finelli}},
  \bibinfo {author} {\bibfnamefont {M.}~\bibnamefont {Luzzi}}, \ and\ \bibinfo
  {author} {\bibfnamefont {G.}~\bibnamefont {Venturi}},\ }\href {\doibase
  10.1103/PhysRevD.72.103516} {\bibfield  {journal} {\bibinfo  {journal} {Phys.
  Rev.}\ }\textbf {\bibinfo {volume} {D72}},\ \bibinfo {pages} {103516}
  (\bibinfo {year} {2005}{\natexlab{c}})},\ \Eprint
  {http://arxiv.org/abs/gr-qc/0510103} {arXiv:gr-qc/0510103 [gr-qc]}
  \BibitemShut {NoStop}%
\bibitem [{\citenamefont {Gong}\ and\ \citenamefont
  {Stewart}(2001)}]{Gong:2001he}%
  \BibitemOpen
  \bibfield  {author} {\bibinfo {author} {\bibfnamefont {J.-O.}\ \bibnamefont
  {Gong}}\ and\ \bibinfo {author} {\bibfnamefont {E.~D.}\ \bibnamefont
  {Stewart}},\ }\href {\doibase 10.1016/S0370-2693(01)00616-5} {\bibfield
  {journal} {\bibinfo  {journal} {Phys. Lett.}\ }\textbf {\bibinfo {volume}
  {B510}},\ \bibinfo {pages} {1} (\bibinfo {year} {2001})},\ \Eprint
  {http://arxiv.org/abs/astro-ph/0101225} {arXiv:astro-ph/0101225 [astro-ph]}
  \BibitemShut {NoStop}%
\bibitem [{\citenamefont {Choe}\ \emph {et~al.}(2004)\citenamefont {Choe},
  \citenamefont {Gong},\ and\ \citenamefont {Stewart}}]{Choe:2004zg}%
  \BibitemOpen
  \bibfield  {author} {\bibinfo {author} {\bibfnamefont {J.}~\bibnamefont
  {Choe}}, \bibinfo {author} {\bibfnamefont {J.-O.}\ \bibnamefont {Gong}}, \
  and\ \bibinfo {author} {\bibfnamefont {E.~D.}\ \bibnamefont {Stewart}},\
  }\href {\doibase 10.1088/1475-7516/2004/07/012} {\bibfield  {journal}
  {\bibinfo  {journal} {JCAP}\ }\textbf {\bibinfo {volume} {0407}},\ \bibinfo
  {pages} {012} (\bibinfo {year} {2004})},\ \Eprint
  {http://arxiv.org/abs/hep-ph/0405155} {arXiv:hep-ph/0405155 [hep-ph]}
  \BibitemShut {NoStop}%
\bibitem [{\citenamefont {Lorenz}\ \emph
  {et~al.}(2008{\natexlab{a}})\citenamefont {Lorenz}, \citenamefont {Martin},\
  and\ \citenamefont {Ringeval}}]{Lorenz:2008et}%
  \BibitemOpen
  \bibfield  {author} {\bibinfo {author} {\bibfnamefont {L.}~\bibnamefont
  {Lorenz}}, \bibinfo {author} {\bibfnamefont {J.}~\bibnamefont {Martin}}, \
  and\ \bibinfo {author} {\bibfnamefont {C.}~\bibnamefont {Ringeval}},\ }\href
  {\doibase 10.1103/PhysRevD.78.083513} {\bibfield  {journal} {\bibinfo
  {journal} {Phys. Rev.}\ }\textbf {\bibinfo {volume} {D78}},\ \bibinfo {pages}
  {083513} (\bibinfo {year} {2008}{\natexlab{a}})},\ \Eprint
  {http://arxiv.org/abs/0807.3037} {arXiv:0807.3037 [astro-ph]} \BibitemShut
  {NoStop}%
\bibitem [{\citenamefont {Martin}\ \emph {et~al.}(2013)\citenamefont {Martin},
  \citenamefont {Ringeval},\ and\ \citenamefont {Vennin}}]{Martin:2013uma}%
  \BibitemOpen
  \bibfield  {author} {\bibinfo {author} {\bibfnamefont {J.}~\bibnamefont
  {Martin}}, \bibinfo {author} {\bibfnamefont {C.}~\bibnamefont {Ringeval}}, \
  and\ \bibinfo {author} {\bibfnamefont {V.}~\bibnamefont {Vennin}},\ }\href
  {\doibase 10.1088/1475-7516/2013/06/021} {\bibfield  {journal} {\bibinfo
  {journal} {JCAP}\ }\textbf {\bibinfo {volume} {1306}},\ \bibinfo {pages}
  {021} (\bibinfo {year} {2013})},\ \Eprint {http://arxiv.org/abs/1303.2120}
  {arXiv:1303.2120 [astro-ph.CO]} \BibitemShut {NoStop}%
\bibitem [{\citenamefont {Kosowsky}(1996)}]{Kosowsky:1994cy}%
  \BibitemOpen
  \bibfield  {author} {\bibinfo {author} {\bibfnamefont {A.}~\bibnamefont
  {Kosowsky}},\ }\href {\doibase 10.1006/aphy.1996.0020} {\bibfield  {journal}
  {\bibinfo  {journal} {Annals Phys.}\ }\textbf {\bibinfo {volume} {246}},\
  \bibinfo {pages} {49} (\bibinfo {year} {1996})},\ \Eprint
  {http://arxiv.org/abs/astro-ph/9501045} {arXiv:astro-ph/9501045 [astro-ph]}
  \BibitemShut {NoStop}%
\bibitem [{\citenamefont {Martin}\ \emph
  {et~al.}(2014{\natexlab{b}})\citenamefont {Martin}, \citenamefont {Ringeval},
  \citenamefont {Trotta},\ and\ \citenamefont {Vennin}}]{Martin:2014lra}%
  \BibitemOpen
  \bibfield  {author} {\bibinfo {author} {\bibfnamefont {J.}~\bibnamefont
  {Martin}}, \bibinfo {author} {\bibfnamefont {C.}~\bibnamefont {Ringeval}},
  \bibinfo {author} {\bibfnamefont {R.}~\bibnamefont {Trotta}}, \ and\ \bibinfo
  {author} {\bibfnamefont {V.}~\bibnamefont {Vennin}},\ }\href {\doibase
  10.1103/PhysRevD.90.063501} {\bibfield  {journal} {\bibinfo  {journal} {Phys.
  Rev.}\ }\textbf {\bibinfo {volume} {D90}},\ \bibinfo {pages} {063501}
  (\bibinfo {year} {2014}{\natexlab{b}})},\ \Eprint
  {http://arxiv.org/abs/1405.7272} {arXiv:1405.7272 [astro-ph.CO]} \BibitemShut
  {NoStop}%
\bibitem [{\citenamefont {Martin}\ \emph
  {et~al.}(2014{\natexlab{c}})\citenamefont {Martin}, \citenamefont {Ringeval},
  \citenamefont {Trotta},\ and\ \citenamefont {Vennin}}]{Martin:2013nzq}%
  \BibitemOpen
  \bibfield  {author} {\bibinfo {author} {\bibfnamefont {J.}~\bibnamefont
  {Martin}}, \bibinfo {author} {\bibfnamefont {C.}~\bibnamefont {Ringeval}},
  \bibinfo {author} {\bibfnamefont {R.}~\bibnamefont {Trotta}}, \ and\ \bibinfo
  {author} {\bibfnamefont {V.}~\bibnamefont {Vennin}},\ }\href {\doibase
  10.1088/1475-7516/2014/03/039} {\bibfield  {journal} {\bibinfo  {journal}
  {JCAP}\ }\textbf {\bibinfo {volume} {1403}},\ \bibinfo {pages} {039}
  (\bibinfo {year} {2014}{\natexlab{c}})},\ \Eprint
  {http://arxiv.org/abs/1312.3529} {arXiv:1312.3529 [astro-ph.CO]} \BibitemShut
  {NoStop}%
\bibitem [{\citenamefont {Trotta}(2008)}]{Trotta:2008qt}%
  \BibitemOpen
  \bibfield  {author} {\bibinfo {author} {\bibfnamefont {R.}~\bibnamefont
  {Trotta}},\ }\href {\doibase 10.1080/00107510802066753} {\bibfield  {journal}
  {\bibinfo  {journal} {Contemp. Phys.}\ }\textbf {\bibinfo {volume} {49}},\
  \bibinfo {pages} {71} (\bibinfo {year} {2008})},\ \Eprint
  {http://arxiv.org/abs/0803.4089} {arXiv:0803.4089 [astro-ph]} \BibitemShut
  {NoStop}%
\bibitem [{\citenamefont {Trotta}(2017)}]{Trotta:2017wnx}%
  \BibitemOpen
  \bibfield  {author} {\bibinfo {author} {\bibfnamefont {R.}~\bibnamefont
  {Trotta}}\ }(\bibinfo {year} {2017})\ \Eprint
  {http://arxiv.org/abs/1701.01467} {arXiv:1701.01467 [astro-ph.CO]}
  \BibitemShut {NoStop}%
\bibitem [{\citenamefont {Kunz}\ \emph {et~al.}(2006)\citenamefont {Kunz},
  \citenamefont {Trotta},\ and\ \citenamefont {Parkinson}}]{Kunz:2006mc}%
  \BibitemOpen
  \bibfield  {author} {\bibinfo {author} {\bibfnamefont {M.}~\bibnamefont
  {Kunz}}, \bibinfo {author} {\bibfnamefont {R.}~\bibnamefont {Trotta}}, \ and\
  \bibinfo {author} {\bibfnamefont {D.}~\bibnamefont {Parkinson}},\ }\href
  {\doibase 10.1103/PhysRevD.74.023503} {\bibfield  {journal} {\bibinfo
  {journal} {Phys. Rev.}\ }\textbf {\bibinfo {volume} {D74}},\ \bibinfo {pages}
  {023503} (\bibinfo {year} {2006})},\ \Eprint
  {http://arxiv.org/abs/astro-ph/0602378} {arXiv:astro-ph/0602378 [astro-ph]}
  \BibitemShut {NoStop}%
\bibitem [{\citenamefont {Lorenz}\ \emph
  {et~al.}(2008{\natexlab{b}})\citenamefont {Lorenz}, \citenamefont {Martin},\
  and\ \citenamefont {Ringeval}}]{Lorenz:2007ze}%
  \BibitemOpen
  \bibfield  {author} {\bibinfo {author} {\bibfnamefont {L.}~\bibnamefont
  {Lorenz}}, \bibinfo {author} {\bibfnamefont {J.}~\bibnamefont {Martin}}, \
  and\ \bibinfo {author} {\bibfnamefont {C.}~\bibnamefont {Ringeval}},\ }\href
  {\doibase 10.1088/1475-7516/2008/04/001} {\bibfield  {journal} {\bibinfo
  {journal} {JCAP}\ }\textbf {\bibinfo {volume} {0804}},\ \bibinfo {pages}
  {001} (\bibinfo {year} {2008}{\natexlab{b}})},\ \Eprint
  {http://arxiv.org/abs/0709.3758} {arXiv:0709.3758 [hep-th]} \BibitemShut
  {NoStop}%
\bibitem [{\citenamefont {Lorenz}\ \emph
  {et~al.}(2008{\natexlab{c}})\citenamefont {Lorenz}, \citenamefont {Martin},\
  and\ \citenamefont {Ringeval}}]{Lorenz:2008je}%
  \BibitemOpen
  \bibfield  {author} {\bibinfo {author} {\bibfnamefont {L.}~\bibnamefont
  {Lorenz}}, \bibinfo {author} {\bibfnamefont {J.}~\bibnamefont {Martin}}, \
  and\ \bibinfo {author} {\bibfnamefont {C.}~\bibnamefont {Ringeval}},\ }\href
  {\doibase 10.1103/PhysRevD.78.063543} {\bibfield  {journal} {\bibinfo
  {journal} {Phys. Rev.}\ }\textbf {\bibinfo {volume} {D78}},\ \bibinfo {pages}
  {063543} (\bibinfo {year} {2008}{\natexlab{c}})},\ \Eprint
  {http://arxiv.org/abs/0807.2414} {arXiv:0807.2414 [astro-ph]} \BibitemShut
  {NoStop}%
\bibitem [{\citenamefont {Martin}\ \emph {et~al.}(2011)\citenamefont {Martin},
  \citenamefont {Ringeval},\ and\ \citenamefont {Trotta}}]{Martin:2010hh}%
  \BibitemOpen
  \bibfield  {author} {\bibinfo {author} {\bibfnamefont {J.}~\bibnamefont
  {Martin}}, \bibinfo {author} {\bibfnamefont {C.}~\bibnamefont {Ringeval}}, \
  and\ \bibinfo {author} {\bibfnamefont {R.}~\bibnamefont {Trotta}},\ }\href
  {\doibase 10.1103/PhysRevD.83.063524} {\bibfield  {journal} {\bibinfo
  {journal} {Phys. Rev.}\ }\textbf {\bibinfo {volume} {D83}},\ \bibinfo {pages}
  {063524} (\bibinfo {year} {2011})},\ \Eprint {http://arxiv.org/abs/1009.4157}
  {arXiv:1009.4157 [astro-ph.CO]} \BibitemShut {NoStop}%
\bibitem [{\citenamefont {Baumann}\ and\ \citenamefont
  {McAllister}(2015)}]{Baumann:2014nda}%
  \BibitemOpen
  \bibfield  {author} {\bibinfo {author} {\bibfnamefont {D.}~\bibnamefont
  {Baumann}}\ and\ \bibinfo {author} {\bibfnamefont {L.}~\bibnamefont
  {McAllister}},\ }\href
  {http://inspirehep.net/record/1289899/files/arXiv:1404.2601.pdf} {\emph
  {\bibinfo {title} {{Inflation and String Theory}}}}\ (\bibinfo  {publisher}
  {Cambridge University Press},\ \bibinfo {year} {2015})\ \Eprint
  {http://arxiv.org/abs/1404.2601} {arXiv:1404.2601 [hep-th]} \BibitemShut
  {NoStop}%
\bibitem [{\citenamefont {Grain}\ and\ \citenamefont
  {Vennin}(2017)}]{Grain:2017dqa}%
  \BibitemOpen
  \bibfield  {author} {\bibinfo {author} {\bibfnamefont {J.}~\bibnamefont
  {Grain}}\ and\ \bibinfo {author} {\bibfnamefont {V.}~\bibnamefont {Vennin}},\
  }\href {\doibase 10.1088/1475-7516/2017/05/045} {\bibfield  {journal}
  {\bibinfo  {journal} {JCAP}\ }\textbf {\bibinfo {volume} {1705}},\ \bibinfo
  {pages} {045} (\bibinfo {year} {2017})},\ \Eprint
  {http://arxiv.org/abs/1703.00447} {arXiv:1703.00447 [gr-qc]} \BibitemShut
  {NoStop}%
\bibitem [{\citenamefont {Steigman}\ and\ \citenamefont
  {Turner}(1983)}]{Steigman:1983hr}%
  \BibitemOpen
  \bibfield  {author} {\bibinfo {author} {\bibfnamefont {G.}~\bibnamefont
  {Steigman}}\ and\ \bibinfo {author} {\bibfnamefont {M.~S.}\ \bibnamefont
  {Turner}},\ }\href {\doibase 10.1016/0370-2693(83)90263-0} {\bibfield
  {journal} {\bibinfo  {journal} {Phys. Lett.}\ }\textbf {\bibinfo {volume}
  {128B}},\ \bibinfo {pages} {295} (\bibinfo {year} {1983})}\BibitemShut
  {NoStop}%
\bibitem [{\citenamefont {Anninos}\ \emph {et~al.}(1991)\citenamefont
  {Anninos}, \citenamefont {Matzner}, \citenamefont {Rothman},\ and\
  \citenamefont {Ryan}}]{Anninos:1991ma}%
  \BibitemOpen
  \bibfield  {author} {\bibinfo {author} {\bibfnamefont {P.}~\bibnamefont
  {Anninos}}, \bibinfo {author} {\bibfnamefont {R.~A.}\ \bibnamefont
  {Matzner}}, \bibinfo {author} {\bibfnamefont {T.}~\bibnamefont {Rothman}}, \
  and\ \bibinfo {author} {\bibfnamefont {M.~P.}\ \bibnamefont {Ryan}},\ }\href
  {\doibase 10.1103/PhysRevD.43.3821} {\bibfield  {journal} {\bibinfo
  {journal} {Phys. Rev.}\ }\textbf {\bibinfo {volume} {D43}},\ \bibinfo {pages}
  {3821} (\bibinfo {year} {1991})}\BibitemShut {NoStop}%
\bibitem [{\citenamefont {Turner}\ and\ \citenamefont
  {Widrow}(1986)}]{Turner:1986gj}%
  \BibitemOpen
  \bibfield  {author} {\bibinfo {author} {\bibfnamefont {M.~S.}\ \bibnamefont
  {Turner}}\ and\ \bibinfo {author} {\bibfnamefont {L.~M.}\ \bibnamefont
  {Widrow}},\ }\href {\doibase 10.1103/PhysRevLett.57.2237} {\bibfield
  {journal} {\bibinfo  {journal} {Phys. Rev. Lett.}\ }\textbf {\bibinfo
  {volume} {57}},\ \bibinfo {pages} {2237} (\bibinfo {year}
  {1986})}\BibitemShut {NoStop}%
\bibitem [{\citenamefont {Goldwirth}\ and\ \citenamefont
  {Piran}(1990)}]{Goldwirth:1989pr}%
  \BibitemOpen
  \bibfield  {author} {\bibinfo {author} {\bibfnamefont {D.~S.}\ \bibnamefont
  {Goldwirth}}\ and\ \bibinfo {author} {\bibfnamefont {T.}~\bibnamefont
  {Piran}},\ }\href {\doibase 10.1103/PhysRevLett.64.2852} {\bibfield
  {journal} {\bibinfo  {journal} {Phys. Rev. Lett.}\ }\textbf {\bibinfo
  {volume} {64}},\ \bibinfo {pages} {2852} (\bibinfo {year}
  {1990})}\BibitemShut {NoStop}%
\bibitem [{\citenamefont {Goldwirth}\ and\ \citenamefont
  {Piran}(1992)}]{Goldwirth:1991rj}%
  \BibitemOpen
  \bibfield  {author} {\bibinfo {author} {\bibfnamefont {D.~S.}\ \bibnamefont
  {Goldwirth}}\ and\ \bibinfo {author} {\bibfnamefont {T.}~\bibnamefont
  {Piran}},\ }\href {\doibase 10.1016/0370-1573(92)90073-9} {\bibfield
  {journal} {\bibinfo  {journal} {Phys. Rept.}\ }\textbf {\bibinfo {volume}
  {214}},\ \bibinfo {pages} {223} (\bibinfo {year} {1992})}\BibitemShut
  {NoStop}%
\bibitem [{\citenamefont {Stein-Schabes}(1987)}]{SteinSchabes:1986sy}%
  \BibitemOpen
  \bibfield  {author} {\bibinfo {author} {\bibfnamefont {J.~A.}\ \bibnamefont
  {Stein-Schabes}},\ }\href {\doibase 10.1103/PhysRevD.35.2345} {\bibfield
  {journal} {\bibinfo  {journal} {Phys. Rev.}\ }\textbf {\bibinfo {volume}
  {D35}},\ \bibinfo {pages} {2345} (\bibinfo {year} {1987})}\BibitemShut
  {NoStop}%
\bibitem [{\citenamefont {Calzetta}\ and\ \citenamefont
  {Sakellariadou}(1992)}]{Calzetta:1992gv}%
  \BibitemOpen
  \bibfield  {author} {\bibinfo {author} {\bibfnamefont {E.}~\bibnamefont
  {Calzetta}}\ and\ \bibinfo {author} {\bibfnamefont {M.}~\bibnamefont
  {Sakellariadou}},\ }\href {\doibase 10.1103/PhysRevD.45.2802} {\bibfield
  {journal} {\bibinfo  {journal} {Phys. Rev.}\ }\textbf {\bibinfo {volume}
  {D45}},\ \bibinfo {pages} {2802} (\bibinfo {year} {1992})}\BibitemShut
  {NoStop}%
\bibitem [{\citenamefont {Perez}\ and\ \citenamefont
  {Pinto-Neto}(2011)}]{Perez:2012pn}%
  \BibitemOpen
  \bibfield  {author} {\bibinfo {author} {\bibfnamefont {R.~S.}\ \bibnamefont
  {Perez}}\ and\ \bibinfo {author} {\bibfnamefont {N.}~\bibnamefont
  {Pinto-Neto}},\ }\href {\doibase 10.1134/S0202289311020174} {\bibfield
  {journal} {\bibinfo  {journal} {Grav. Cosmol.}\ }\textbf {\bibinfo {volume}
  {17}},\ \bibinfo {pages} {136} (\bibinfo {year} {2011})},\ \Eprint
  {http://arxiv.org/abs/1205.3790} {arXiv:1205.3790 [gr-qc]} \BibitemShut
  {NoStop}%
\bibitem [{\citenamefont {Goldwirth}\ and\ \citenamefont
  {Piran}(1989)}]{Goldwirth:1989vz}%
  \BibitemOpen
  \bibfield  {author} {\bibinfo {author} {\bibfnamefont {D.~S.}\ \bibnamefont
  {Goldwirth}}\ and\ \bibinfo {author} {\bibfnamefont {T.}~\bibnamefont
  {Piran}},\ }\href {\doibase 10.1103/PhysRevD.40.3263} {\bibfield  {journal}
  {\bibinfo  {journal} {Phys. Rev.}\ }\textbf {\bibinfo {volume} {D40}},\
  \bibinfo {pages} {3263} (\bibinfo {year} {1989})}\BibitemShut {NoStop}%
\bibitem [{\citenamefont {Goldwirth}(1991)}]{Goldwirth:1990pm}%
  \BibitemOpen
  \bibfield  {author} {\bibinfo {author} {\bibfnamefont {D.~S.}\ \bibnamefont
  {Goldwirth}},\ }\href {\doibase 10.1103/PhysRevD.43.3204} {\bibfield
  {journal} {\bibinfo  {journal} {Phys. Rev.}\ }\textbf {\bibinfo {volume}
  {D43}},\ \bibinfo {pages} {3204} (\bibinfo {year} {1991})}\BibitemShut
  {NoStop}%
\bibitem [{\citenamefont {Albrecht}\ \emph {et~al.}(1985)\citenamefont
  {Albrecht}, \citenamefont {Brandenberger},\ and\ \citenamefont
  {Matzner}}]{Albrecht:1985yf}%
  \BibitemOpen
  \bibfield  {author} {\bibinfo {author} {\bibfnamefont {A.}~\bibnamefont
  {Albrecht}}, \bibinfo {author} {\bibfnamefont {R.~H.}\ \bibnamefont
  {Brandenberger}}, \ and\ \bibinfo {author} {\bibfnamefont {R.}~\bibnamefont
  {Matzner}},\ }\href {\doibase 10.1103/PhysRevD.32.1280} {\bibfield  {journal}
  {\bibinfo  {journal} {Phys. Rev.}\ }\textbf {\bibinfo {volume} {D32}},\
  \bibinfo {pages} {1280} (\bibinfo {year} {1985})}\BibitemShut {NoStop}%
\bibitem [{\citenamefont {Kung}\ and\ \citenamefont
  {Brandenberger}(1989)}]{Kung:1989xz}%
  \BibitemOpen
  \bibfield  {author} {\bibinfo {author} {\bibfnamefont {J.~H.}\ \bibnamefont
  {Kung}}\ and\ \bibinfo {author} {\bibfnamefont {R.~H.}\ \bibnamefont
  {Brandenberger}},\ }\href {\doibase 10.1103/PhysRevD.40.2532} {\bibfield
  {journal} {\bibinfo  {journal} {Phys. Rev.}\ }\textbf {\bibinfo {volume}
  {D40}},\ \bibinfo {pages} {2532} (\bibinfo {year} {1989})}\BibitemShut
  {NoStop}%
\bibitem [{\citenamefont {Brandenberger}\ and\ \citenamefont
  {Kung}(1990)}]{Brandenberger:1990wu}%
  \BibitemOpen
  \bibfield  {author} {\bibinfo {author} {\bibfnamefont {R.~H.}\ \bibnamefont
  {Brandenberger}}\ and\ \bibinfo {author} {\bibfnamefont {J.~H.}\ \bibnamefont
  {Kung}},\ }\href {\doibase 10.1103/PhysRevD.42.1008} {\bibfield  {journal}
  {\bibinfo  {journal} {Phys. Rev.}\ }\textbf {\bibinfo {volume} {D42}},\
  \bibinfo {pages} {1008} (\bibinfo {year} {1990})}\BibitemShut {NoStop}%
\bibitem [{\citenamefont {Kurki-Suonio}\ \emph {et~al.}(1987)\citenamefont
  {Kurki-Suonio}, \citenamefont {Matzner}, \citenamefont {Centrella},\ and\
  \citenamefont {Wilson}}]{KurkiSuonio:1987pq}%
  \BibitemOpen
  \bibfield  {author} {\bibinfo {author} {\bibfnamefont {H.}~\bibnamefont
  {Kurki-Suonio}}, \bibinfo {author} {\bibfnamefont {R.~A.}\ \bibnamefont
  {Matzner}}, \bibinfo {author} {\bibfnamefont {J.}~\bibnamefont {Centrella}},
  \ and\ \bibinfo {author} {\bibfnamefont {J.~R.}\ \bibnamefont {Wilson}},\
  }\href {\doibase 10.1103/PhysRevD.35.435} {\bibfield  {journal} {\bibinfo
  {journal} {Phys. Rev.}\ }\textbf {\bibinfo {volume} {D35}},\ \bibinfo {pages}
  {435} (\bibinfo {year} {1987})}\BibitemShut {NoStop}%
\bibitem [{\citenamefont {Laguna}\ \emph {et~al.}(1991)\citenamefont {Laguna},
  \citenamefont {Kurki-Suonio},\ and\ \citenamefont {Matzner}}]{Laguna:1991zs}%
  \BibitemOpen
  \bibfield  {author} {\bibinfo {author} {\bibfnamefont {P.}~\bibnamefont
  {Laguna}}, \bibinfo {author} {\bibfnamefont {H.}~\bibnamefont
  {Kurki-Suonio}}, \ and\ \bibinfo {author} {\bibfnamefont {R.~A.}\
  \bibnamefont {Matzner}},\ }\href {\doibase 10.1103/PhysRevD.44.3077}
  {\bibfield  {journal} {\bibinfo  {journal} {Phys. Rev.}\ }\textbf {\bibinfo
  {volume} {D44}},\ \bibinfo {pages} {3077} (\bibinfo {year}
  {1991})}\BibitemShut {NoStop}%
\bibitem [{\citenamefont {Kurki-Suonio}\ \emph {et~al.}(1993)\citenamefont
  {Kurki-Suonio}, \citenamefont {Laguna},\ and\ \citenamefont
  {Matzner}}]{KurkiSuonio:1993fg}%
  \BibitemOpen
  \bibfield  {author} {\bibinfo {author} {\bibfnamefont {H.}~\bibnamefont
  {Kurki-Suonio}}, \bibinfo {author} {\bibfnamefont {P.}~\bibnamefont
  {Laguna}}, \ and\ \bibinfo {author} {\bibfnamefont {R.~A.}\ \bibnamefont
  {Matzner}},\ }\href {\doibase 10.1103/PhysRevD.48.3611} {\bibfield  {journal}
  {\bibinfo  {journal} {Phys. Rev.}\ }\textbf {\bibinfo {volume} {D48}},\
  \bibinfo {pages} {3611} (\bibinfo {year} {1993})},\ \Eprint
  {http://arxiv.org/abs/astro-ph/9306009} {arXiv:astro-ph/9306009 [astro-ph]}
  \BibitemShut {NoStop}%
\bibitem [{\citenamefont {East}\ \emph {et~al.}(2016)\citenamefont {East},
  \citenamefont {Kleban}, \citenamefont {Linde},\ and\ \citenamefont
  {Senatore}}]{East:2015ggf}%
  \BibitemOpen
  \bibfield  {author} {\bibinfo {author} {\bibfnamefont {W.~E.}\ \bibnamefont
  {East}}, \bibinfo {author} {\bibfnamefont {M.}~\bibnamefont {Kleban}},
  \bibinfo {author} {\bibfnamefont {A.}~\bibnamefont {Linde}}, \ and\ \bibinfo
  {author} {\bibfnamefont {L.}~\bibnamefont {Senatore}},\ }\href {\doibase
  10.1088/1475-7516/2016/09/010} {\bibfield  {journal} {\bibinfo  {journal}
  {JCAP}\ }\textbf {\bibinfo {volume} {1609}},\ \bibinfo {pages} {010}
  (\bibinfo {year} {2016})},\ \Eprint {http://arxiv.org/abs/1511.05143}
  {arXiv:1511.05143 [hep-th]} \BibitemShut {NoStop}%
\bibitem [{\citenamefont {Clough}\ \emph {et~al.}(2017)\citenamefont {Clough},
  \citenamefont {Lim}, \citenamefont {DiNunno}, \citenamefont {Fischler},
  \citenamefont {Flauger},\ and\ \citenamefont {Paban}}]{Clough:2016ymm}%
  \BibitemOpen
  \bibfield  {author} {\bibinfo {author} {\bibfnamefont {K.}~\bibnamefont
  {Clough}}, \bibinfo {author} {\bibfnamefont {E.~A.}\ \bibnamefont {Lim}},
  \bibinfo {author} {\bibfnamefont {B.~S.}\ \bibnamefont {DiNunno}}, \bibinfo
  {author} {\bibfnamefont {W.}~\bibnamefont {Fischler}}, \bibinfo {author}
  {\bibfnamefont {R.}~\bibnamefont {Flauger}}, \ and\ \bibinfo {author}
  {\bibfnamefont {S.}~\bibnamefont {Paban}},\ }\href {\doibase
  10.1088/1475-7516/2017/09/025} {\bibfield  {journal} {\bibinfo  {journal}
  {JCAP}\ }\textbf {\bibinfo {volume} {1709}},\ \bibinfo {pages} {025}
  (\bibinfo {year} {2017})},\ \Eprint {http://arxiv.org/abs/1608.04408}
  {arXiv:1608.04408 [hep-th]} \BibitemShut {NoStop}%
\bibitem [{\citenamefont {Martin}\ and\ \citenamefont
  {Brandenberger}(2001)}]{Martin:2000xs}%
  \BibitemOpen
  \bibfield  {author} {\bibinfo {author} {\bibfnamefont {J.}~\bibnamefont
  {Martin}}\ and\ \bibinfo {author} {\bibfnamefont {R.~H.}\ \bibnamefont
  {Brandenberger}},\ }\href {\doibase 10.1103/PhysRevD.63.123501} {\bibfield
  {journal} {\bibinfo  {journal} {Phys. Rev.}\ }\textbf {\bibinfo {volume}
  {D63}},\ \bibinfo {pages} {123501} (\bibinfo {year} {2001})},\ \Eprint
  {http://arxiv.org/abs/hep-th/0005209} {arXiv:hep-th/0005209 [hep-th]}
  \BibitemShut {NoStop}%
\bibitem [{\citenamefont {Brandenberger}\ and\ \citenamefont
  {Martin}(2001)}]{Brandenberger:2000wr}%
  \BibitemOpen
  \bibfield  {author} {\bibinfo {author} {\bibfnamefont {R.~H.}\ \bibnamefont
  {Brandenberger}}\ and\ \bibinfo {author} {\bibfnamefont {J.}~\bibnamefont
  {Martin}},\ }\href {\doibase 10.1142/S0217732301004170} {\bibfield  {journal}
  {\bibinfo  {journal} {Mod. Phys. Lett.}\ }\textbf {\bibinfo {volume} {A16}},\
  \bibinfo {pages} {999} (\bibinfo {year} {2001})},\ \Eprint
  {http://arxiv.org/abs/astro-ph/0005432} {arXiv:astro-ph/0005432 [astro-ph]}
  \BibitemShut {NoStop}%
\bibitem [{\citenamefont {Martin}\ and\ \citenamefont
  {Brandenberger}(2003)}]{Martin:2003kp}%
  \BibitemOpen
  \bibfield  {author} {\bibinfo {author} {\bibfnamefont {J.}~\bibnamefont
  {Martin}}\ and\ \bibinfo {author} {\bibfnamefont {R.}~\bibnamefont
  {Brandenberger}},\ }\href {\doibase 10.1103/PhysRevD.68.063513} {\bibfield
  {journal} {\bibinfo  {journal} {Phys. Rev.}\ }\textbf {\bibinfo {volume}
  {D68}},\ \bibinfo {pages} {063513} (\bibinfo {year} {2003})},\ \Eprint
  {http://arxiv.org/abs/hep-th/0305161} {arXiv:hep-th/0305161 [hep-th]}
  \BibitemShut {NoStop}%
\bibitem [{\citenamefont {Brandenberger}\ and\ \citenamefont
  {Martin}(2002)}]{Brandenberger:2002hs}%
  \BibitemOpen
  \bibfield  {author} {\bibinfo {author} {\bibfnamefont {R.~H.}\ \bibnamefont
  {Brandenberger}}\ and\ \bibinfo {author} {\bibfnamefont {J.}~\bibnamefont
  {Martin}},\ }\href {\doibase 10.1142/S0217751X02010765} {\bibfield  {journal}
  {\bibinfo  {journal} {Int. J. Mod. Phys.}\ }\textbf {\bibinfo {volume}
  {A17}},\ \bibinfo {pages} {3663} (\bibinfo {year} {2002})},\ \Eprint
  {http://arxiv.org/abs/hep-th/0202142} {arXiv:hep-th/0202142 [hep-th]}
  \BibitemShut {NoStop}%
\bibitem [{\citenamefont {Brandenberger}\ and\ \citenamefont
  {Martin}(2013)}]{Brandenberger:2012aj}%
  \BibitemOpen
  \bibfield  {author} {\bibinfo {author} {\bibfnamefont {R.~H.}\ \bibnamefont
  {Brandenberger}}\ and\ \bibinfo {author} {\bibfnamefont {J.}~\bibnamefont
  {Martin}},\ }\href {\doibase 10.1088/0264-9381/30/11/113001} {\bibfield
  {journal} {\bibinfo  {journal} {Class. Quant. Grav.}\ }\textbf {\bibinfo
  {volume} {30}},\ \bibinfo {pages} {113001} (\bibinfo {year} {2013})},\
  \Eprint {http://arxiv.org/abs/1211.6753} {arXiv:1211.6753 [astro-ph.CO]}
  \BibitemShut {NoStop}%
\bibitem [{\citenamefont {Brandenberger}\ and\ \citenamefont
  {Martin}(2005)}]{Brandenberger:2004kx}%
  \BibitemOpen
  \bibfield  {author} {\bibinfo {author} {\bibfnamefont {R.~H.}\ \bibnamefont
  {Brandenberger}}\ and\ \bibinfo {author} {\bibfnamefont {J.}~\bibnamefont
  {Martin}},\ }\href {\doibase 10.1103/PhysRevD.71.023504} {\bibfield
  {journal} {\bibinfo  {journal} {Phys. Rev.}\ }\textbf {\bibinfo {volume}
  {D71}},\ \bibinfo {pages} {023504} (\bibinfo {year} {2005})},\ \Eprint
  {http://arxiv.org/abs/hep-th/0410223} {arXiv:hep-th/0410223 [hep-th]}
  \BibitemShut {NoStop}%
\bibitem [{\citenamefont {Lemoine}\ \emph {et~al.}(2002)\citenamefont
  {Lemoine}, \citenamefont {Lubo}, \citenamefont {Martin},\ and\ \citenamefont
  {Uzan}}]{Lemoine:2001ar}%
  \BibitemOpen
  \bibfield  {author} {\bibinfo {author} {\bibfnamefont {M.}~\bibnamefont
  {Lemoine}}, \bibinfo {author} {\bibfnamefont {M.}~\bibnamefont {Lubo}},
  \bibinfo {author} {\bibfnamefont {J.}~\bibnamefont {Martin}}, \ and\ \bibinfo
  {author} {\bibfnamefont {J.-P.}\ \bibnamefont {Uzan}},\ }\href {\doibase
  10.1103/PhysRevD.65.023510} {\bibfield  {journal} {\bibinfo  {journal} {Phys.
  Rev.}\ }\textbf {\bibinfo {volume} {D65}},\ \bibinfo {pages} {023510}
  (\bibinfo {year} {2002})},\ \Eprint {http://arxiv.org/abs/hep-th/0109128}
  {arXiv:hep-th/0109128 [hep-th]} \BibitemShut {NoStop}%
\bibitem [{\citenamefont {Hassan}\ and\ \citenamefont
  {Sloth}(2003)}]{Hassan:2002qk}%
  \BibitemOpen
  \bibfield  {author} {\bibinfo {author} {\bibfnamefont {S.~F.}\ \bibnamefont
  {Hassan}}\ and\ \bibinfo {author} {\bibfnamefont {M.~S.}\ \bibnamefont
  {Sloth}},\ }\href {\doibase 10.1016/j.nuclphysb.2003.09.041} {\bibfield
  {journal} {\bibinfo  {journal} {Nucl. Phys.}\ }\textbf {\bibinfo {volume}
  {B674}},\ \bibinfo {pages} {434} (\bibinfo {year} {2003})},\ \Eprint
  {http://arxiv.org/abs/hep-th/0204110} {arXiv:hep-th/0204110 [hep-th]}
  \BibitemShut {NoStop}%
\bibitem [{\citenamefont {Martin}\ and\ \citenamefont
  {Ringeval}(2004{\natexlab{a}})}]{Martin:2003sg}%
  \BibitemOpen
  \bibfield  {author} {\bibinfo {author} {\bibfnamefont {J.}~\bibnamefont
  {Martin}}\ and\ \bibinfo {author} {\bibfnamefont {C.}~\bibnamefont
  {Ringeval}},\ }\href {\doibase 10.1103/PhysRevD.69.083515} {\bibfield
  {journal} {\bibinfo  {journal} {Phys. Rev.}\ }\textbf {\bibinfo {volume}
  {D69}},\ \bibinfo {pages} {083515} (\bibinfo {year} {2004}{\natexlab{a}})},\
  \Eprint {http://arxiv.org/abs/astro-ph/0310382} {arXiv:astro-ph/0310382
  [astro-ph]} \BibitemShut {NoStop}%
\bibitem [{\citenamefont {Martin}\ and\ \citenamefont
  {Ringeval}(2005)}]{Martin:2004yi}%
  \BibitemOpen
  \bibfield  {author} {\bibinfo {author} {\bibfnamefont {J.}~\bibnamefont
  {Martin}}\ and\ \bibinfo {author} {\bibfnamefont {C.}~\bibnamefont
  {Ringeval}},\ }\href {\doibase 10.1088/1475-7516/2005/01/007} {\bibfield
  {journal} {\bibinfo  {journal} {JCAP}\ }\textbf {\bibinfo {volume} {0501}},\
  \bibinfo {pages} {007} (\bibinfo {year} {2005})},\ \Eprint
  {http://arxiv.org/abs/hep-ph/0405249} {arXiv:hep-ph/0405249 [hep-ph]}
  \BibitemShut {NoStop}%
\bibitem [{\citenamefont {Martin}\ and\ \citenamefont
  {Ringeval}(2004{\natexlab{b}})}]{Martin:2004iv}%
  \BibitemOpen
  \bibfield  {author} {\bibinfo {author} {\bibfnamefont {J.}~\bibnamefont
  {Martin}}\ and\ \bibinfo {author} {\bibfnamefont {C.}~\bibnamefont
  {Ringeval}},\ }\href {\doibase 10.1103/PhysRevD.69.127303} {\bibfield
  {journal} {\bibinfo  {journal} {Phys. Rev.}\ }\textbf {\bibinfo {volume}
  {D69}},\ \bibinfo {pages} {127303} (\bibinfo {year} {2004}{\natexlab{b}})},\
  \Eprint {http://arxiv.org/abs/astro-ph/0402609} {arXiv:astro-ph/0402609
  [astro-ph]} \BibitemShut {NoStop}%
\bibitem [{\citenamefont {Vennin}\ and\ \citenamefont
  {Starobinsky}(2015)}]{Vennin:2015hra}%
  \BibitemOpen
  \bibfield  {author} {\bibinfo {author} {\bibfnamefont {V.}~\bibnamefont
  {Vennin}}\ and\ \bibinfo {author} {\bibfnamefont {A.~A.}\ \bibnamefont
  {Starobinsky}},\ }\href {\doibase 10.1140/epjc/s10052-015-3643-y} {\bibfield
  {journal} {\bibinfo  {journal} {Eur. Phys. J.}\ }\textbf {\bibinfo {volume}
  {C75}},\ \bibinfo {pages} {413} (\bibinfo {year} {2015})},\ \Eprint
  {http://arxiv.org/abs/1506.04732} {arXiv:1506.04732 [hep-th]} \BibitemShut
  {NoStop}%
\bibitem [{\citenamefont {Starobinsky}(1986)}]{Starobinsky:1986fx}%
  \BibitemOpen
  \bibfield  {author} {\bibinfo {author} {\bibfnamefont {A.~A.}\ \bibnamefont
  {Starobinsky}},\ }\bibfield  {booktitle} {\emph {\bibinfo {booktitle} {{In
  *De Vega, H.j. ( Ed.), Sanchez, N. ( Ed.): Field Theory, Quantum Gravity and
  Strings*, 107-126. Proceedings of a Seminar Series Held at DAPHE,
  Observatoire de Meudon, and LPTHE, Université Pierre et Marie Curie, Paris,
  Between October 1984 and October 1985.}}},\ }\href {\doibase
  10.1007/3-540-16452-9_6} {\bibfield  {journal} {\bibinfo  {journal} {Lect.
  Notes Phys.}\ }\textbf {\bibinfo {volume} {246}},\ \bibinfo {pages} {107}
  (\bibinfo {year} {1986})}\BibitemShut {NoStop}%
\bibitem [{\citenamefont {Starobinsky}\ and\ \citenamefont
  {Yokoyama}(1994)}]{Starobinsky:1994bd}%
  \BibitemOpen
  \bibfield  {author} {\bibinfo {author} {\bibfnamefont {A.~A.}\ \bibnamefont
  {Starobinsky}}\ and\ \bibinfo {author} {\bibfnamefont {J.}~\bibnamefont
  {Yokoyama}},\ }\href {\doibase 10.1103/PhysRevD.50.6357} {\bibfield
  {journal} {\bibinfo  {journal} {Phys. Rev.}\ }\textbf {\bibinfo {volume}
  {D50}},\ \bibinfo {pages} {6357} (\bibinfo {year} {1994})},\ \Eprint
  {http://arxiv.org/abs/astro-ph/9407016} {arXiv:astro-ph/9407016 [astro-ph]}
  \BibitemShut {NoStop}%
\bibitem [{\citenamefont {Martin}\ and\ \citenamefont
  {Musso}(2006{\natexlab{a}})}]{Martin:2005hb}%
  \BibitemOpen
  \bibfield  {author} {\bibinfo {author} {\bibfnamefont {J.}~\bibnamefont
  {Martin}}\ and\ \bibinfo {author} {\bibfnamefont {M.}~\bibnamefont {Musso}},\
  }\href {\doibase 10.1103/PhysRevD.73.043517} {\bibfield  {journal} {\bibinfo
  {journal} {Phys. Rev.}\ }\textbf {\bibinfo {volume} {D73}},\ \bibinfo {pages}
  {043517} (\bibinfo {year} {2006}{\natexlab{a}})},\ \Eprint
  {http://arxiv.org/abs/hep-th/0511292} {arXiv:hep-th/0511292 [hep-th]}
  \BibitemShut {NoStop}%
\bibitem [{\citenamefont {Martin}\ and\ \citenamefont
  {Musso}(2006{\natexlab{b}})}]{Martin:2005ir}%
  \BibitemOpen
  \bibfield  {author} {\bibinfo {author} {\bibfnamefont {J.}~\bibnamefont
  {Martin}}\ and\ \bibinfo {author} {\bibfnamefont {M.}~\bibnamefont {Musso}},\
  }\href {\doibase 10.1103/PhysRevD.73.043516} {\bibfield  {journal} {\bibinfo
  {journal} {Phys. Rev.}\ }\textbf {\bibinfo {volume} {D73}},\ \bibinfo {pages}
  {043516} (\bibinfo {year} {2006}{\natexlab{b}})},\ \Eprint
  {http://arxiv.org/abs/hep-th/0511214} {arXiv:hep-th/0511214 [hep-th]}
  \BibitemShut {NoStop}%
\bibitem [{\citenamefont {Lorenz}\ \emph {et~al.}(2010)\citenamefont {Lorenz},
  \citenamefont {Martin},\ and\ \citenamefont {Yokoyama}}]{Lorenz:2010vf}%
  \BibitemOpen
  \bibfield  {author} {\bibinfo {author} {\bibfnamefont {L.}~\bibnamefont
  {Lorenz}}, \bibinfo {author} {\bibfnamefont {J.}~\bibnamefont {Martin}}, \
  and\ \bibinfo {author} {\bibfnamefont {J.}~\bibnamefont {Yokoyama}},\ }\href
  {\doibase 10.1103/PhysRevD.82.023515} {\bibfield  {journal} {\bibinfo
  {journal} {Phys. Rev.}\ }\textbf {\bibinfo {volume} {D82}},\ \bibinfo {pages}
  {023515} (\bibinfo {year} {2010})},\ \Eprint {http://arxiv.org/abs/1004.3734}
  {arXiv:1004.3734 [hep-th]} \BibitemShut {NoStop}%
\bibitem [{\citenamefont {Martin}\ and\ \citenamefont
  {Vennin}(2012)}]{Martin:2011ib}%
  \BibitemOpen
  \bibfield  {author} {\bibinfo {author} {\bibfnamefont {J.}~\bibnamefont
  {Martin}}\ and\ \bibinfo {author} {\bibfnamefont {V.}~\bibnamefont
  {Vennin}},\ }\href {\doibase 10.1103/PhysRevD.85.043525} {\bibfield
  {journal} {\bibinfo  {journal} {Phys. Rev.}\ }\textbf {\bibinfo {volume}
  {D85}},\ \bibinfo {pages} {043525} (\bibinfo {year} {2012})},\ \Eprint
  {http://arxiv.org/abs/1110.2070} {arXiv:1110.2070 [astro-ph.CO]} \BibitemShut
  {NoStop}%
\bibitem [{\citenamefont {Assadullahi}\ \emph {et~al.}(2016)\citenamefont
  {Assadullahi}, \citenamefont {Firouzjahi}, \citenamefont {Noorbala},
  \citenamefont {Vennin},\ and\ \citenamefont {Wands}}]{Assadullahi:2016gkk}%
  \BibitemOpen
  \bibfield  {author} {\bibinfo {author} {\bibfnamefont {H.}~\bibnamefont
  {Assadullahi}}, \bibinfo {author} {\bibfnamefont {H.}~\bibnamefont
  {Firouzjahi}}, \bibinfo {author} {\bibfnamefont {M.}~\bibnamefont
  {Noorbala}}, \bibinfo {author} {\bibfnamefont {V.}~\bibnamefont {Vennin}}, \
  and\ \bibinfo {author} {\bibfnamefont {D.}~\bibnamefont {Wands}},\ }\href
  {\doibase 10.1088/1475-7516/2016/06/043} {\bibfield  {journal} {\bibinfo
  {journal} {JCAP}\ }\textbf {\bibinfo {volume} {1606}},\ \bibinfo {pages}
  {043} (\bibinfo {year} {2016})},\ \Eprint {http://arxiv.org/abs/1604.04502}
  {arXiv:1604.04502 [hep-th]} \BibitemShut {NoStop}%
\bibitem [{\citenamefont {Polarski}\ and\ \citenamefont
  {Starobinsky}(1996)}]{Polarski:1995jg}%
  \BibitemOpen
  \bibfield  {author} {\bibinfo {author} {\bibfnamefont {D.}~\bibnamefont
  {Polarski}}\ and\ \bibinfo {author} {\bibfnamefont {A.~A.}\ \bibnamefont
  {Starobinsky}},\ }\href {\doibase 10.1088/0264-9381/13/3/006} {\bibfield
  {journal} {\bibinfo  {journal} {Class. Quant. Grav.}\ }\textbf {\bibinfo
  {volume} {13}},\ \bibinfo {pages} {377} (\bibinfo {year} {1996})},\ \Eprint
  {http://arxiv.org/abs/gr-qc/9504030} {arXiv:gr-qc/9504030 [gr-qc]}
  \BibitemShut {NoStop}%
\bibitem [{\citenamefont {Martin}\ and\ \citenamefont
  {Vennin}(2016)}]{Martin:2015qta}%
  \BibitemOpen
  \bibfield  {author} {\bibinfo {author} {\bibfnamefont {J.}~\bibnamefont
  {Martin}}\ and\ \bibinfo {author} {\bibfnamefont {V.}~\bibnamefont
  {Vennin}},\ }\href {\doibase 10.1103/PhysRevD.93.023505} {\bibfield
  {journal} {\bibinfo  {journal} {Phys. Rev.}\ }\textbf {\bibinfo {volume}
  {D93}},\ \bibinfo {pages} {023505} (\bibinfo {year} {2016})},\ \Eprint
  {http://arxiv.org/abs/1510.04038} {arXiv:1510.04038 [astro-ph.CO]}
  \BibitemShut {NoStop}%
\bibitem [{\citenamefont {Mukhanov}(2015)}]{Mukhanov:2014uwa}%
  \BibitemOpen
  \bibfield  {author} {\bibinfo {author} {\bibfnamefont {V.}~\bibnamefont
  {Mukhanov}},\ }\href {\doibase 10.1002/prop.201400074} {\bibfield  {journal}
  {\bibinfo  {journal} {Fortsch. Phys.}\ }\textbf {\bibinfo {volume} {63}},\
  \bibinfo {pages} {36} (\bibinfo {year} {2015})},\ \Eprint
  {http://arxiv.org/abs/1409.2335} {arXiv:1409.2335 [astro-ph.CO]} \BibitemShut
  {NoStop}%
\bibitem [{\citenamefont {Winitzki}(2001)}]{Winitzki:2001fc}%
  \BibitemOpen
  \bibfield  {author} {\bibinfo {author} {\bibfnamefont {S.}~\bibnamefont
  {Winitzki}},\ }in\ \href@noop {} {\emph {\bibinfo {booktitle} {{5th
  International Conference on Particle Physics and the Early Universe (COSMO
  2001) Rovaniemi, Finland, August 30-September 4, 2001}}}}\ (\bibinfo {year}
  {2001})\ \Eprint {http://arxiv.org/abs/gr-qc/0111109} {arXiv:gr-qc/0111109
  [gr-qc]} \BibitemShut {NoStop}%
\bibitem [{\citenamefont {Vachaspati}(2003)}]{Vachaspati:2003de}%
  \BibitemOpen
  \bibfield  {author} {\bibinfo {author} {\bibfnamefont {T.}~\bibnamefont
  {Vachaspati}},\ }in\ \href@noop {} {\emph {\bibinfo {booktitle} {{Cosmic
  inflation. Proceedings, Meeting, Davis, USA, March 22-25, 2003}}}}\ (\bibinfo
  {year} {2003})\ \Eprint {http://arxiv.org/abs/astro-ph/0305439}
  {arXiv:astro-ph/0305439 [astro-ph]} \BibitemShut {NoStop}%
\bibitem [{\citenamefont {Guth}\ \emph {et~al.}()\citenamefont {Guth},
  \citenamefont {Vachaspati},\ and\ \citenamefont {Winitzki}}]{GVW}%
  \BibitemOpen
  \bibfield  {author} {\bibinfo {author} {\bibfnamefont {A.}~\bibnamefont
  {Guth}}, \bibinfo {author} {\bibfnamefont {T.}~\bibnamefont {Vachaspati}}, \
  and\ \bibinfo {author} {\bibfnamefont {S.}~\bibnamefont {Winitzki}},\ }\href
  {\doibase
  https://pdfs.semanticscholar.org/59aa/c67aaa1873ff0860a879ffc94814789ec813.pdf}
  {\
  https://pdfs.semanticscholar.org/59aa/c67aaa1873ff0860a879ffc94814789ec813.pdf}\BibitemShut
  {NoStop}%
\bibitem [{\citenamefont {Coule}\ and\ \citenamefont
  {Martin}(2000)}]{Coule:1999wg}%
  \BibitemOpen
  \bibfield  {author} {\bibinfo {author} {\bibfnamefont {D.~H.}\ \bibnamefont
  {Coule}}\ and\ \bibinfo {author} {\bibfnamefont {J.}~\bibnamefont {Martin}},\
  }\href {\doibase 10.1103/PhysRevD.61.063501} {\bibfield  {journal} {\bibinfo
  {journal} {Phys. Rev.}\ }\textbf {\bibinfo {volume} {D61}},\ \bibinfo {pages}
  {063501} (\bibinfo {year} {2000})},\ \Eprint
  {http://arxiv.org/abs/gr-qc/9905056} {arXiv:gr-qc/9905056 [gr-qc]}
  \BibitemShut {NoStop}%
\bibitem [{\citenamefont {Gross}(2005)}]{Gross:2005qm}%
  \BibitemOpen
  \bibfield  {author} {\bibinfo {author} {\bibfnamefont {D.~J.}\ \bibnamefont
  {Gross}},\ }\bibfield  {booktitle} {\emph {\bibinfo {booktitle}
  {{Proceedings, Nobel Symposium 127 on String theory and cosmology: Sigtuna,
  Sweden, August 14-19, 2003}}},\ }\href {\doibase
  10.1238/Physica.Topical.117a00102} {\bibfield  {journal} {\bibinfo  {journal}
  {Phys. Scripta}\ }\textbf {\bibinfo {volume} {T117}},\ \bibinfo {pages} {102}
  (\bibinfo {year} {2005})}\BibitemShut {NoStop}%
\bibitem [{\citenamefont {Gangui}\ \emph {et~al.}(1996)\citenamefont {Gangui},
  \citenamefont {Durrer},\ and\ \citenamefont {Sakellariadou}}]{Gangui:1995vp}%
  \BibitemOpen
  \bibfield  {author} {\bibinfo {author} {\bibfnamefont {A.}~\bibnamefont
  {Gangui}}, \bibinfo {author} {\bibfnamefont {R.}~\bibnamefont {Durrer}}, \
  and\ \bibinfo {author} {\bibfnamefont {M.}~\bibnamefont {Sakellariadou}},\
  }\bibfield  {booktitle} {\emph {\bibinfo {booktitle} {{Proceedings,
  Conference on Mapping, Measuring and Modelling the Universe: Valencia, Spain,
  September 18-22, 1995}}},\ }\href@noop {} {\bibfield  {journal} {\bibinfo
  {journal} {ASP Conf. Ser.}\ }\textbf {\bibinfo {volume} {94}},\ \bibinfo
  {pages} {335} (\bibinfo {year} {1996})},\ \Eprint
  {http://arxiv.org/abs/astro-ph/9602018} {arXiv:astro-ph/9602018 [astro-ph]}
  \BibitemShut {NoStop}%
\bibitem [{\citenamefont {Durrer}\ \emph {et~al.}(1996)\citenamefont {Durrer},
  \citenamefont {Gangui},\ and\ \citenamefont {Sakellariadou}}]{Durrer:1995ni}%
  \BibitemOpen
  \bibfield  {author} {\bibinfo {author} {\bibfnamefont {R.}~\bibnamefont
  {Durrer}}, \bibinfo {author} {\bibfnamefont {A.}~\bibnamefont {Gangui}}, \
  and\ \bibinfo {author} {\bibfnamefont {M.}~\bibnamefont {Sakellariadou}},\
  }\href {\doibase 10.1103/PhysRevLett.76.579} {\bibfield  {journal} {\bibinfo
  {journal} {Phys. Rev. Lett.}\ }\textbf {\bibinfo {volume} {76}},\ \bibinfo
  {pages} {579} (\bibinfo {year} {1996})},\ \Eprint
  {http://arxiv.org/abs/astro-ph/9507035} {arXiv:astro-ph/9507035 [astro-ph]}
  \BibitemShut {NoStop}%
\bibitem [{\citenamefont {Ringeval}\ \emph {et~al.}(2007)\citenamefont
  {Ringeval}, \citenamefont {Sakellariadou},\ and\ \citenamefont
  {Bouchet}}]{Ringeval:2005kr}%
  \BibitemOpen
  \bibfield  {author} {\bibinfo {author} {\bibfnamefont {C.}~\bibnamefont
  {Ringeval}}, \bibinfo {author} {\bibfnamefont {M.}~\bibnamefont
  {Sakellariadou}}, \ and\ \bibinfo {author} {\bibfnamefont {F.}~\bibnamefont
  {Bouchet}},\ }\href {\doibase 10.1088/1475-7516/2007/02/023} {\bibfield
  {journal} {\bibinfo  {journal} {JCAP}\ }\textbf {\bibinfo {volume} {0702}},\
  \bibinfo {pages} {023} (\bibinfo {year} {2007})},\ \Eprint
  {http://arxiv.org/abs/astro-ph/0511646} {arXiv:astro-ph/0511646 [astro-ph]}
  \BibitemShut {NoStop}%
\bibitem [{\citenamefont {Linde}\ and\ \citenamefont
  {Mezhlumian}(1995)}]{Linde:1995rv}%
  \BibitemOpen
  \bibfield  {author} {\bibinfo {author} {\bibfnamefont {A.~D.}\ \bibnamefont
  {Linde}}\ and\ \bibinfo {author} {\bibfnamefont {A.}~\bibnamefont
  {Mezhlumian}},\ }\href {\doibase 10.1103/PhysRevD.52.6789} {\bibfield
  {journal} {\bibinfo  {journal} {Phys. Rev.}\ }\textbf {\bibinfo {volume}
  {D52}},\ \bibinfo {pages} {6789} (\bibinfo {year} {1995})},\ \Eprint
  {http://arxiv.org/abs/astro-ph/9506017} {arXiv:astro-ph/9506017 [astro-ph]}
  \BibitemShut {NoStop}%
\bibitem [{\citenamefont {Starobinsky}(2005)}]{Starobinsky:2005ab}%
  \BibitemOpen
  \bibfield  {author} {\bibinfo {author} {\bibfnamefont {A.~A.}\ \bibnamefont
  {Starobinsky}},\ }\href {\doibase 10.1134/1.2121807} {\bibfield  {journal}
  {\bibinfo  {journal} {JETP Lett.}\ }\textbf {\bibinfo {volume} {82}},\
  \bibinfo {pages} {169} (\bibinfo {year} {2005})},\ \bibinfo {note} {[Pisma
  Zh. Eksp. Teor. Fiz.82,187(2005)]},\ \Eprint
  {http://arxiv.org/abs/astro-ph/0507193} {arXiv:astro-ph/0507193 [astro-ph]}
  \BibitemShut {NoStop}%
\bibitem [{\citenamefont {Silverstein}(2017)}]{Silverstein:2017zfk}%
  \BibitemOpen
  \bibfield  {author} {\bibinfo {author} {\bibfnamefont {E.}~\bibnamefont
  {Silverstein}},\ }\href@noop {} {\  (\bibinfo {year} {2017})},\ \Eprint
  {http://arxiv.org/abs/1706.02790} {arXiv:1706.02790 [hep-th]} \BibitemShut
  {NoStop}%
\end{thebibliography}%

\end{document}